\documentstyle[12pt]{article}
\begin{document}
\begin{titlepage}
\renewcommand{\thefootnote}{\fnsymbol{footnote}}

\begin{flushright}
CERN-TH.7250/94\\
TPI-MINN-94/12-T\\
UMN-TH-1250-94\\
UND-HEP-94-BIG\hspace*{.2em}05\\
hep-ph/9405410\\
revised
\end{flushright}
\vspace{.3cm}
\begin{center} \LARGE
{\bf Sum Rules for Heavy Flavor Transitions
in the SV Limit}
\end{center}
\vspace*{.3cm}
\begin{center} {\Large
I. Bigi $^{a,b}$, M. Shifman $^c$, N.G. Uraltsev $^{a,c,d}$,
A. Vainshtein $^{c,e}$} \\
\vspace{.4cm}
{\normalsize $^a${\it TH Division, CERN, CH-1211 Geneva 23,
Switzerland}\footnote{During the academic year 1993/94}\\
$^b${\it Dept.of Physics,
Univ. of Notre Dame du
Lac, Notre Dame, IN 46556, U.S.A.}\footnote{Permanent address}\\
$^c$ {\it  Theoretical Physics Institute, Univ. of Minnesota,
Minneapolis, MN 55455}\\
$^d$ {\it St.Petersburg Nuclear Physics Institute,
Gatchina, St.Petersburg 188350, Russia}$^\dagger$\\
$^e$ {\it Budker Institute of Nuclear Physics, Novosibirsk
630090, Russia}\\
\vspace{.3cm}
e-mail addresses:\\
{\it bigi@undhep.hep.nd.edu, shifman@vx.cis.umn.edu,
vainshte@vx.cis.umn.edu}
\vspace*{.4cm}
}
{\Large{\bf Abstract}}
\end{center}

\vspace*{.2cm}
We show how
sum rules for the weak decays of heavy flavor
hadrons can be derived as the moments of spectral distributions
in the small velocity (SV) limit. This systematic approach allows us
to
determine
corrections to these sum rules, to obtain new sum rules and
it provides us with a transparent physical interpretation;  it also
opens
a new perspective on the notion of the heavy quark mass.
Applying these sum rules we derive a {\it lower} bound on the
deviation of
the exclusive form factor $F_{B\rightarrow D^*}$ from unity at zero
recoil;
likewise
we give a field-theoretical  derivation of a previously formulated
inequality
between the expectation value for the kinetic energy operator of the
heavy
quark and for the chromomagnetic operator. We analyze
how the known results on
nonperturbative corrections must be understood
when one takes into account the
normalization point dependence of the low scale parameters.
Relation between the field-theoretic derivation of the sum rules and
the quantum-mechanical approach is elucidated.

\end{titlepage}
\addtocounter{footnote}{-2}

\newpage

\tableofcontents

\newpage

\section{Introduction}

The theory of preasymptotic effects in inclusive weak decays of
heavy flavor hadrons has been developing from the early
eighties on \cite{1} -- \cite{6b} and is entering now
a rather mature stage
\cite{7} -- \cite{mannel}. In recent papers \cite{9,11,10a} it was
shown in particular
how the effects due to the motion of the heavy quarks inside the
hadrons can be incorporated in a systematic way, namely
through distribution functions which crucially depend on the ratio
$\gamma = m_q/m_Q$
where $m_Q$ and $m_q$ are the masses of the initial and final state
quarks, respectively. Among other things, it was mentioned that the
formalism developed in Ref. \cite{9} automatically ensures  the
Bjorken
sum rule \cite{12,13,12a} in the small velocity (SV) limit
\cite{14} (Sect. 4 of Ref. \cite{9}). In the present  work we discuss
in more
detail
the sum rules of Refs. \cite{14}, \cite{12}  -- \cite{13}, the so-called
optical
sum rules derived later by Voloshin \cite{voloshin} (see also
\cite{grozin,grozin1}), as well as
other similar sum rules including those
considered
by Lipkin \cite{lipkin}  within a quantum-mechanical approach.

These are the main observations of our paper:

$\bullet$
We demonstrate that these apparently isolated sum rules
represent merely different moments of observable spectral
distributions. Their physical meaning becomes absolutely
transparent
within the formulation of the  problem suggested in Ref. \cite{9}.
Actually many crucial elements are already included, implicitly and
explicitly, in Ref. \cite{9}, so our task is  to combine them. This
approach will allow us to get new sum rules and to obtain corrections
to the old ones in a systematic and comprehensive way.

$\bullet$
The sum rules are used to derive a lower bound on the deviation
of
the exclusive form factor $F_{B\rightarrow D^*}$ from unity at zero
recoil.
The lower bound is essentially determined by the average value
of the chromomagnetic operator $\mu_G^2$, familiar from
previous studies. We also obtain a new field-theoretic derivation
of the previously formulated
inequality $\mu_\pi^2> \mu_G^2$, where
\begin{equation}
\mu_G^2 =\frac{1}{2M_B}\langle B|
\bar b\,\frac{i}{2}\sigma_{\mu\nu}G^{\mu\nu}\,b|B\rangle
\;\;,\;\;\;
\mu_\pi^2 =\frac{1}{2M_B}\langle B|
\bar b\,(i\vec{D})^2 \,b|B\rangle\;\; .
\label{defme}
\end{equation}
Its validity has been questioned by some authors
under the pretext that the original line of reasoning was purely
quantum-mechanical.

$\bullet$
The analysis of the sum rules gives us an opportunity to
discuss the notion of the heavy quark mass from a point of view
complementary to  recent investigations \cite{15,beneke}. It is
shown
that the key theoretical parameter $\overline{\Lambda} (\mu )$ is
directly related
to quantities measurable in inclusive semileptonic decays of
$B$ mesons
in a certain kinematical regime. The relation obtained makes
absolutely
explicit the fact that $\overline{\Lambda}$ does {\em not} exist as a
universal
constant, as
had been previously believed. Any consistent treatment must deal
with a $\mu$ dependent function, $\overline{\Lambda} (\mu )$,
where
$\mu$
is a normalization point.

The assertions formulated above for $\overline{\Lambda} (\mu )$
are
applicable
in full to the expectation value of the kinetic energy operator
$\mu_\pi^2$.

$\bullet$
The conventional derivation of the sum rules in the SV limit based on
QCD
and the heavy quark expansion is shown to be equivalent to Lipkin's
approach
\cite{lipkin} in all aspects where the ultraviolet domain (i.e.  virtual
momenta
of order of the heavy quark masses) is unimportant. In Lipkin's
formalism the
problem is treated quantum-mechanically from the very beginning;
the
$b\rightarrow c$ transition is treated as an instantaneous
perturbation. These two formalisms are complimentary with respect
to each
other.

\vspace{0.3cm}

Generically the sum rules considered yield the moments of energy of
the final
hadronic state in the $b\rightarrow c$ transition as an expansion in
different
theoretical parameters: the perturbative expansion in $\alpha_s$, the
nonperturbative expansion in $1/m_{b,c}$, and, finally, the
expansion
in $\vec q /m_c$ where $q$ is the four-momentum carried away by
the
current (the current which induces the $b\rightarrow c$ transition
under
consideration). This latter expansion is nothing else than the SV limit.
In the point of the zero recoil $\vec q \equiv 0$. Near zero recoil
$|\vec q |/m_c$ can be treated as a small parameter and all
quantities
can be consistently expanded in this parameter. Each particular term
in this
expansion generates its own set of sum rules. We will be mostly
dealing with
the sum rules at zero recoil and those emerging in the order
 ${(\vec q /m_c)}^2$.

In the focus of our investigation are the  nonperturbative
corrections
proportional to powers of $1/m_{b,c}$. The Wilson operator product
expansion
(OPE) \cite{wilson} is the theoretical foundation for calculation of
these
corrections. The version of OPE used in the problem at hand is
somewhat
peculiar and is supplemented by a systematic expansion in the
inverse
heavy quark masses which is close, both conceptually and
technically,
to what is currently known as Heavy Quark Effective Theory (HQET)
\cite{HQET}. In many works devoted to the subject it is not realized
that HQET
can be consistently formulated only in the context of OPE, with all
attributes
needed for implementation of Wilson's idea of separation of large
and short
distance contributions. In particular, the theory of the $1/m_{b,c}$
corrections
{\em cannot} be built without explicitly introducing a normalization point
$\mu$, the boundary between the large and short distance domains.
All
theoretical parameters appearing in the expansion are $\mu$-dependent,
generally speaking; at the same time, the predictions for measurable
quantities are
$\mu$-independent, of course. The issue of the $\mu$ dependence is
implicit
as long as one neglects perturbative corrections. Simultaneous
analysis of
the perturbative and nonperturbative terms calls for explicit
introduction of
the normalization point. This aspect is discussed in a special section,
and we
repeatedly emphasize, throughout the paper, that all HQET-type
expansions
become well-defined only provided they are formulated in the
framework of
the operator product expansion.

The organization of the paper is as follows. In Sect.~2 we outline the
specifics of the general
OPE-based approach applied to heavy quark decays and derive the expansion for
the heavy flavor hadron masses in QCD.
In Sect. 3  basic elements
of our approach are demonstrated in a simplified model where
the heavy quarks are deprived of their spins and the external
``weak'' current
considered is scalar. Sections 4 and 5 are devoted to
real QCD; here we derive a set of sum rules describing the inclusive
semileptonic decay $B\rightarrow X_c l\nu$. The sum rules are
considered in detail  at
zero recoil in Sect.~4 and in the general situation with the emphasis on the
SV kinematics, $\vec q \ll m_c$,  in Sect.~5. We  find a
lower bound
on $1-F_{B\rightarrow D^*}^2$ at zero recoil and
present expressions for $\overline{\Lambda} (\mu )$ in terms
of the  differential distributions observable in
semileptonic $B$ meson decays. In
Sect.~6 we establish a relation between field-theoretic and quantum
mechanical derivations of the sum rules in the SV limit.
Sect.~7 addresses practical implications of ``running'' of basic
low-energy parameters of the heavy quark
theory. Sect. 8
briefly summarizes the main results. In Appendix we give a
quantum-mechanical derivation of a sum rule relating $\overline
\Lambda$ to observable quantities.

\section{General OPE Approach and Mass Formulae}

Inclusive heavy flavor decays are closely related to
deep inelastic scattering (DIS). While the
latter was in the focus of theoretical investigations in the early
days
of QCD, heavy flavor decays received marginal attention. Recently
it has been shown that many elements of the theory of DIS find their
parallels in heavy flavor decays. In this paper we discuss
one more aspect with an apparent analogy in the theoretical
treatments,
namely the sum rules.

Let us recall that the standard analysis of DIS \cite{ioffe}
proceeds as follows.
One starts from
the operator product expansion
(OPE) for the $T$ product
\begin{equation}
{\hat T}_{\mu\nu} = i\int d^4 x \;{\rm e}^{-iqx} \;T\,\{ j_\mu^\dagger (x)
j_\nu (0)\}=\sum_r\;c_{\mu\nu\,\gamma_1...\gamma_n}^{(r)}(q)\cdot
O_{\gamma_1...\gamma_n}^{(r)}(0)
\label{OPE}
\end{equation}
where $j_\mu$ is the electromagnetic or some other current of
interest and $O^{(r)}(0)$ are local operators.
The average of ${\hat T}_{\mu\nu}$ over the
nucleon state with momentum $p$ presents a forward scattering
amplitude
of the Compton type.
This amplitude depends on two kinematic variables $q^2$ and  $\nu=qp$.
For large Euclidean $q^2$ OPE leads to a set of predictions for the
coefficients of the Taylor expansion in $\nu$ of the Compton amplitude at
$\nu=0$. These predictions are formulated
in terms of the expectation values of local operators
$\langle N|O^{(r)}|N\rangle$
over the nucleon state.
The coefficients of the expansion are related, {\em via} dispersion
relations, to  integrals over the imaginary part of the
amplitude at hand (moments of the structure functions).

The strategy, as well as the results obtained, are quite general. At
the
same time, certain moments play a distinguished role due to the fact
that they turn out to be proportional to operators whose matrix
elements
between the nucleon state are known on general grounds.
Relations emerging in this way are called sum rules proper and
possess
particular names. In the case of unpolarized targets we deal with
the Adler sum rule for neutrino scattering and the
Gross-Lewellyn-Smith sum rule \cite{ioffe}.

Conceptually a very similar description can be applied to the
weak decays of heavy flavor hadrons $H_Q$. One relates the
observable
quantities to a non-local transition operator $\hat T$, expands
the latter into a series of local operators and determines
their expectation values between the state $H_Q$. There exist,
of course, several technical differences relative to the case of
DIS: one deals with heavy quark currents, uses the heavy quark
mass as the expansion parameter instead of the square of the
momentum transfer, and forms the expectation values for the heavy
flavor hadron.
Moreover, much larger number of these expectation values
is known (in the leading approximation in $1/m_Q$), as compared to light
hadrons. It is due to the fact that the
distribution of the heavy quark inside hadron is trivial to leading order,
in contrast with the case of light quarks. As a consequence one can predict,
for
example, absolute decay rates, not only their evolution as the scale changes.
\vspace*{.2cm}

Now we
make a digression of a general nature concerning
the heavy quark expansion for the hadron masses.
The corresponding expressions will appear in many instances
in the sum rules below.

For any hadron its mass can be written as a matrix element of the trace of
full energy-momentum tensor of the theory, in particular,
\begin{equation}
M_{H_Q} = \frac{1}{2M_{H_Q}}\langle H_Q |\theta_{\mu\mu}
|H_Q\rangle ,
\label{1m}
\end{equation}
where $\theta_{\mu\nu}$ denotes the energy-momentum tensor;
we use the relativistic normalization
of the states,
$$
\langle H_Q|H_Q\rangle = 2M_{H_Q}V
$$
for the hadron at rest.
Our goal is to obtain the expansion for $M_{H_Q}$ in the inverse powers of
$m_Q$. In particular, we need to construct $1/m_Q$ expansion
for $\theta_{\mu\mu}$ which follows from the expansion of the Lagrangian.

The original QCD Lagrangian has the form
\begin{equation}
{\cal L}_{\rm QCD}=-\frac{1}{4g_s^2} (G_{\mu\nu}^a)^2 + \sum_q\;\bar q
(i \not\!\!D -m_q)q
\label{N1}
\end{equation}
where the sum runs over all existing quarks. In the field theory one has to
specify exactly the normalization point $\mu$ where all operators are
defined; thus all couplings, viz. $g_s$ and $m_i$ are functions of $\mu$. The
standard Eq.~(\ref{N1}) assumes that the normalization point is much higher
than all masses in the theory, $\mu\gg m_i$. In particular, no terms $\sim
(m/\mu)^n $ are kept.

Constructing the effective theory designated to describe the low energy
properties of heavy flavor hadrons we need to have the normalization point
$\mu$ below mass of the heavy quark $m_Q$. This changes the generic form of the
Lagrangian and a series of operators of higher dimension can appear. Their
coefficients contain inverse powers of the heavy quark mass $m_Q$. It
is important that all these effective operators are bound to be
Lorentz scalars.

In what follows we are not interested in terms of order higher than $1/m_Q^2$.
To this accuracy the general form of the Lagrangian is
$$
{\cal L}=-\frac{1}{4g_s^2} (G_{\mu\nu}^a)^2 + \bar Q (i\not\!\!D -m_Q)Q +
\sum_q\;\bar q(i\not\!\!D - m_q) q\;+
\; \frac{c_G}{4m_Q}\bar Q i\sigma_{\mu\nu}G_{\mu\nu} Q\;+
$$
$$
+\;
\sum_q \frac{f_q}{m_Q^2}m_q \bar q i\sigma_{\mu\nu}G_{\mu\nu}q \;+\;
\sum_{\Gamma,\;q}
\frac{d_{Qq}^{(\Gamma)}}{m_Q^2} \bar Q \Gamma Q \bar q \Gamma q \;+
\sum_{\Gamma,q,q'}\;\frac{d_{qq'}^{(\Gamma)}}{m_Q^2}\bar q \Gamma q \bar q'
\Gamma q'
\;+
$$
\begin{equation}
+\;\frac{h}{m_Q^2}\,{\rm Tr}\,G_{\mu\nu}G_{\nu\rho}G_{\rho\mu}\;+
\;{\cal O}\left(\frac{1}{m_Q^3}\right)
\label{N2}
\end{equation}
where $G_{\mu\nu}^a$ is the gluon field strength tensor and $G_{\mu\nu}\equiv
t^a G_{\mu\nu}^a$ is its matrix representation (${\rm
Tr}\,t^at^b=\delta_{ab}/2$); below we will
often use the short-hand notation
$i\sigma G=i\sigma_{\mu\nu}G_{\mu\nu}=i\gamma_{\mu} \gamma_{\nu} G_{\mu\nu}$.
The sum over light quark flavors is shown explicitly as well as the sum over
possible structures $\Gamma$ of the four fermion operators. Again, all masses,
couplings and coefficients of higher dimension operators
depend on the renormalization point. For example, the coefficient
$c_G$ has the form
\begin{equation}
c_G(\mu)=\left(\frac{\alpha_s(\mu)}{\alpha_s(m_Q)}\right)^{-\frac{3}{b}}-1
+{\cal O}(\alpha_s)\;\;\;,\;\;\;\;b=11-\frac{2}{3}n_f\;\;\;.
\label{N3}
\end{equation}
Expansion in the parameter $1/m_Q$ leads to the fact that the logarithmic
evolution of operators
generally mixes them with the ones of {\em higher} dimension, in contrast
with the expansion in $m_q$ for light quarks. The
above relation is the effect of such mixing of the operator $\bar Q Q$ with the
chromomagnetic one, $\bar Qi\sigma G Q$.

Evolving the parameters of the effective Lagrangian to the normalization point
$\mu$ one can use the values of $c,\,d,\,f,\,h$ obtained at $\mu\sim m_Q$ as
the
initial conditions. In fact all operators with $d\ge 5$ appear at this scale
only as a result of quantum corrections and contain explicit powers of
$\alpha_s(m_Q)/\pi$.
For the sake of simplicity we will generally neglect them in
our subsequent analysis, although it is easy to keep track of them if
necessary. This neglection does not mean that we disregard all effects of the
order of $1/m_Q^2$; they appear from the leading ``tree level'' operators as
well and will be accounted for consistently. Note that neglecting higher order
terms in Eq.~(\ref{N2}) gives us the standard QCD equations of motion for both
the gluon and quark fields. If we were interested in quantum corrections at the
order of $1/m_Q$ or $1/m_Q^2$ we would need to incorporate additional terms
in the equation of motions or, alternatively, consistently consider these
higher order terms in the Lagrangian as perturbation and expand in them to
necessary order.

In the theory with Lagrangian Eq.~(\ref{N1}) the trace of the energy-momentum
tensor has the form
\begin{equation}
\theta_{\mu\mu} =- m_Q\frac{\rm d}{{\rm d}\it m_Q}\,{\cal L}\;+\;
\left(\mu\frac{\rm d}{{\rm d}\mu}-
m_q\frac{\rm d}{{\rm d}\it  m_q}\right)\,{\cal L}
\label{theta}
\end{equation}
\begin{equation}
- m_Q\frac{\rm d}{{\rm d}\it m_Q}\,{\cal L}\;=\;m_Q \,\bar QQ \;\;,
\label{mQQ}
\end{equation}
\begin{equation}
\left(\mu\frac{\rm d}{{\rm d}\mu}-
m_q\frac{\rm d}{{\rm d}\it m_q}\right)\,{\cal L}\;\equiv\;{\cal D}=\;
\left[\frac{\beta(\alpha_s)}{16\pi\alpha_s^2}(G_{\mu\nu}^a)^2
+\sum_q\:m_q(1+\gamma_m)\,\bar qq
-\mu \frac{{\rm d}\it m_Q(\mu)}{{\rm d}\mu}\,\bar Q Q\right]
\label{N4}
\end{equation}
where $\gamma_m$ is
the anomalous dimension of light quark mass, $\mu \,{\rm d}m_q(\mu)/{\rm d}\mu=
 -\gamma_m\,m_q$ and
$\beta (\alpha_s)$ is the Gell-Mann-Low function, $\mu \,{\rm d}\alpha_s(\mu)
/{\rm d}\mu=\beta (\alpha_s)$.
Note that the scale dependence of
mass enters $\theta_{\mu\mu}$ in the same way for light and heavy quarks; the
explicit form of this dependence is, of course, different. The modification of
$\theta_{\mu\mu}$ for
the case when higher dimension operators  are included is
straightforward using the general expression Eq.~(\ref{theta}).
In Eq.~(\ref{N4}) ${\cal D}$ has the meaning of the part associated with light
degrees of freedom.

In the case we consider, thus, the hadron mass can be expressed in terms
of two expectation values,
\begin{equation}
M_{H_Q} = \frac{1}{2M_{H_Q}}\langle H_Q | m_Q\bar QQ
|H_Q\rangle \;+\;
\frac{1}{2M_{H_Q}}\langle H_Q |{\cal D}|H_Q\rangle \;\;.
\label{N5}
\end{equation}
The advantage of this formula over a more familiar representation
\cite{HQET} where the mass is calculated through
the expectation value of the Hamiltonian is rather transparent --
mass is a Lorentz scalar quantity, and
in Eq. (\ref{N5}) it is given as the  expectation value of the Lorentz
scalar operators. Of course, if one wants to develop $1/m_Q$
expansion for $M_{H_Q} $ one eventually arrives at the
standard expansion involving the same nonrelativistic operators.
All relevant terms in the effective Hamiltonian appear to be directly related
to the $1/m_Q$ expansion of the operator $\bar Q Q$.
Let us elucidate
this point in more detail dealing with the two operators in Eq. (\ref{N5})
in turn.

To calculate the matrix element of $\bar QQ$ we, following Refs.
\cite{6,6a}, start from the heavy quark current.
In the rest frame of $H_Q$ we have:
\begin{equation}
\frac{1}{2M_{H_Q}} \langle H_Q|\bar Q \,\gamma_0 Q|H_Q\rangle\;=\;1\;\;.
\label{4m}
\end{equation}
Using the decomposition
\begin{equation}
iD_\mu = m_Qv_\mu +\pi_\mu\;\;\;,
\;\;\;\;\;\;v_\mu=\left(p_{H_Q}\right)_\mu/\,M_{H_Q}
\label{11a}
\end{equation}
and the equation of motion
$ i\not\!\! D\, Q\;=\; m_Q\,Q$,
which imply
\begin{equation}
\frac{1-\gamma_0}{2}\,Q\;=\;\frac{\not\!\pi}{2m_Q}\,Q
\;\;\;\;,\;\;\;\; \pi_0\,Q\;=\;-\frac{\pi^2+\frac{i}{2}\sigma G}{2m_Q}\,Q
\label{N7}
\end{equation}
we arrive at the following identity
$$
\bar Q\,Q\;=\;\bar Q\gamma_0Q\;+\;2\bar Q\left(\frac{1-\gamma_0}{2}\right)^2 Q
\;=\;\bar Q\gamma_0Q\;-\;2\bar
Q\,\frac{\stackrel{\leftarrow}{\not\!\pi}}{2m_Q}\cdot
\frac{\stackrel{\rightarrow}{\not\!\pi}}{2m_Q}\,Q\;=
$$
\begin{equation}
=\;\bar Q\gamma_0Q\;+\;\bar
Q\frac{\not\!\pi^2}{2m_Q^2}
Q\;+\;\mbox{ total derivative }\;\;;
\label{5m}
\end{equation}
in the first relation above the operators $\stackrel{\leftarrow}{\pi}_\mu$
act on the $\bar Q$ field. We are considering the forward matrix
elements with zero momentum transfer and thus can drop all terms with total
derivatives. Eq.~(\ref{5m}) supplemented with equations of motion (\ref{N7})
generates then the complete $1/m_Q$ expansion for
the scalar density:
$$
\bar Q\,Q\;=\;\bar Q\,\gamma_0\,Q\;
+\;\frac{1}{2 m_Q^2}
\bar Q\,(\pi^2+\frac{i}{2}\sigma G)\,Q\;=\;
\bar Q\,\gamma_0\,Q\;
-\;\frac{1}{2 m_Q^2}
\bar Q\,(\vec\pi\vec\sigma)^2 Q\;-\;
$$
\begin{equation}
-\;\frac{1}{4 m_Q^3}\,\bar Q\left(-(\vec D \vec E)\,+\,2\vec\sigma\cdot\vec
E\times\vec\pi\right) Q
\;+\;{\cal O}\left(\frac{1}{m_Q^{4}}\right)\;\;.
\label{N9}
\end{equation}
Here $E_i=G_{i0}$ is chromoelectric field and its covariant derivative is
defined as\hspace*{.15em}\footnote{Note that in our notation $\vec D=
-\partial/\partial \vec x\,-\,i\vec A$, therefore $(\vec D\vec E)
=-{\rm div}\, {\bf E}$ in the Abelian case.} $D_jE_k=-i[\pi_j,E_k]$;
we have omitted the term
$\bar Q ([\pi_k,[\pi_0,\pi_i]]-[\pi_i,[\pi_0,\pi_k]])Q$ using the Jacobi
identity.
Moreover, $ \vec D \vec E=g_s^2 t^aJ_0^a$ by virtue of the QCD equation of
motion, where $J_\mu^a=\sum_q \bar q \gamma_\mu t^a q$ is the color quark
current; therefore the first of the
$1/m_Q^3$ terms can be rewritten as the local four fermion
interaction \cite{9}.

We see that
the $m_Q\bar Q Q$ part of $\theta_{\mu\mu}$ in Eq.~(\ref{N5})
generates
\begin{equation}
m_Q -
\frac{1}{2 M_{H_Q}} \langle H_Q|\;\bar Q \left[
\frac{(\vec\sigma\vec\pi)^2}{2m_Q}+ \frac{1}{4m_Q^2}\,
\left(-(\vec D\vec E)+2\vec\sigma\cdot\vec E\times\vec\pi\right)
\right]Q\;|H_Q \rangle\;+
\;{\cal O}\left(\frac{1}{m_Q^{3}}\right)\;\;.
\label{6m}
\end{equation}

Eq.~(\ref{6m}) shows the explicit dependence of $\theta_{\mu\mu}$ on the heavy
quark mass. However, it is not the only source of $m_Q$ dependence. Indeed, the
hadronic state $H_Q$ also depends on $m_Q$. To account for this dependence one
needs to introduce the basis of asymptotic ($m_Q=\infty$) states, $|H_Q
\rangle_{m_Q=\infty}$ and develop the perturbation theory in explicit $1/m_Q$
terms in
the Largangian. Let us emphasize that the explicit $1/m_Q^n$ terms above come
from short distances of order $1/m_Q$ whereas effects due to $m_Q$ dependence
of states $H_Q$ are large distance ($\sim \Lambda_{\rm QCD}^{-1}$) ones.

We are interested in terms $\sim \Lambda_{\rm QCD}$, $\Lambda^2_{\rm QCD}/m_Q$
and $\Lambda^3_{\rm QCD}/m_Q^2$ in mass. Therefore we must account for $1/m_Q$
effects in the matrix elements of $\bar Q (\vec\sigma\vec\pi)^2 Q$ as well as
terms $m_Q^{-1}$ and $m_Q^{-2}$ in the expectation value of ${\cal D}$.

The term $m_Q$ in Eq. (\ref{6m}) is the only one in $M_{H_Q}$ which is linear
in mass. To zeroth order in  $m_Q$ the $m_Q \bar Q Q$ part gives no
contribution and the only $m_Q$-independent term comes from the expectation
value of ${\cal D}$. To leading order in $m_Q$ this expectation value
presents what is usually called $\overline\Lambda$,
$$
\frac{1}{2M_{H_Q}}\langle H_Q |\;{\cal D}\;|H_Q\rangle \;=
\;\frac{1}{2M_{H_Q}}\langle H_Q |\,
\frac{\beta (\alpha_s)}{16\pi\alpha_s^2}G^2\,
|H_Q\rangle\;+
$$
\begin{equation}
+\;\frac{1}{2M_{H_Q}}\langle H_Q |\,m_q(1+\gamma_m)\bar q\,q\;
|H_Q\rangle\;
-\;\mu\,\frac{{\rm d}\it m_Q}{{\rm d}\mu}\;
=\;\overline\Lambda +{\cal O}(m_Q^{-1})\;\;.
\label{7m}
\end{equation}
In a sense, the first term is an analog of the gluon condensate in the
QCD vacuum \cite{SVZ}.  Instead of the QCD vacuum we deal now,
however, with the ground state in the sector with the heavy quark
charge equal to unity.

The analogy continues further and allows us to derive low energy
theorems very similar to those taking place \cite{NSVZlow}
for the vacuum correlation functions involving $G^2$ and other
operators. In the present context the low energy theorems take the
form
$$
-i\int d^4x\,\,
\frac{1}{2M_{H_Q}}\langle H_Q |
T\left\{ {\cal D}(0),
\bar Q(x) O Q(x)\right\}
|H_Q\rangle \;=
$$
\begin{equation}
=\;d\cdot \frac{1}{2M_{H_Q}}\langle H_Q |\bar Q O Q
|H_Q\rangle
\left( 1+{\cal O}(m_Q^{-1})\right)
\label{8m}
\end{equation}
where the operator $\bar Q(x) O Q(x)$ is bilinear in $Q$ and $\bar Q$ and $O$
contains derivatives and light fields $A_\mu,\:q$; the factor
$d$ is equal to
the dimension \footnote{In fact it includes the anomalous dimension of
$\bar Q O Q$
which appears when radiative corrections are incorporated.}
of $O$. Derivation of these low energy relations
merely parallels that given in Ref. \cite{NSVZlow}.

Armed with these theorems we will be able to calculate
subleading terms in Eq. (\ref{7m}). But first we need the expansion
of the QCD Lagrangian to the
necessary order in $1/m_Q$. We obtain in the standard
way
$$
\bar Q (i\not\!\!D -m_Q)Q\;=\;\bar Q
\frac{1+\gamma_0}{2}\left(1+\frac{(\vec\sigma\vec\pi)^2}{8m_Q^2}\right)\left[
\,\pi_0\;-\;\frac{1}{2m_Q}\,(\vec\pi\vec\sigma)^2\;-\right.
$$
$$
\left.
-\;\frac{1}{8 m_Q^2}\,\left(-(\vec D\vec E)+2\vec\sigma\cdot\vec
E\times\vec\pi\right)\,
\right]\left(1+\frac{(\vec\sigma\vec\pi)^2}{8m_Q^2}
\right)\frac{1+\gamma_0}{2}\,Q
\;+\;{\cal O}\left(\frac{1}{m_Q^{3}}\right)\;=
$$
\begin{equation}
=\;
\varphi^+(\pi_0-{\cal H}\,)
\varphi \;+\;{\cal O}\left(\frac{1}{m_Q^{3}}\right)
\label{lagr}
\end{equation}
where
\begin{equation}
\varphi=\left(1+\frac{(\vec\sigma\vec\pi)^2}{8m_Q^2}
\right)\frac{1+\gamma_0}{2}\,Q
\label{18a}
\end{equation}
is nonrelativistic two-component spinor field. This substitution is nothing but
the Foldy-Wouthuysen transformation which is necessary to keep term linear in
$\pi_0$ in its canonical form to ensure the mass independence of the field
$\varphi$. If it is not accomplished there would be implicit dependence
of the heavy quark fields on mass, and the low energy theorems will be
modified. In the last equation (\ref{lagr})
\begin{equation}
{\cal H}\,=\,\frac{1}{2m_Q}\,(\vec\sigma\vec\pi)^2\,+\,\frac{1}{8
m_Q^2}\,\left(-(\vec D\vec E)+2\vec\sigma\cdot\vec
E\times\vec\pi\right)\;\;\;,\;\;\;\;(\vec\sigma\vec\pi)^2\,=\,\vec\pi\,^2\,+\,
\vec\sigma\vec B
\label{hamil}
\end{equation}
is the nonrelativistic Hamiltonian through second order in $1/m_Q$. It
coincides with the text-book expression \cite{BD} in the Abelian
case. Notice that $1/m_Q^2$ and $1/m_Q^3$ terms in Eq.~(\ref{N9}) are just
${\rm d}{\cal H}/{\rm d}\it m_Q$.

Now we can calculate power in $1/m_Q$ corrections to matrix elements.
Indeed, the next-to-leading term in $(2M_{H_Q})^{-1}\langle H_Q |\,{\cal D}\,
|H_Q\rangle $ is given by
$$
m_Q\cdot \left\{ \frac{1}{2M_{H_Q}}\langle H_Q |
\,{\cal D}\,|H_Q\rangle \right\}_{1/m_Q} =
-i\int d^4x\,\,
\frac{1}{2M_{H_Q}}\langle H_Q |
T\left\{ {\cal D}(0),
\bar Q(x) \frac{(\vec\sigma\vec\pi)^2}{2} Q(x)\right\}
|H_Q\rangle =
$$
\begin{equation}
=\;
2\times\left\{
\frac{1}{2M_{H_Q}}\langle H_Q |\bar Q \frac{(\vec\sigma\vec\pi)^2}{2}
 Q|H_Q\rangle \right\}_{m_Q=\infty}\; .
\label{9m}
\end{equation}
Combining now Eqs. (\ref{9m}) and (\ref{N9}) we arrive at a standard formula
\cite{HQET,BUVpr} for the hadron mass,
\begin{equation}
M_{H_Q} = m_Q +\overline\Lambda
+\frac{1}{m_Q}\left\{\frac{1}{2M_{H_Q}}
\langle H_Q |\bar Q \frac{(\vec\sigma\vec\pi)^2}{2}
Q|H_Q\rangle\right\}_{m_Q=\infty}\;
+\;{\cal O}(m_Q^{-2}) \;\;.
\label{10m}
\end{equation}

Using the same technology it is not difficult
to develop the expansion one step further to include the
${\cal O}(m_Q^{-2}) $ term. To this end we must find and take into
account the $1/m_Q$ contribution in $\langle H_Q |\bar Q (\vec \sigma\vec\pi)^2
Q|H_Q\rangle$ in Eq.~(\ref{N9})
as well as the $1/m_Q^2$ contribution
in $\langle H_Q |G^2 |H_Q\rangle$. For the first matrix element
we have
$$
m_Q\cdot \left\{
\frac{1}{2M_{H_Q}}\langle H_Q |\bar Q
(\vec \sigma\vec\pi)^2 Q|H_Q\rangle \right\}_{1/m_Q} \;=
$$
\begin{equation}
=\;
-i \int d^4 x\,
\; \frac{1}{4M_{H_Q}}\langle H_Q|T\{\bar Q (x)(\vec\sigma\vec\pi)^2Q(x),
\bar Q (0)(\vec\sigma\vec\pi)^2Q(0)\}|H_Q\rangle '_{m_Q=\infty}\;
\equiv -\rho^3
\label{11m}
\end{equation}
where $\rho^3$ is a positive parameter of the order of $\Lambda_{\rm
QCD}^3$ measuring the correlation function above.
The prime in Eq. (\ref{11m}) indicates
that the diagonal transitions have to be removed
from the correlation function
(analogously to elimination of the disconnected
parts in the vacuum correlators).

By the same token using low energy theorems, the $1/m_Q^2$ piece
in $(2M_{H_Q})^{-1}\cdot$ $ \langle H_Q |\,{\cal D}\, |H_Q\rangle $
can be written in terms of a similar correlation function
$$
m_Q^2\cdot \left\{ \frac{1}{2M_{H_Q}}\langle H_Q |
\,{\cal D}\,|H_Q\rangle
\right\}_{1/m_Q^2}\;=
$$
$$
=\;\frac{1}{2M_{H_Q}}(-i)\int d^4x\,
\langle H_Q |T\left\{\,{\cal D}(0),\,\frac{1}{8}
\bar Q(x) \left(-(\vec D\vec E)+2\vec\sigma\cdot\vec
E\times\vec\pi\right)
Q(x)\right\}|H_Q\rangle \;-
$$
$$
-\;\frac{1}{8}\,\frac{1}{2M_{H_Q}}\int d^4x\, d^4y\,
\langle H_Q |T\left\{
\,{\cal D}(0),
\bar Q(x)(\vec\sigma\vec\pi)^2Q(x), \bar Q(y)(\vec\sigma\vec\pi)^2Q(y)\right\}
|H_Q\rangle \;=
$$
\begin{equation}
=\;
\frac{3}{8}\,\frac{1}{2M_{H_Q}}\langle H_Q | \,
\bar Q(x) \left(-(\vec D\vec E)+2\vec\sigma\cdot\vec
E\times\vec\pi\right)
Q(x)|H_Q\rangle \;-\;
\frac{3}{4}\rho^3
\label{12m}
\end{equation}
where the proper analog of Eq. (\ref{8m}) is used.

Now we have at our disposal everything needed to write down the
mass formula including $1/m_Q^2$ corrections. Combining
Eqs.(\ref{6m}), (\ref{11m}) and (\ref{12m}) we obtain
$$
M_{H_Q} = m_Q
+\overline\Lambda
+\frac{1}{m_Q}\,\left\{
\frac{1}{2M_{H_Q}}\langle H_Q |\bar Q \frac{(\vec\sigma\vec\pi)^2}{2}
Q|H_Q\rangle\right\}_{m_Q=\infty}\;+
$$
\begin{equation}
+\;
\frac{1}{8m_Q^2}\left\{\frac{1}{2M_{H_Q}}\langle H_Q |\bar Q
\left(-(\vec D\vec E) +2\vec\sigma\cdot\vec E\times\vec\pi\right)
Q|H_Q\rangle\right\}_{m_Q=\infty}
-\;\frac{\rho^3}{4m_Q^2}\;
+\;{\cal O}(m_Q^{-3})\;.
\label{13m}
\end{equation}
It is clear that the $m_Q$ dependence obtained has the usual quantum mechanical
interpretation: the third and fourth terms in  Eq.~(\ref{13m}) present the
expectation values of the Hamiltonian ${\cal H}$, Eq.~(\ref{hamil}), over the
asymptotic state $|H_Q\rangle_{m_Q=\infty}$, and the term $-\rho^3/4m_Q^2$
presents the second order iteration of the Hamiltonian.

For further usage we introduce here the following notations for the expectation
values of the two terms in $1/m_Q^2$ Hamiltonian, which are the Darwin and the
convection current (spin-orbital) interactions, respectively:
$$
\rho_D^3\;=\;\frac{1}{2M_B}\,
\langle B |\,\bar b (-\frac{1}{2}\vec D \vec E) b\,|B\rangle\;=\;
\frac{1}{2M_{B^*}}\,
\langle B^* |\,\bar b (-\frac{1}{2}\vec D \vec E) b\,|B^*\rangle
$$
\begin{equation}
\rho_{LS}^3\;=\;\frac{1}{2M_B}\,
\langle B |\,\bar b\; \vec\sigma \cdot \vec E \times \vec \pi\; b\,|B\rangle
\;=\;
-\frac{3}{2M_{B^*}}\,
\langle B^* |\,\bar b\; \vec\sigma \cdot \vec E \times \vec \pi\;
b\,|B^*\rangle\;\;.
\label{24a}
\end{equation}
$\rho_D^3$ is directly related to the third moment of the heavy quark
distribution function and has been estimated in Ref.~\cite{9} (see also
\cite{mannel2}).
On the other hand, the value of $\rho_{LS}^3$ is
expected to be suppressed for $L=0$ states like $B$ and $B^*$.

A similar decomposition can be obviously made
for the non-local correlators whose expectation value has been generically
denoted by $\rho^3$; the four unknown parameters appearing here can be
related to the $1/m_Q$ part of the expectation values of $\bar Q
\vec\pi\,^2 Q$ and $\bar Q \vec\sigma\vec B Q$ in pseudoscalar and vector
ground states:
$$
\rho_{\pi\pi}^3=
i \int d^4 x\,
\; \frac{1}{4M_{B}}\langle B|T\{\bar b \vec\pi\,^2b(x),
\;\bar b\vec\pi\,^2b(0)\}|B\rangle '
$$
$$
\rho_{\pi G}^3=
i \int d^4 x\,
\; \frac{1}{2M_{B}}\langle B|T\{\bar b \vec\pi\,^2b(x),
\;\bar b\vec\sigma \vec B b(0)\}|B\rangle '
$$
\begin{equation}
\frac{1}{3}\rho_{S}^3\delta_{ij}\delta_{kl}+\frac{1}{6}
\rho_{A}^3(\delta_{ik}\delta_{jl}-
\delta_{il}\delta_{jk})=
i \int d^4 x\,
\; \frac{1}{4M_{B}}\langle B|T\{\bar b \sigma_i B_kb(x),\;
\bar b \sigma_j B_l b(0)\}|B\rangle '\;\;.
\label{24b}
\end{equation}
Then one has for the parameters $\rho^3$ in $B$ and $B^*$, respectively
\begin{equation}
\left(\rho^3\right)_B=\rho_{\pi\pi}^3+\rho_{\pi G}^3+\rho_{S}^3+
\rho_{A}^3\;\;\;,\;\;\;
\left(\rho^3\right)_{B^*}=\rho_{\pi\pi}^3-\frac{1}{3}\rho_{\pi G}^3+
\rho_{S}^3 - \frac{1}{3}\rho_{A}^3\;\;.
\label{24c}
\end{equation}
The $1/m_Q$ corrections to the expectation values of $\vec\pi\,^2$ and
$\vec\sigma \vec B$ are
$$\frac{1}{2M_B}\langle
\vec\pi\,^2\rangle_B=\mu_\pi^2-\frac{2\rho_{\pi\pi}^3+\rho_{\pi G}^3}{2m_b}+
{\cal O}(m_b^{-2})
$$
$$\frac{1}{2M_{B^*}}\langle
\vec\pi\,^2\rangle_{B^*}=\mu_\pi^2-\frac{2\rho_{\pi\pi}^3-\frac{1}{3}
\rho_{\pi G}^3}{2m_b}+ {\cal O}(m_b^{-2})
$$
$$\frac{1}{2M_{B}}\langle
\vec\sigma\vec B\rangle_{B}=-\mu_G^2-\frac{\rho_{\pi G}^3+2\rho_S^3+
2\rho_A^3}{2m_b}+ {\cal O}(m_b^{-2})
$$
\begin{equation}
\frac{1}{2M_{B^*}}\langle\vec\sigma\vec B\rangle_{B^*}=
\frac{1}{3}\mu_G^2-\frac{-\rho_{\pi G}^3+6\rho_S^3-
2\rho_A^3}{6m_b}+ {\cal O}(m_b^{-2})\;\;.
\label{24d}
\end{equation}
Very similar parameters have been introduced in Ref.~\cite{mannel2} where the
mass formulae were discussed in the standard HQET
approach\hspace*{.15em}\footnote{We do not agree with some numerical
coefficients of Ref.~\cite{mannel2}. In particular, Eqs. (50-55) of that paper
are not quite consistent. Also, our result for the contribution of the Darwin
term in mass is
positive in the factorization approximation and smaller by a factor of two.}.

In the subsequent section we will consider in detail the toy model where
heavy quarks are spinless and light quark masses vanish \cite{9}. The absence
of spin simplifies the analysis above. Let us briefly
review the changes. In this toy model the trace of
energy-momentum tensor is given by
\begin{equation}
\theta_{\mu\mu} = 2m_Q^2\bar QQ +
\frac{\beta (\alpha_s)}{16\pi\alpha_s^2}G_{\mu\nu}^aG_{\mu\nu}^a-
2\mu\frac{dm_Q^2}{d\mu} \bar Q Q \;\;;
\label{2ms}
\end{equation}
the heavy quark field is still denoted by $Q$ but its dimension is $m$ now, not
$m^{3/2}$. Moreover, this expression, as well as the most of other
ones can be obtained
from the spinor case by substitution
$$
Q_{\rm spinor} \rightarrow \sqrt {2m_Q} \;Q_{\rm scalar} \;\;.
$$
Spin matrices are now absent, i.e.
$ (\vec \sigma \vec \pi)^2_{\rm spinor} \rightarrow (\vec \pi)^2_{\rm scalar}$.
In particular, the equation of motion and the expansion of the expectation
value
for $\bar Q Q $ take the form:
$$
\pi_0 Q = -\frac{\pi^2}{2m_Q} Q\;\;
$$
\begin{equation}
 \frac{1}{2M_{H_Q}}\langle H_Q|2m_Q\bar Q Q|H_Q \rangle =
 1-\frac{1}{2 M_{H_Q}} \langle H_Q|\bar Q \frac{{\vec\pi}^2}{m_Q}Q|H_Q
\rangle +{\cal O}(m_Q^{-4}) \;\;.
\label{5ms}
\end{equation}
Notice that in the scalar case there is no explicit $1/m_Q^3$ terms;
it corresponds to the absence of $1/m_Q^2$ terms in the nonrelativistic
Hamiltonian for scalar particles:
$$
{\cal H}=\frac{{\vec\pi}^2}{2m_Q}+{\cal O}(m_Q^{-3})\;\;.
$$
Low energy theorems have exactly the same form as before with
$$
{\cal D}= \frac{\beta (\alpha_s)}{16\pi\alpha_s^2}G^2 -
2\mu \frac{{\rm d} \it m_Q^{\rm 2}}{{\rm d}\mu}\, \bar Q Q \;\;.
$$
As a result the expansion for the hadron mass
in the scalar case takes the form
\begin{equation}
M_{H_Q} = m_Q +\overline\Lambda
+\frac{1}{2m_{Q}}\left\{\langle H_Q |\bar Q {\vec\pi}^2
Q|H_Q\rangle\right\}_{m_Q=\infty}
-\frac{\rho^3}{4m_Q^2}
+{\cal O}(m_Q^{-3})
\label{13ms}
\end{equation}
with
$$
\overline\Lambda=
\left\{\frac{1}{2 M_{H_Q}} \langle H_Q|{\cal D}
|H_Q\rangle\right\}_{m_Q=\infty}\;\;,
$$
\begin{equation}
\rho^3=2m_Q^2 i \int d^4 x \frac{1}{2M_{H_Q}}
\langle H_Q|T\{\bar Q (x){\vec\pi}^2Q(x),
\bar Q (0){\vec\pi}^2Q(0)\}|H_Q\rangle '
\label{11ms}
\end{equation}
and
\begin{equation}
\langle H_Q |\bar Q {\vec\pi}^2 Q|H_Q\rangle=\frac{M_{H_Q}}{m_Q}\,
\left\{\langle H_Q |\bar Q {\vec\pi}^2 Q|H_Q\rangle\right\}_{m_Q=\infty}\;-\;
\frac{\rho^3}{m_Q}\;+\;{\cal O}(m_Q^{-2})\;\;.
\label{28a}
\end{equation}

Let us return to ordinary QCD and briefly
discuss the normalization point dependence of
$\bar\Lambda$. In this context we can neglect terms of order $1/m_Q$ and
higher; Eq.~(\ref{7m}) gives the following definition of
$\overline\Lambda(\mu)$:
$$
\overline\Lambda(\mu)\equiv \left\{M_{H_Q}-m_Q(\mu)\right\}_{m_Q=\infty} =
\frac{1}{2M_{H_Q}}\langle H_Q|\frac{\beta (\alpha_s)}{16\pi\alpha_s^2}G^2+
\sum_q
m_q(1+\gamma_m)\bar q \,q|H_Q\rangle\;-
$$
\begin{equation}
-\;\mu\frac{{\rm d} \it m_Q}{{\rm d}\mu}\;\;.
\label{S1}
\end{equation}
The last term is specific for the field theory and is absent in the naive
quantum mechanical approach.

In the perturbation theory the $\mu$-dependence of $m_Q$ appears in the order
$\alpha_s$,
\begin{equation}
\frac{{\rm d}\it m_Q}{{\rm d} \mu}\;=
\;-c_m\frac{\alpha_s}{\pi}\;+\;{\cal O}(\alpha_s^2)\;\;.
\label{S2}
\end{equation}
The matrix element in the right hand side of Eq.~(\ref{S1}) constituting the
first part of $\overline\Lambda$, does not undergo perturbative
renormalization~\footnote{The term with the anomalous dimension of the light
quark mass {\em per se} is not renorm-invariant in the order $\alpha_s^2$, but
its $\mu$-dependence
is compensated by the mixing with the $G^2$ term.}
in order $\alpha_s$. Nevertheless it is $\mu$-dependent in the theory with
heavy quarks; its scale dependence
appears in terms $\sim \alpha_s^2$. It might be tempting to drop the
last term in Eq.~(\ref{S1}) and consider
\begin{equation}
\overline\lambda=
\frac{1}{2M_{H_Q}}\langle H_Q|\frac{\beta (\alpha_s)}{16\pi\alpha_s^2}G^2+
\sum_q
m_q(1+\gamma_m)\bar q \,q|H_Q\rangle=
M_{H_Q}-m_Q(\mu)+
\mu\frac{{\rm d}\it m_Q}{d\mu}
\label{S4}
\end{equation}
as an option for $\overline\Lambda$. However, it would not help defining a
``purely
nonperturbative'' $\overline\Lambda$; in particular, an attempt to put
$\mu\rightarrow 0$ would lead to the same infrared renormalon as in the pole
mass \cite{15,beneke}.
The transparent physical interpretation of this phenomenon will be discussed
below in Sects. 3.4 and 5.1. Though it is worth noting that the combination
$m_Q(\mu) - \mu \,{\rm d}\it m_Q/{\rm d}\mu $
in Eq.~(\ref{S4}) has the meaning of the {\em one loop} pole mass of the heavy
quark and therefore it
naturally enters the ``practical OPE'' calculations performed at the one loop
level (for the detailed discussion see Sect.~7).

The $1/m_Q^2$ terms in mass $M_{H_Q}$ obtained in this section will be used
 in derivation of the second sum rules at zero recoil.
Eq. (\ref{S1}) might be interesting by itself, though, as an alternative --
compared to the standard HQET analysis -- definition of one of its
key parameters, $\overline\Lambda$.

\section{The  Bjorken, `Optical' and Other Sum Rules -- Toy Model}

To introduce the reader to the range of questions to be considered
below in the most straightforward way we first ``peel off''
all inessential technicalities, like the quark
spins, and  resort to a simplified model  which has been previously
discussed
in Ref. \cite{9}. It will be a rather simple exercise to return
afterwards to standard QCD which will be done in Sect.~4.

\subsection{Description of the model}

We consider a toy example where all quarks are spinless;
two, denoted by $Q$ and $q$, couple
to a massless real scalar field $\phi$:
\begin{equation}
 {\cal L}_\phi = h \bar{Q} \phi q \; + \; {\rm h.c.} \; ,
\label{Qqphi}
\end{equation}
where $h$ is the coupling constant and
$\bar{Q}=Q^{\dagger}$.
The masses of the quarks
$Q$ and $q$ are both {\em large}. Later on we will
analyze the SV
limit
where
\begin{equation}
\Lambda_{\rm QCD}\ll m_Q - m_q\ll m_Q
\label{SV}
\end{equation}
has to hold,
but for the time being
the ratio $\gamma = m_q/m_Q$  can be
arbitrary provided that $m_Q-m_q\gg\Lambda_{\rm QCD}$
\footnote{To reach the SV limit in the $B\rightarrow X_ce\nu$
transitions
there is no need to assume that $m_b-m_c\ll m_c$. The small
velocity regime for
the $c$ quark can be ensured by adjusting $q^2$ appropriately.}.
The field $\phi$
carries color charge zero;
the  reaction $Q\rightarrow q+\phi$ is thus a toy model for the
radiative
decays of the type $B\rightarrow X_s\gamma$.

The total  width for the free quark decay
$Q\rightarrow q +\phi$ is given by the following expression:
\begin{equation}
\Gamma (Q\rightarrow q\phi ) =\frac{h^2E_0}{8\pi m_Q^2}\equiv
\Gamma_0
\label{gamma0}
\end{equation}
where
\begin{equation}
E_0 = \frac{m_Q^2-m_q^2}{2m_Q} .
\label{E0}
\end{equation}

As explained in Refs. \cite{1} -- \cite{9} the theory of
preasymptotic effects in inclusive decays is based on
introducing the transition operator,
\begin{equation}
\hat T = i\int d^4 x \,{\rm e}^{-iqx} T\{ \bar Q(x) q(x) \, , \,\bar
q (0)
Q (0)\} .
\label{trans}
\end{equation}
Then the energy spectrum of the $\phi$ particle  in the inclusive
decay is
obtained from
$\hat T$ in the following way:
\begin{equation}
\frac{d\Gamma}{dE} = \frac{h^2 E}{4\pi^2M_{H_Q}}
  {\rm Im}\, \langle H_Q|\hat T |H_Q \rangle
\label{spectrum}
\end{equation}
where $H_Q$ denotes a hadron built from the heavy quark $Q$ and
the light cloud (including the light antiquark). If not stated
otherwise, $H_Q$ will denote the ground state in a given channel.
Moreover, one can (and must) apply the  Wilson operator product
expansion \cite{wilson} (OPE) to express the non-local
operator $\hat T$ through an infinite series  of local operators with
calculable coefficients.

\subsection{OPE and predictions for observable quantities}

In the Born approximation the transition operator has the form (Fig.
1)
\begin{equation}
\hat T = - \int d^4 x (x|\bar Q\frac{1}{(P_0 - q +\pi )^2-
m_q^2}Q|0)\;\;,
\label{born}
\end{equation}
$$
iD_\mu \equiv (P_0)_\mu +\pi_\mu \equiv m_Q v_\mu + \pi_\mu
$$
which is
particularly suitable for constructing the
OPE by expanding Eq. (\ref{born})
in powers of $\pi$ (see Ref. \cite{9} where all notations have been
introduced).
The operators appearing as a
result of this expansion are ordered according to their dimension.
The leading operator
$\bar Q Q$ has  dimension 2 (let us recall that the scalar $Q$ field
has
dimension $1$ in contrast to the real quark fields of dimension
$3/2$, which
leads in particular to different normalization factors; we still use
relativistic normalization in the bulk of the paper, except Sect.~6). Its
expectation value for the
state $H_Q$ reduces
to unity to leading order in $1/m_Q$; this contribution
gives rise to the  parton result (\ref{gamma0}).  Beyond the leading
approximation, according to Eq.~(\ref{5ms}),  it takes the form
\begin{equation}
 \frac{1}{2M_{H_Q}}\langle H_Q|\bar Q Q|H_Q \rangle =
\frac{1}{2m_Q} \left( 1
-\frac{1}{2m_Q^2} \langle H_Q|\bar Q {\vec\pi}^2Q|H_Q
\rangle  + ... \right) ;
\label{20}
\end{equation}
thus, one gets a correction of order $1/m_Q^2$.

Corrections of the same order come from higher-dimensional
operators
in the expansion of the transition operator. There are no relevant
operators of the next-to-leading dimension three
(more exactly, as first noted in
Ref. \cite{4}
in the framework of HQET,
they vanish because of the equations of motion \footnote{This
statement is sometimes erroneously interpreted as a proof of the
absence of a term linear in
$1/m_Q$ in the total width, see below. As a matter of fact the
authors of Ref. \cite{4} believed that the linear term {\em may}
appear in
the
total width from
the overall normalization of $\langle H_Q|\bar h_v h_v|H_Q \rangle$,
as
explicitly stated, e.g., on page 404 of Ref. \cite{4}. Also, ``matching''
to QCD, the necessary step in the HQET approach, was not considered. In fact,
the question of absence or presence of the $1/m_Q$ correction
cannot
be solved  in HQET {\em per se} since the problem requires a
full-QCD analysis
of the total decay rate to perform such a matching, see
the
corresponding discussion in Ref. \cite{15}.}).
The only operator of  dimension four in our toy model has the form
$\bar Q {\vec\pi}^2Q$, and  it generates a $1/m_Q^2$ correction after
taking the matrix element over $H_Q$.

After some simple algebra one finds
$$
\frac{1}{\pi}{\rm Im}\,{\langle\hat T\rangle}
=\left( \frac{\langle\bar QQ\rangle }{2m_Q}
-\frac{\langle\bar Q {\vec\pi}^2 Q\rangle}{12m_Q^3}\right)
\delta (E-E_0 )\; -
$$
\begin{equation}
-\;\frac{E_0 \langle\bar Q {\vec\pi}^2 Q\rangle}{12m_Q^3}\delta '(E-
E_0) +
\frac{E_0^2 \langle\bar Q {\vec\pi}^2 Q\rangle}{12m_Q^3}\delta ''(E-
E_0) + ...
\label{ImT}
\end{equation}
where operators of higher dimension have been ignored, and
we have used a shorthand notation
for the expectation value over $H_Q$:
$\langle ...\rangle \equiv \langle H_Q |...|H_Q\rangle$.

The expansion of ${\rm Im}\, {\hat T}$
into local operators  generates more and more
singular terms at the point where the $\phi$ spectrum would be
concentrated in the free quark approximation. The
physical spectrum on the other hand
is a smooth function of  $E$. In principle,
one could derive a smooth spectrum by summing up an infinite set of
operators to all orders (for more detail see Refs. \cite{9,11,10a}).
There is
no need to carry out this summation here, however, since we are
interested only in
integral characteristics (we will discuss certain sum rules), and as
far
as they are concerned, the expansion in Eq. (\ref{ImT})
is perfectly legitimate.

At first, we calculate the total width by substituting eq.
(\ref{ImT})
into eq. (\ref{spectrum}) and integrating over $E$. As a matter of
fact,
the result has already been given in Ref. \cite{9},
\begin{equation}
\Gamma = \int dE \frac{d\Gamma}{dE}=\Gamma_0
\left( 1-\frac{\mu_\pi^2}{2m_Q^2} +...\right)
\label{gamma}
\end{equation}
where the integration runs from 0 to the physical boundary
$E_0^{phys}$, expressed in the hadron masses
\begin{equation}
E_0^{phys} = \frac{M_Q^2 -M_q^2}{2M_Q}\;\;, \;\;\; M_Q\equiv M_{H_Q} ,
\label{E0phys}
\end{equation}
and the same convention for $M_q$. We use here the standard
notation for the expectation value of ${\vec\pi}^2$,
\begin{equation}
\mu_\pi^2 = \frac{1}{2M_{H_Q}}\langle H_Q|2m_Q\bar Q{\vec\pi}^2Q|H_Q\rangle
\;\;.
\label{mupi}
\end{equation}

The expression in the brackets in Eq. (\ref{gamma}) is nothing else
than
the (corrected)
matrix element of the operator $\bar Q Q$;  the only possible
effective
operator $\bar Q {\vec\pi}^2Q$ of
dimension 4 is not a
Lorentz scalar and thus cannot appear
in $\Gamma$ \cite{9} . The
meaning of this term in $\langle \bar Q Q\rangle$ is quite
transparent: it reflects time dilation for the moving quark, and
the coefficient $(-1/2)$ could, therefore,
have been guessed from the very
beginning. The absence of a correction of order $1/m_Q$ in the total
width in the toy  example under consideration is a manifestation of
the general theorem established in Refs.~\cite{6,6a} and discussed in
more
detail recently in Ref. \cite{15}.

As we will see shortly, Eq. (\ref{gamma}) treated in the SV limit
is equivalent to two results simultaneously: that of Ref. \cite{14}
and the
Bjorken sum
rule \cite{12,13}, with appropriate corrections due to
terms which were not considered in Refs. \cite{12,13,14}. All
sum rules analogous to that of eq. (\ref{gamma}) below will
generically be referred to as {\em the first sum rule}.

Eq. (\ref{gamma}) is the first example of the sum rules we
will be dealing with throughout the paper. Its derivation (as well as
that
of all similar sum rules) intuitively is perfectly
clear -- the integrated spectrum of the decay obtained at the
quark-gluon level using the OPE is equated to the integrated
physical
spectrum, saturated by the genuine hadronic states. We will
elaborate
on the justification for this procedure in Sect.~4.1 where, among
other
things,
we discuss the accuracy one may expect
from relations of this type; the general idea lying behind all
such relations is wide-spread in QCD.
For the moment we just adopt the heuristic attitude outlined
above  without submerging into further, less pragmatic,
questions.

Next, we calculate the average energy of the $\phi$ particle. The
corresponding sum rule  in the SV limit is
just a version of Voloshin's optical sum rule!

To be more exact, let us define moments $I_n$
\begin{equation}
I_n =\int_0^{E_0^{phys}} dE \, (E_0^{phys} - E)^n\, \frac{1}{\Gamma_0}
\frac{d\Gamma}{dE}
\label{I1def}
\end{equation}
and consider the first moment $I_1$.
Notice that in the SV limit $E_0^{phys} - E$ reduces to the
excitation energy of the final hadron produced in the decay. The
factor $E_0^{phys} - E$ in the integrand
eliminates the ``elastic" peak, so
that the integral is saturated only by the inelastic contributions.
Let us note in passing (we will discuss this point later in more
detail)
that the optical sum rule \cite{voloshin} in its original formulation
is
actually divergent; a cutoff must be introduced by hand. In our
formulation, of course, there is no place for divergences; the
expectation value of $E_0^{phys} - E$ is defined {\em via}
a convolution with the physical spectrum, and since the $\phi$
energy
in the decay is finite, the expectation value of $E_0^{phys} - E$,
as well as higher moments, are certainly finite. The decay kinematics
provides us with a natural cutoff at the scale of the energy release.

Explicit calculation of $I_1$ using Eq. (\ref{ImT}) yields
\begin{equation}
I_1 = \Delta - \frac{\mu_\pi^2E_0^{phys}}{2m_Q^2}
\label{I1res}
\end{equation}
where $\Delta$ is defined as follows
\begin{equation}
\Delta = E_0^{phys} - E_0 ,
\label{delta}
\end{equation}
see Eqs. (\ref{E0}) and (\ref{E0phys}).
The sum rule (\ref{I1res}) contains Voloshin's result, again with
corrections left aside previously \cite{voloshin,grozin}.
The sum rules analogous to that of Eq. (\ref{I1res}) below will
be generically referred to as {\em the second sum rule}.

It is quite evident that the series of such sum rules can readily be
continued further. For the
next moment, for instance, we get
\begin{equation}
I_2 =\int_0^{E_0^{phys}} dE \, (E_0^{phys} - E)^2
\,\frac{1}{\Gamma_0}\frac{d\Gamma}{dE}\;
= \;\Delta^2
+\frac{\mu_\pi^2E_0^2}{3m_Q^2}\;\;.
\label{I2}
\end{equation}
Relations of this type will be referred to as
{\em the third sum rule}.
Analyzing this sum rule in the SV limit one obtains, in principle,
additional information, not included in the results of Refs.
\cite{12,13,14,voloshin}. It is worth emphasizing that in Eqs.
(\ref{gamma}), (\ref{I1res}) and (\ref{I2}) we have collected all
terms
through order $\Lambda_{\rm QCD}^2$, whereas
those of order $\Lambda_{\rm QCD}^3$ are systematically
omitted. Predictions for higher moments would require
calculating terms ${\cal O}(\Lambda_{\rm QCD}^3)$ and higher.

\subsection{The SV limit and the sum rules}

We proceed now to discussing the sum rules (\ref{gamma}),
(\ref{I1res}) and (\ref{I2}) in the SV limit, i.e. in the limit of small
velocity $\vec v= -\vec q/E$ where $ -\vec q$ and $E$ are momentum and energy
of the final hadronic system. The assumption
Eq. (\ref{SV}) means that $|\vec v\,| \ll 1$.
The availability of the extra expansion parameter, $|\vec v |\ll 1$,
makes the SV limit a very interesting theoretical laboratory.
We have already mentioned that in this limit the Bjorken sum rule
relates the distortion of the ``elastic" peak to the integral
over
the inelastic contributions.  Here we will elaborate on this issue.
First, we briefly recall what is already known and then present
new results.

For technical reasons we will assume that
\begin{equation}
\Lambda_{\rm QCD}\ll m_Q-m_q \ll \sqrt{m_Q\Lambda_{\rm QCD}} .
\label{SVprim}
\end{equation}
The inequality on the right hand side is not essential and can
easily be lifted;
it helps, however, to make all formulae more concise, and we will
accept it for a while,  thus  replacing  Eq. (\ref{SV}) by the
stronger
condition of
Eq. (\ref{SVprim}). We now supplement the expansion in
$\Lambda_{\rm QCD}$ carried out above by an expansion in $\vec
v$.
The natural
hierarchy of parameters in the domain (\ref{SVprim}) is as follows
$$
{\vec v}^2 ,\,\,\, |\vec v| (\Lambda_{\rm QCD} /m_Q), \,\,\,
\Lambda_{\rm
QCD}^2
/m_Q^2 \;\;.
$$
Terms of order $\vec v\,^4 \ll\Lambda_{\rm QCD}^2/m_Q^2$ will be
omitted.

Expanding the hadron masses in terms of the heavy quark masses
$m_Q$ and $m_q$ one finds for scalar quarks
\begin{equation}
M_Q=m_Q+\overline{\Lambda} +\frac{\mu_\pi^2}{2m_Q} + {\cal
O}(1/m_Q^2),\;\;\;
M_q=m_q+\overline{\Lambda} +\frac{\mu_\pi^2}{2m_q} + {\cal
O}(1/m_q^2)\;\; .
\label{mass}
\end{equation}
Eq. (\ref{mass}) implies, in turn, that
\begin{equation}
\Delta = E_0^{phys}-E_0=\frac{1}{2}v_0^2
\left( \overline{\Lambda} +\frac{\overline{\Lambda}^2 +\frac
{1}{2}\mu_\pi^2}{M_Q}\right)
-v_0\frac{\mu_\pi^2}{2M_Q} + ...
\label{deltaSV}
\end{equation}
where for convenience we have introduced an auxiliary parameter
$v_0$,
\begin{equation}
v_0 = (M_Q-M_q)/M_Q \simeq \frac{E_0}{M_Q}.
\label{v0}
\end{equation}
This parameter approximately coincides in the SV limit    with the velocity
of the heavy hadron produced in the transition $Q\rightarrow
q\phi$.
Indeed,
if the mass of the produced excited hadron is $M_q+\epsilon$ then
its velocity is
\begin{equation}
|\vec v| =v_0 - \frac{\epsilon}{M_Q} - \frac{v_0 \epsilon}{M_Q}
\label{v}
\end{equation}
plus terms of the higher order in $\epsilon$ and/or $v_0$.

With all these definitions the sum rules (\ref{gamma}), (\ref{I1res}) and
(\ref{I2}) take the form
\begin{equation}
I_0 = 1- \frac{\mu_\pi^2}{2m_Q^2} + ...\;\;,
\label{53a}
\end{equation}
\begin{equation}
I_1 = \frac{1}{2}v_0^2\overline{\Lambda}
-v_0\frac{\mu_\pi^2}{M_Q} + ...\;\;,
\label{I1prim}
\end{equation}
\begin{equation}
I_2 = \,\frac{1}{3}v_0^2\mu_\pi^2 + ...
\label{I2prim}
\end{equation}
where the ellipses denote the systematically omitted terms of order
$\Lambda_{\rm QCD}^3$, as well as ${\cal O}(v_0^2\Lambda_{\rm
QCD}^2)$
terms for $I_1$ and
${\cal
O}(v_0^4\Lambda_{\rm QCD}^2)$ ones for $I_2$. The first term on
the
right hand
side of Eq. (\ref{I1prim}) corresponds to Voloshin's relation, the
second
term is a correction whose physical meaning will soon become clear.

To interpret the sum rules derived it will be instructive
to consider the transition operator off the physical cut.
Expressing Eq.~(\ref{ImT}) in terms of $E-E_0^{phys}$ we then get
$$
M_Q^{-1}\langle H_Q |{\hat T}|H_Q\rangle =\frac{1}{2m_Q^2}
\left\{ \left( 1-\frac{2\mu_\pi^2}{3m_Q^2}\right)
\frac{1}{E-E_0^{phys}}
\right . +
$$
\begin{equation}
\left . \left( -\Delta +\frac{E_0\mu_\pi^2}{6m_Q^2}\right)
\frac{1}{(E-E_0^{phys})^2} +\frac{E_0^2\mu_\pi^2}{3m_Q^2}
\frac{1}{(E-E_0^{phys})^3}+...\right\} \, .
\label{T}
\end{equation}

To zeroth order in
$\vec v$ or $\Lambda_{\rm QCD}$ there remains only the first term on
the
right hand side of Eq. (\ref{T}). The
inclusive decay rate is then totally saturated by a single
``elastic" channel, the production of the ground state meson
containing $q$.
This is the perfect inclusive-exclusive duality noted in
Ref. \cite{14}. The peak in the $\phi$ spectrum obtained in the quark
transition $Q\rightarrow \phi q$ survives hadronization in this
approximation; at the hadronic level
$H_Q\rightarrow H_q \phi$ we still have the same peak at the
same energy.
Eqs. (\ref{I1prim}) and (\ref{I2prim}) are, of course, trivially
satisfied in
this case because in  the absence of the inelastic contributions both
sum rules yield vanishing numbers.

Furthermore, the terms with $\Lambda_{\rm QCD}$ come with $v$.
To order $\overline{\Lambda}$ there is only one such
term, appearing in eq. (\ref{I1prim}) for $I_1$. This term shows that
the inelastic production {\em must} already be present
at this level. It corresponds to the
production of a meson $H^*_q$ with excitation energy
$\sim\Lambda_{\rm QCD}$ and a residue $\propto \vec v\,^2_0$.
Then eq. (\ref{53a}) implies that the height of the elastic peak is
reduced by $\vec v\,^2_0$. A rough model exemplifying this picture
can be obtained from
Eq. (\ref{T}).
In this approximation, one can rewrite it as follows,
omitting all numerical factors:
\begin{equation}
M_Q^{-1}\langle H_Q |{\hat T}|H_Q\rangle =\frac{1}{2m_Q^2}
\left\{ ( 1-v_0^2) \frac{1}{E-E_0^{phys}} +
v_0^2\frac{1}{(E-E_0^{phys}+\overline{\Lambda})} +...\right\} \, .
\label{T1}
\end{equation}
The first term is the elastic peak while the second is an inelastic
contribution. Of course, Eq. (\ref{T1}) is an illustration
and not a unique solution.

Technically, the term $\vec v\,^2\overline{\Lambda}$ in Eq. (\ref{I1prim})
arises
because
the ``elastic" peak of the quark transition situated
at $E=E_0$ is slightly shifted when we pass to the hadronic
transition;
the genuine hadronic elastic peak is situated at $E=E_0^{phys}$, to
the right of the quark peak (see Fig. 2).

Including ${\cal O}(\overline{\Lambda}^2)$ contributions we get
correction terms in Eqs. (\ref{53a}) and  (\ref{I1prim}), and Eq.
(\ref{I2prim}) for $I_2$ becomes non-trivial.
The new term in
Eq. (\ref{I1prim}) has a simple meaning. In the toy example at hand
the excited
mesons produced in the decay at order $\vec v\,^2$ are spin-1
mesons, with the vertex proportional to $(\vec v\vec\epsilon )$
where $\vec\epsilon$ is the polarization vector and $\vec v$ is the
velocity of the given meson. This velocity, differs, however, from
$v_0$ by terms of order $\Lambda_{\rm QCD} /m_Q$, see Eq.
(\ref{v}).
This rather
trivial shift in velocity nicely explains all the sum rules above.
Indeed, let us take into account that the physical $\vec v\,^2$
is reduced by the amount $\sim v_0(\overline{\Lambda} /M_Q)$.
The height of the
inelastic contribution is proportional to the square of the physical
velocity, while the size of the inelastic domain is
$\sim \overline{\Lambda}$. Hence,
with
our accuracy the right-hand
side of Eq. (\ref{I1prim}) is expected to be
$\sim v_0^2\overline{\Lambda}
-v_0(\overline{\Lambda}^2 /m_Q)\,$, while the
right-hand side of Eq. (\ref{I2prim}) is expected to be
$\sim \overline{\Lambda}^2 v_0^2\;$, in full accord with what we
actually have.
A qualitative model of saturation now takes the form
$$
M_Q^{-1}\langle H_Q |{\hat T}|H_Q\rangle =\frac{1}{2m_Q^2}
\left\{ \left( 1-\frac{2\mu_\pi^2}{3m_Q^2}
-\left(\frac{\Delta}{\lambda} -\frac{E_0\mu_\pi^2}{6\lambda m_Q^2})\right)
\right)
\frac{1}{E-E_0^{phys}}
\right . \;+
$$
\begin{equation}
+\;
\left . \left( \frac{\Delta}{\lambda} -\frac{E_0\mu_\pi^2}{6\lambda m_Q^2}
\right)
\frac{1}{E-E_0^{phys}+\lambda} \right\} \, ,
\label{T2}
\end{equation}
where
$$
\lambda \simeq \frac{2}{3}\frac{\mu_\pi^2}{\overline{\Lambda}}
\sim \Lambda_{\rm QCD}\;\;.
$$

\subsection{Perturbative gluon corrections}

So far we have assigned the gluon field to play the role
of a soft medium to incorporate the effects of long distance
dynamics and have completely ignored
perturbative gluon corrections. Yet those have to be included;
among other things the emission of hard gluons generates
the spectral density outside the end-point region which is
very relevant for our analysis.

In calculating radiative gluon correction we can disregard,
in the leading approximation,
nonperturbative effects, like the difference between $m_Q$
and $M_Q$ or the `Fermi' motion of the initial quark. Thus we deal
with the  decay of the free quark $Q$ at rest into $q +\phi +$
gluon.  The virtual gluon contribution merely
renormalizes the constant $h$  in the analysis  presented above.
The analogous renormalization for the spinor case has
been calculated in Ref. \cite{14}.

The effects from real gluon emission are most
simply calculated in the
Coulomb gauge, where only the graph shown in Fig. 3 contributes.
A straightforward computation yields to leading order in
$\vec v\,^2$
\begin{equation}
\frac{d\Gamma^{(1)}}{dE}=\Gamma_0\frac{8\alpha_s}{9\pi}
\frac{E^3}{E_0m_Q^2}\frac{1}{E_0-E}\;\;.
\label{rad}
\end{equation}
It is well-known that the logarithmic singularity in the integral
over
$E$ for $\Gamma_{tot}$ is canceled
by a contribution of soft virtual gluons to
the renormalization
of $h$. For the second sum rule we are going to discuss here,
this infrared range is not singular.

Gluon emission obviously contributes to the spectrum in its
entire domain $0<E<E_0$. In this order $\alpha_s$ does not run, of
course.
To demonstrate its scale dependence one has to carry out a two-loop
calculation; it is quite evident, however, that
it is $\alpha_s (E-E_0)$ that enters. Therefore, strictly speaking,
one cannot apply Eq. (\ref{rad}) too close to $E_0$. Even
leaving aside the blowing up of $\alpha_s (E-E_0)$, there exists
another
reason not to use Eq. (\ref{rad}) in the vicinity of $E_0$: if
$E$ is close to
$E_0$, the emitted gluon is soft; such gluons are to be treated
as belonging to the soft gluon medium in order to avoid double
counting. The separation between soft and hard gluons
is achieved by explicitly
introducing a normalization point $\mu$. The value of $\mu$ should
be large enough to justify a small value for $\alpha_s(\mu)$.
On the other hand
we would like to choose $\mu$ as small as possible. The possible
choice
is to have $\mu$
proportional to $\Lambda_{\rm QCD}$, but with a constant of
proportionality
that is much larger than unity
$$
\mu = C\Lambda_{\rm QCD} ,\,\,\,  C\gg 1 .
$$
Then we draw a line: to the left of $E_0-\mu$ the gluon is considered
to be hard, to the right  soft. Of course, the consistent
introduction of
the
infrared renormalization point $\mu$ requires that the purely
perturbative
corrections to the weak vertex,
even if they happen to be infrared convergent,
have to be calculated using this explicit cutoff as well (see,
e.g.
the
discussion in Ref. \cite{15}) -- which is almost never done in
practice.
The corresponding modifications will be discussed below in Sect. 7.

Let us discuss now the sum rule corresponding to the first moment
of $ E_0^{phys} -E$, i.e. an analog of Eq. (\ref{I1prim})
with radiative corrections now included. Since our main
purpose in this section is
methodical, we will limit ourselves to the first order in
$\overline{\Lambda}$ and the second order in $v$. A qualitative
sketch of how
$d\Gamma/dE$ looks like is presented in Fig. 4.
Then the prediction for $I_1$ can obviously be rewritten as
$$
I_1= \int_0^{E_0^{phys}}(E_0^{phys} -E)\frac{d\Gamma}{dE}dE=
$$
\begin{equation}
=\Gamma_0\frac{1}{2}v_0^2\left[ \overline{\Lambda} (\mu )
+v_0^{-2}\int_0^{E_0-\mu} \frac{16\alpha_s}{9\pi}
\frac{E^3}{E_0m_Q^2}dE \right]
\label{rad1}
\end{equation}
where by definition
\begin{equation}
\overline{\Lambda} (\mu ) = \int_{E_0^{phys}-\mu}^{E_0^{phys}}\,
\frac{2}{v_0^2}\,\frac{1}{\Gamma_0}
\frac{d\Gamma}{dE} (E_0^{phys}-E)\, dE .
\label{lambdaphys}
\end{equation}
Without the radiative tail the prediction for $I_1$
could be obtained by integrating the theoretical expression
(\ref{ImT})
over a very narrow domain near $E_0$.
Clearly, there is  no
way to switch off the radiative corrections in QCD: one has to deal with
the perturbative and nonperturbative contributions simultaneously.
The introduction of the parameter $\mu$ thus becomes
mandatory. Eq. (\ref{lambdaphys}) then can provide us with one
possible
physical
definition of $\overline{\Lambda} (\mu)$
(among others) relating this quantity to an integral over
a physically  measurable spectral density. One may rephrase this
statement as follows. Since quarks are permanently confined the
notion of the heavy quark mass becomes ambiguous.
To eliminate this
ambiguity one must explicitly specify the procedure of measuring
``the heavy quark mass''. Any conceivable procedure will necessarily
involve a cut-off parameter $\mu$ much in the same way as the
procedure
defined above, and then $\overline{\Lambda} (\mu ) = M_Q-
m_Q(\mu )$.
In the
``most
inclusive'' procedure when one does not try to separate out any kind
of
effects, one integrates the tail to the kinematical bound $E\simeq
m_Q-m_q$
and, therefore, obtains $\overline{\Lambda}$ normalized at the
scale of energy
release.

Since this question is very important let us look at  it
from a slightly different angle. It had widely been believed that
$\overline{\Lambda}$ can be defined as a universal constant. The
standard
definition, being applied to our example, would involve three
steps:
(i)~Take the radiative perturbative
tail to the left of the shoulder and extrapolate it all the way to
the point $E=E_0$; (ii)~subtract the result from the measured
spectrum; (iii)~integrate the difference over $dE$
with the weight function $(E-E_0^{phys})$. The elastic peak
drops out and the remaining integral is equal to $\Gamma_0
(\vec v\,^2/2)\overline{\Lambda}$. It is quite clear that this procedure
cannot be
carried out consistently -- there exists no unambiguous way to
extrapolate the perturbative tail too close to $E_0^{phys}\,$,
the end-point of the spectrum \footnote{The standard
routine corresponds to using the literal perturbative
expression for this tail with the non-running strong coupling (for
one-loop
calculations), or accounting for the first term in the expansion of
$\alpha_s(k)$ in terms of $\alpha_s(m)$ (in the
two-loop ones).}.
Our procedure, with the normalization point $\mu$
introduced explicitly, is free from this ambiguity. We will further
comment on this issue in Sect. 5.1 where we discuss the
possibility of measuring $\overline{\Lambda} (\mu )$ in the
inclusive
$B\rightarrow X_c l\nu$ decays.

In practice, the $\mu$ dependence of $\overline{\Lambda} (\mu )$
may
turn out to
be rather weak.
This is the case if the spectral density is such as shown in Fig. 4,
where the contribution of the first excitations (lying within
$\sim \Lambda_{\rm QCD}$ from $E_0^{phys}$) is numerically much
larger
than the
radiative tail representing (at least in the sense of duality) high
excitations. It is quite clear that if the physical spectral density
resembles that of Fig. 4 and $\mu$ = several units
$\times\:\Lambda_{\rm QCD}$,
the running $\overline{\Lambda} (\mu)$ is rather insensitive to the
particular choice
of
$\mu$. As known from QCD sum rules it is just this situation
which occurs for
the standard quark and gluon condensates (the so-called
practical version of Wilson's OPE).

Still, even if the $\mu$
dependence of $\overline{\Lambda}$ is practically weak,
conceptually
it is
impossible to define $\overline{\Lambda}$ in the limit
$\mu\rightarrow 0$. Physically it is
quite clear from the discussion presented
above. This consideration can be thought
of as an illustration to a  more formal argument presented recently
\cite{15,beneke}.

If one replaces $E$ in the numerator of Eq. (\ref{rad1}) by $E_0$
(the non-relativistic
approximation) one arrives at the formula obtained previously in
Ref. \cite{voloshin}. Notice that the total prediction for $I_1$ is,
of course,
$\mu$ independent. Eq. (\ref{rad1}) shows how
$\overline{\Lambda}
(\mu )$
changes under
the variation of the normalization point,
\begin{equation}
\delta\overline{\Lambda} =\delta\mu
\frac{16}{9}\frac{\alpha_s(\mu
)}{\pi}
{}.
\label{della}
\end{equation}
The numerical coefficient in front of $\delta\mu$ is slightly
different
from that found in Ref. \cite{15} (16/9 versus $2\pi /3$). The reason
is
obvious -- we use here a different procedure for defining
$\overline{\Lambda} (\mu )$ compared to that suggested in
\cite{15}:
introducing a
gluon mass
$\lambda$ ``switches off'' the perturbative tail in a  `soft' rather
than
`hard' way at
$\mu=\lambda$. The fact itself of the
presence of a linear (in $\mu$) renormalization is obviously common
for all proper
procedures.
In other words, running of
$\overline{\Lambda}$ is not logarithmic,
but power-like. This fact can be traced back to the mixing
between
the operators $\bar Q (ivD) Q$ and $\bar Q Q$ established in
\cite{15}.
Numerically $\delta\overline{\Lambda} \sim 0.1 \,\mbox{GeV}$ if
$\mu$
changes from $1$ to
$1.5\,\mbox{GeV}$, i.e. $\delta\overline{\Lambda} (\mu
)/\overline{\Lambda} (\mu
)\sim 0.2$.

By the same token, using the
third sum rule, Eq. (\ref{I2prim}), one
can give
a physical definition of $\mu_\pi^2 (\mu )$,
$$
I_2 = \int_0^{E_0^{phys}}(E_0^{phys} -E)^2\frac{d\Gamma}{dE}dE=
$$
\begin{equation}
=\Gamma_0\frac{1}{3}v_0^2\left[ \mu_\pi^2 (\mu )
+v_0^{-2}\int_0^{E_0-\mu} \frac{8\alpha_s}{3\pi}
\frac{E^3(E_0-E)}{E_0m_Q^2}dE \right]
\label{rad2}
\end{equation}
where by definition
\footnote{We hope that this rather clumsy notation will not cause
confusion: $\mu_\pi^2$ is the matrix element of the
kinetic energy operator while $\mu$ (with no subscript)
is a normalization point.}
\begin{equation}
\mu_\pi^2 (\mu ) = 3\Gamma_0^{-1}v_0^{-2}\int_{E_0^{phys}-
\mu}^{E_0^{phys}}
\frac{d\Gamma}{dE} (E_0^{phys}-E)^2\, dE .
\label{muphys}
\end{equation}
The $\mu$ dependence is then obtained as follows:
$$\delta\mu_\pi^2\sim\frac{4\alpha_s}{3\pi}\delta\mu^2
\sim 0.1\,\mbox{GeV}^2\;\;\; \mbox{ if } \;\;\; \delta\mu^2\sim
1\,\mbox{GeV}^2 \;\;.
$$
Because the integral of the perturbative tail does not depend on the
particular
heavy flavor hadron in the initial state, this equation is equivalent
to
the
corresponding power mixing of the kinetic energy operator with the
leading one:
\begin{equation}
\frac{d}{d\mu^2} \left[ \bar Q\, (\vec{\pi})^2\,Q\right]_\mu =
\frac{4\alpha_s(\mu)}{3\pi}\;
\left[ \bar Q Q\right]_\mu
\label{mixing}
\end{equation}
where the numerical value of the coefficient quoted above
$(4/3\pi )$ refers to  this particular way of
introducing the
renormalization point.

The fact that the operators $\left[ \bar Q\,
(\vec{\pi})^2\,Q\right]_\mu $
and $\left[ \bar Q Q\right]_\mu $ mix with each other, and the
mixing
is quadratic in $\mu$, has been noted some time ago in lattice
calculations (G.~Martinelli raised this question in a private
discussion with M.S. in 1991; see \cite{mart}).

\subsection{Sum rules for the form factors at zero recoil}

We continue investigating our toy model with the aim of establishing
sum rules for the form factors at zero recoil. These sum rules will
allow us to find corrections to the elastic form factor at {\em zero
recoil} in terms of inelastic contributions.

The starting  idea is to consider the kinematical point $\vec q = 0$
and $q_0$ close to $\Delta M = M_B - M_D$. In other words
we abandon the case $\vec q\,^2 = 0$ and turn to $\vec q\,^2\neq 0$
where $q$ is
now the
momentum transfer carried away by our hypothetical $\phi$
quantum.
Since $\vec q = 0$ the ground state $B$ meson cannot decay
into a P-wave state. Unlike the case of the Bjorken sum rule there is
no ``external" vector ${\vec v}$. The fact that we are at zero recoil
implies that
inelastic contributions are  suppressed as $\Lambda^2/m^2$, i.e.
produce effect of the same order of magnitude as the corrections to
the
elastic form factor.

In the case when $\vec q \neq 0$ the strength of the contribution of
the excited resonance is proportional to ${\vec q}\,^2/M_D^2$
\cite{13}.
In our kinematics $\vec q =0$ and the only relevant velocity is
provided
by the primordial motion of the $c$ quark inside $D$ (more exactly,
it is
the `difference' between the quark motions in $B$ and $D$ that
counts). The
limit
$\vec q \rightarrow 0$  is obtained, qualitatively, if instead of the
``external" velocity we use that of the primordial
motion,
$$
{\vec q}\,^2/M_D^2 \rightarrow \Lambda_{\rm QCD}^2 /M_D^2\;\;.
$$
This analogy
nicely illustrates why the residues of the excited
resonances
are proportional to $\Lambda_{\rm QCD}^2/M_D^2$ and match
the picture of Fig.~4 where the contribution from the
excited resonances is proportional to ${\vec q}\,^2 /M_D^2$.

Let us sketch
how the sum rules at zero recoil emerge technically. To this end
 we again consider the $T$ product (\ref{trans})
sandwiched between the $B$ meson state. The contribution of  a
given hadronic state in this amplitude is proportional to
\begin{equation}
\frac{1}{(M_Q -q_0)^2- M_q^2}
=\frac{1}{(\Delta M - q_0)(\Delta M - q_0 +2M_q)}\;\;.
\label{2pole}
\end{equation}
It is convenient to introduce the variable
\begin{equation}
\epsilon =\Delta M - q_0 \;\;.
\label{epsilon}
\end{equation}
We are interested in singularities of the amplitude at
$\epsilon\ll M_q$. The second pole in Eq. (\ref{2pole}) at $\epsilon  =
-2M_q$ is a reflection of a
distant singularity corresponding
to the transition $B+\bar D \rightarrow
\phi^*$ where the asterisk marks the virtual $\phi$ quantum. We
assume, as usual, the hierarchy
$ \Lambda_{\rm QCD} \ll
\epsilon \ll m_{Q,q}$. Correspondingly, we will expand in
$\Lambda_{\rm QCD}/\epsilon$ and in $\epsilon /m$. In particular,
the
factor
responsible for the second pole in eq. (\ref{2pole}) becomes
$$
\frac{1}{\Delta M - q_0 +2M_q} = \frac{1}{2M_q}(1-
\frac{\epsilon}{2M_q} + ...)\;\;.
$$
The second and higher order terms in the brackets can be omitted
since
they  lead to a non-singular  (at $\epsilon = 0$) expression; the
non-singular expressions have no imaginary part and are irrelevant
for
our purposes.

Next one considers the theoretical expression for the quark level
transition operator,
\begin{equation}
-\hat T =
\bar Q\frac{1}{(m_Q-m_q-q_0)(m_Q+m_q-q_0)+(\pi^2+2m_Q\pi_0
-2q_0\pi_0)} Q\;\;.
\label{transq}
\end{equation}

Our task is to expand the transition operator in $\Lambda_{\rm QCD}
/\epsilon$
and in $\epsilon /m_{Q,q}$ and then to compare the terms singular
in
$1/\epsilon$ in this theoretical expansion with the
phenomenological expression obtained in the language of the
resonance saturation. A technical point which deserves mentioning
right at the beginning is a mismatch in the definitions of
$\epsilon$.
The theoretical expression (\ref{transq}) is phrased in terms of
the quark mass differences without reference to mass differences
of mesons.
Since we would like to
get the sum rules written in terms of the physical excitation
energies
(measured from the mass of the lowest-lying meson state) we have
to express Eq. (\ref{transq}) in terms of $\Delta M$ rather than
$\Delta m$
before expanding it. As we will see shortly, for our purposes it
is necessary to keep all effects through order $\Lambda^3 /m^3$;
those
of order $\Lambda^4 /m^4$ and higher can be neglected.

The expansion for the meson mass in inverse powers of the
heavy quark mass
for
spinless quarks considered here is given in Eq.~(\ref{13ms}).
It is worth reminding    that $\rho^3$ defined in Eq.~(\ref{11ms})
is the second order perturbation in $1/m_Q$, and as such is
positive-definite.
Using this mass formula one gets
\begin{equation}
\Delta M = \Delta m \left( 1
-\frac{\langle{\vec\pi}^2\rangle_0}{2m_Qm_q}
+\rho^3\frac{m_Q+m_q}{4m_Q^2m_q^2}\right)
\label{DMmu}
\end{equation}
where the following
short-hand notation is used:
$$
\langle {\vec\pi}^2\rangle_0 \equiv \langle H_Q|\bar Q
{\vec\pi}^2Q|H_Q\rangle_{m_Q=\infty} \;\; .
$$

Substituting Eq. (\ref{DMmu}) into Eq. (\ref{transq}) and expanding
in $\Lambda / \epsilon$ and $\epsilon /m_{Q,q}$ using Eq.~(\ref{28a}) we
arrive at
\begin{equation}
-\langle H_Q|\hat T |H_Q\rangle =
\frac{1}{\epsilon} \frac{1}{2m_q}
\left( \langle \bar QQ\rangle -\frac{\langle{\bar Q\vec\pi}^2 Q \rangle}{2
m_q^2}\right)
+\frac{1}{\epsilon^2}\rho^3\frac{(\Delta m)^2}{8m_Q^2m_q^3}
\label{tepsilon}
\end{equation}
plus terms of higher order in $1/\epsilon$  and in
$\Lambda_{\rm QCD}$.

This theoretical expression is to be confronted now with what one
obtains from saturating the amplitude at hand with meson poles.
 From Eq. (\ref{2pole}) it is not difficult to see that
\begin{equation}
-\langle H_Q |\hat T|H_Q \rangle = \frac{M_Q}{2m_qm_Q}\sum_{i=0,1,...}
\left( \frac{1}{\epsilon} \,F_i^2  \;
+\;\frac{1}{\epsilon^2}\,
\epsilon_i F_i^2 + ...\right)
\label{res}
\end{equation}
where $F_i$ is the form factor for the transition $H_Q\rightarrow
H_q^i$ induced by the vertex (\ref{Qqphi}),
\begin{equation}
\langle H_q^{(i)}|\bar qQ|H_Q\rangle \;= \;
\left(\frac{M_QM_q^{(i)}}{m_Qm_q}\right)^{\frac{1}{2}}
F_i\;\;,
\end{equation}
$M_q^{(i)}$ is the mass of $i$-th state ($\,M_q^{(0)}=M_q$ is the mass of
ground state)  and
\begin{equation}
\epsilon_i=M_q^{(i)} - M_q^{(0)}
\end{equation}
is the excitation energy of the $i$-th state; all form factors are
taken
at the zero recoil point, where the meson produced in the transition
$Q\rightarrow q$ is at rest in the rest frame of $H_Q$.

Comparing Eqs. (\ref{res}) and (\ref{tepsilon}) and using Eq.~(\ref{5ms}) we
find that
\begin{equation}
\sum_{i=0,1,...} \,F_i^2 \;=\; 1\;-\;\frac{1}{2M_{Q}} \langle H_Q |2m_Q\bar Q
\vec\pi\,^2 Q|H_Q\rangle\left(\frac{1}{2m_Q^2}+\frac{1}{2m_q^2}\right)\;+\;
{\cal O}\left(\frac{1}{m^4}\right)
\label{sr1}
\end{equation}
and
\begin{equation}
\sum_{i=1,...}\, \epsilon_i F_i^2 \;=\;\rho^3
\frac{(\Delta m)^2}{4m_Q^2m_q^2}\;+\;{\cal O}\left(\frac{1}{m^3}\right)\;\;.
\label{sr2}
\end{equation}
Note that the elastic pole gives no contribution
in Eq. (\ref{sr2}).
Hence we conclude that the residues of the excited states
are proportional to $\Lambda^2$. Transferring then all excited states
in Eq. (\ref{sr1}) to the right-hand side we observe that the square
of
the elastic form factor receives corrections of order $\Lambda^2$,
both, of
local nature and due to excited states.
Since all excitation energies and all residues are
positive an obvious inequality holds,
\begin{equation}
\sum_{i=1,2,...} \: F_i^2 < \frac{1}{\epsilon_1}
\rho^3\frac{(\Delta m)^2 }{4m_Q^2m_q^2}
\label{ineq}
\end{equation}
where $\epsilon_1$ is the excitation energy of the first excited
state.
Therefore
$$
1\;-\;\frac{1}{2M_{H_Q}} \langle H_Q |2m_Q\bar Q
\vec\pi\,^2 Q|H_Q\rangle\left(\frac{1}{2m_Q^2}+\frac{1}{2m_q^2}\right)\;-
\; \frac{\rho^3}{4\epsilon_1}
\frac{(\Delta m)^2}{m_Q^2m_q^2}\;
<\;\;F_0^2\;\;<\;
$$
\begin{equation}
<\;\;1\;-\;\frac{1}{2M_{H_Q}} \langle H_Q |2m_Q\bar Q
\vec\pi\,^2 Q|H_Q\rangle\left(\frac{1}{2m_Q^2}+\frac{1}{2m_q^2}\right)
\label{ineq2}
\end{equation}
where $F_0$ is the form factor of the ``elastic"
transition $H_Q\rightarrow H_q$.

It is worth noting that the explicit form of the corrections, in
particular
to the first sum rule (\ref{sr1}) and, therefore, the ``local'' terms
$\propto
\langle {\vec{\pi}^2}\rangle$ in Eq. (\ref{ineq2}) depend on the
structure of
the ``weak''
current considered and refer to the case of the scalar vertex. Should
we use
the vector current, coefficients in the sum rules would take a form
leading to $F_0=1$ at $m_q=m_Q$ in
accord with the
exact conservation of the vector current for equal masses.

Concluding this section let us mention a convenient computational
device. It is helpful to let the initial quark mass go to infinity
and
retain corrections only in $1/m_q$.
In this way one removes nonperturbative corrections
originating in the
initial state. The results referring to finite $m_Q$ can be simply
reconstructed at the very end. On the other hand, in the opposite
limit, $m_q\gg
m_Q$ one suppresses nonperturbative effects in the final state; in
this way it
is convenient to obtain relations for static hadronic quantities.

Similar sum rules at zero recoil in real QCD will be discussed in
Sect. 4.2.

\section{Real QCD}

We proceed now to discuss the sum rules emerging in QCD for
processes of the type $B\rightarrow X_c l \nu$. It is clear that the
approximation $m_b-m_c\ll m_{b,c}$ is not suitable in this case; one
still can reach the SV limit, however, by using the fact that $q^2$
is
not necessarily zero in this transition (from now on $q$ is the
momentum
of the lepton pair). Indeed, if $q^2$ is close to its maximal value,
$$
q^2_{max} = (M_B-M_D)^2\;\;,
$$
the $D$ meson velocity is small. At the maximal value of $q^2$
the velocity vanishes.
The velocity $\vec v = -\vec q /E_D
$
is related to $q^2$ as follows
\begin{equation}
\frac{E_D -M_D}{M_D} =
\frac{1}{\sqrt{1-\vec v\,^2}} - 1 =
\frac{(M_B - M_D)^2 -q^2}{2M_BM_D}\;\;,
\label{veloc}
\end{equation}
whereas at the quark level
\begin{equation}
\frac{1}{\sqrt{1-\vec v\,^2}} - 1 =
\frac{(m_b - m_c)^2 -q^2}{2m_bm_c} \;\;.
\label{veloci}
\end{equation}
For a sizable fraction of events measured in the semi-leptonic $B$
decays the values of $q^2$ are such that these events actually do
belong to the SV limit (i.e. $v$ is small).  An indirect proof of the
relevance of the SV limit to the inclusive
semi-leptonic decays of $B$'s comes
from the fact that about 65\% of the total semi-leptonic rate is
given
by the ``elastic" transitions to $D$ and $D^*$ \cite{ron}.
The analysis below is carried out under the assumption
that the right-hand side in Eq.~(\ref{veloc}) is small.
Even though we do not assume that $m_b-m_c\ll m_{b,c}$ both
quarks,
$b$ and $c$,  will be
treated as heavy, $m_{b,c}\gg\Lambda_{QCD}$.

Needless to say that the proximity of $q^2$ to $q^2_{max}$ can be
realized
in different ways; for instance, one can merely put $\vec q = 0$ --
this is especially convenient if we are interested in the zero recoil
point --
or one can keep  $\vec q\neq  0$, but small and study the terms
proportional to  ${\vec q}\,^2$. This yields a practically realizable
method
of measuring $\overline{\Lambda} (\mu )$ in the semileptonic
decays
$B\rightarrow X_c e\nu$.
We will consider first the  simplest sum rule for the total decay
width
analogous to eq. (\ref{sr1}) in Sect. 2.5. Surprisingly, this
analysis
produces
a lower bound on the deviation from unity in the $B\rightarrow D^*$
elastic form factor at zero recoil which does not quite agree with
the previous estimates obtained by a different method
\cite{Falk-Neubert}
(see also Ref. \cite{mannel2}).
The result is of a
paramount importance for the experimental determination
of $|V_{cb}|$, the CKM matrix element, from the exclusive
decay $B\rightarrow D^*e\nu$. Then we turn to an analog of
Voloshin's
sum
rule which appears to be a promising tool
for extracting $\overline{\Lambda} (\mu)$.

The relation for
$\overline{\Lambda}$ as the quantity
measuring the mass difference between the heavy flavor hadron and the heavy
quark has been obtained
in Ref. \cite{9} by analyzing the heavy quark distribution function
appearing in the SV
limit for the final state quark. It is convenient to rewrite
the expressions obtained in Ref. \cite{9} as follows.

Consider, first, Eq. (75) of Ref. \cite{9}:
$$
\frac{1}{2M_{H_Q}}\; \mbox {Im}\: \langle H_Q|{\hat T}|H_Q\rangle =
\frac{\pi}{4m_Q^2 \overline{\Lambda}}\left[ \delta (x)
\left( 1-\frac{1}{3}\frac{{\vec q}\,^2}{m_Q^2}\int dy \;y^{-2} G(y) \right)
+
\right.
$$
\begin{equation}
\left.
+\delta ' (x)\left( \frac{q_0\langle
{\vec\pi}^2\rangle}{2m_Q^2 \overline{\Lambda}}
+\frac{1}{3}\frac{{\vec q}\,^2}{m_Q^2}\int dy \;y^{-1} G(y) \right) +
\frac{1}{3}\frac{{\vec q}\,^2}{m_Q^2} x^{-2}G(x) \right]\;,
\label{75}
\end{equation}
where $G$ is the temporal distribution function defined there.
The parameter $x$ in this
expression measures the energy with respect to the quark boundary
$E_0$, rather than to $E_0^{phys}\,$:
$$
x=\frac{E-E_0}{\overline{\Lambda}}\;\;,
\;\;\;\;E_0\;=\;\frac{m_b^2-m_c^2+q^2}{2m_b}\;\;.
$$
It is clear then that the $\delta '$
term in Eq.~(\ref{75}) must merely shift the argument of the
$\delta$
term to its physical value, $x^{phys} =
(E-E_0^{phys})/\overline{\Lambda} $.

Limiting
ourselves to the effects linear in $\overline{\Lambda}$ we get, as a
consequence of
this requirement,
\begin{equation}
\int_{-\infty}^1 \frac{dx}{x} G(x) = -\frac{3}{2} .
\label{3/2}
\end{equation}
This sum rule constraining the
temporal distribution function is interesting by itself. We can
transform it further using its integral representation, Eq. (76) of
Ref.~\cite{9}:
\begin{equation}
G(y)= \frac{1}{2\pi \overline{\Lambda}} \int dt \;
{\rm e}^{-iyt \overline{\Lambda}}
\;{\rm e}^{im_bt}
\langle B| \bar b(t=0,\vec{x} =0) \,
\pi_i \;{\rm T}\,{\rm e}^{-i\int_0^t A_0(t',0 )dt'}\;\pi_i \,b(t, \vec x =0)\,
|B\rangle
\label{76}
\end{equation}
(the factor $\exp{(im_bt)}$ accounts for the explicit time
dependence associated
with the rest energy $m_b$, that has not been factored out here).
Multiplying this relation
by $y^{-1}$ and integrating over $y$ we arrive at
\begin{equation}
\overline{\Lambda} =\frac{2}{3}\int_0^\infty
 \, \frac{d\xi}{\xi} \int \frac{dt}{2\pi} {\rm e}^{-i t \xi}
\frac{{\rm e}^{im_bt}}{2M_B} \langle B|{\bar b(t=0,\vec{x} =0) \;
\pi_i
\;{\rm T}\,{\rm e}^{-i\int_0^t A_0(t')dt'}\;\pi_i\,
b(t, \vec x =0)}|B\rangle\;\;.
\label{deflam}
\end{equation}
Thus $\overline{\Lambda}$ is expressed in terms of a
{\em non-local} correlator of heavy quark currents (see also
Ref. \cite{15}). Let us remind that the definition (\ref{S1}) involves local
operators made of light fields.
Its normalization point dependence can easily be
traced formally through the
properties of the path ordered exponent which, being the field
operator, requires specification of the normalization point for the
gauge fields. In Sect. 3.4 we have
illustrated the renormalization point dependence using the saturation
of the
correlator by intermediate states,
which is equivalent to calculating the correlator {\em via}
dispersion relations.

It will be demonstrated in Sect.~6 that matrix elements of the type
entering
Eq.~(\ref{deflam}) have simple interpretation in usual quantum
mechanics using
the first quantized language. Here we illustrate it for the case under
consideration.
Eq.~(\ref{deflam}) is written in the heavy quark limit where all
corrections $1/m$ are neglected. To this accuracy the time ordered
exponent is
nothing but the (time) correlation function
of a nonrelativistic heavy quark in the external gluon field:
\begin{equation}
\langle Q(0)\bar Q(x)\rangle_{A_\mu}\; = \;
{\rm T}\,{\rm e}^{-i\int_0^t A_0(t')dt'}\;\delta^3(\vec x\,)\,{\rm
e}^{im_Qt}\;\;.
\label{path}
\end{equation}
Then the matrix element in Eq.~(\ref{deflam}) can be written as
$$
\frac{{\rm e}^{im_bt}}{2M_B}
\;\langle B|{\bar b(t=0,\vec{x} =0) \;
\pi_i
\;{\rm T\,e}^{-i\int_0^t A_0(t',0)dt'}\;\pi_i\,
b(t, \vec x =0)}|B\rangle\;=
$$
\begin{equation}
=\;\int\:d^3x\;\frac{1}{2M_B}\langle B|{\bar b\,
\pi_i \,Q(t=0,\vec{x})\;\;
\bar Q\,\pi_ib\,(t, \vec x =0)}|B\rangle \cdot {\rm e}^{-i(m_Q-m_b)t}
\label{matel}
\end{equation}
where $Q$ is a heavy quark, $b$ or $c$.
This form is convenient for the transition to quantum mechanical notations
(see Eq.~(\ref{transit}) in Sect.~6) in which this matrix element becomes
\begin{equation}
\langle B|\pi_i(t=0)
\;\pi_i(t)\,|B\rangle_{QM}\;=
\;\sum_n {|\langle B |\pi_i |n\rangle_{QM} |^2}
\cdot {\rm e}^{i\epsilon_n t}\;\;.
\label{QMM}
\end{equation}
Here $\pi$ is the momentum operator of the heavy quark; matrix
elements with
the subscript QM are
understood in the quantum mechanical sense, namely, only the states
with zero total momentum are considered and the nonrelativistic
normalization
is assumed. $\epsilon_n$ are the the excitation energies of the heavy
meson,
$\epsilon_n\simeq M_{B_n}-M_B\simeq M_{D_n}-M_D$.
Integrating the correlator over $t$  and $\xi$ according to
Eq.~(\ref{deflam}) we thus get
\begin{equation}
\overline{\Lambda} = \frac{2}{3}\sum_n\;\frac{|\langle
B|\,\pi_i\,|n\rangle_{QM}| ^2}
{E_n-M_B}\;\;.
\label{QML}
\end{equation}

The expressions (\ref{deflam}), (\ref{QML}) are relations that give an
alternative to Eq.~(\ref{S1}), phenomenological definition
of $\overline{\Lambda}$ in terms of measurable correlators.
In Eq.~(\ref{QML}) the sum over $n$ extends up to excitation energy
$\epsilon_n=\mu$ and in this way yields
$\overline{\Lambda} (\mu )$.

The quantum mechanical derivation of Eq. (\ref{QML}) as a relation
between the
mass of the heavy quark and the total mass of the quantum
mechanical system
containing heavy quark, will be
given in Appendix.

\subsection{Sum rules at zero recoil: generalities}

Our analysis of the $B\rightarrow X_ce\nu$ problem at zero recoil
will parallel
the corresponding consideration carried out in the toy model
of Sect. 3.5. The presence of spin is a technicality which can easily
be
incorporated. As a matter of fact, all formulae necessary for
derivation
of the first and the second sum rules exist in the literature; we
will
borrow them from Ref. \cite{8} as well as   all relevant
notations.

The point $\vec q =0$ represents zero recoil. Then the transition
operator
\begin{equation}
{\hat T}_{\mu\nu} = i\int {\rm e}^{-iqx}dx\;
T\{ j_\mu^\dagger (x) j_\nu (0)\}
\label{jj}
\end{equation}
for the $b\rightarrow c$ transitions, with
$j_\mu =\bar c\Gamma_\mu b \;, \,\,\,\, \Gamma_\mu\equiv
\gamma_\mu
(1-\gamma_5)$,
can be presented (in the tree approximation)
in the form of the following expansion:
\begin{equation}
{\hat T}_{\mu\nu} =\bar b\Gamma_\mu
(k_0\gamma_0 +m_c + \not\!\pi )
\frac{1}{(m_c^2 -k_0^2)}\sum_{n=0}^\infty
\left( \frac{2k_0\pi_0 +\pi^2 +(i/2)\sigma G}{m_c^2-k_0^2}\right)^n
\Gamma_\nu b
\label{tree}
\end{equation}
with
$$
k_0 = m_b - q_0 .
$$

The operator product expansion (\ref{tree}) is justified
provided that
$$
\Lambda_{\rm QCD} \ll | m_c - k_0 |.
$$
In other words, the expansion  (\ref{tree}) is a series
in $\Lambda_{\rm QCD} /(m_c -k_0)$. At the same time, apart from
the
poles  $1/(m_c -k_0)$ it obviously contains powers of
$1/(m_c +k_0)$ which develop ``distant" singularities
at $k_0 =-m_c$. We want these  singularities
corresponding to the propagation of the antiquark $\bar c$
to be
indeed distant so that the dispersion integrals we will be
dealing with do  not stretch up to these
$\bar c$ containing states. To this end we must impose a
second condition on $|m_c - k_0 |$,
namely
$$
|m_c - k_0 |\ll m_c \, .
$$

Once this condition is imposed we expand ${\hat T}_{\mu\nu}$
in powers of $(m_c - k_0)/m_c$ and $ \Lambda_{\rm QCD} /(m_c -
k_0)$.
The
result
is then ordered with respect to the powers of $1/(m_c - k_0)$.
The terms non-singular in $(m_c - k_0)$ are irrelevant and can be
discarded.
Each particular power $1/(m_c - k_0)^{n+1}$ in the expansion
leads to a sum rule with the weight
function $\propto (m_c - k_0)^n$.

Let us sketch the basic elements of the procedure in some detail.
We start from a series in $1/(m_c - k_0)$. The next step is
averaging of ${\hat T}_{\mu\nu}$ over the $B$ meson state,
\begin{equation}
h_{\mu\nu} =\frac{1}{2M_B} \langle B|{\hat T}_{\mu\nu}|B\rangle
{}.
\end{equation}
The hadronic tensor $h_{\mu\nu}$ consists of different kinematical
structures \cite{4,8},
\begin{equation}
h_{\mu\nu} = -h_1g_{\mu\nu}
+h_2v_\mu v_\nu -ih_3\epsilon_{\mu\nu\alpha\beta}v_\alpha
q_\beta
+h_4 q_\mu q_\nu + h_5 (q_\mu v_\nu + q_\nu v_\mu ).
\label{five}
\end{equation}
Moreover, the invariant hadronic functions $h_1$ to $h_5$ depend
on two
variables, $q_0$ and $q^2$, or $q_0$ and $|\vec q \,|$. For
$\vec q =0$
only one variable survives, and only two of five tensor structures in
$h_{\mu\nu}$ are independent.

Each of these hadronic invariant functions
satisfies a dispersion relation in $q_0$,
\begin{equation}
h_i(q_0) =\frac{1}{2\pi}\int\;
\frac{w_i ({\tilde q}_0)d{\tilde q}_0}{{\tilde q}_0 - q_0}
\,\, + \mbox{ polynomial}
\label{disp}
\end{equation}
where $w_i$ are observable structure functions,
$$
w_i =2\,{\rm Im}\, h_i \;\;.
$$

This dispersion representation for $h_i$ assumes, as usual, that the
integral on the right-hand
side runs over all cuts that the transition operator may have. The
general structure of the cuts in the
complex $q_0$ plane is rather sophisticated; the issue deserves a
special discussion since it is not always properly understood.

The structure of the cuts of the functions $h_i (q_0)$ is shown in
Fig.~5.
The part accessible in the decay channel of the B mesons
covers the interval $[0, M_B-M_D]$. The dispersion integral
(\ref{disp}) can be written as a sum of
two integrals,
\begin{equation}
h_i(q_0) =\frac{1}{2\pi}\int_0^{M_B-M_D}
\frac{w_i ({\tilde q}_0)d{\tilde q}_0}{{\tilde q}_0 - q_0}
+
\frac{1}{2\pi}\int_A
\frac{w_i ({\tilde q}_0)d{\tilde q}_0}{{\tilde q}_0 - q_0}
\label{disper}
\end{equation}
where the domain $A$ consists of two subdomains, $q_0<0$ ($A_1$)
and
$q_0>M_B+M_D$ ($A_2$). For real decays we are interested only in
the first
integral since the second one, rather than describing
the $B\rightarrow X_ce\nu$ decay, refers to other physical
processes.
The
subdomain
$A_1$ actually describes a similar $b\rightarrow c $ amplitude, yet
with
negative $q_0$, and can be called the lower cut (at $q_0< -M_B-M_D$ it contains
also the $q^2$ cut and for $q_0 < -3M_B-M_D$
the $u$-channel contribution is present as well).
The integral over $A_2$, on the other hand, will be referred to as
the
integral over
the distant cuts.
Two kinds of problems are encountered in evaluating the total
dispersion
integral.
The first one emerges due to the fact that for real decay kinematics
one has
only
$q_0>0$; therefore, say, for calculating the total width one has not
the
integral over the whole physical cut, but needs to consider the
smaller
interval
without the subdomain $A_1$. The corresponding problems of
separating the
contribution of the same type of intermediate states, but at
different
values
of $q_0$ are usually referred to as ``local'' duality. In the context
of
the
present paper this is however not very important. For in our sum
rules, from a
purely
theoretical point of view, it does not matter whether a particular
transition can be measured in a real experiment or not; e.g. the lower
bounds we
will discuss rely only on the positivity of the corresponding
transition
probabilities.

There is generally another complication associated with the integral
over
the distant cuts (in particular, subdomain $A_2$)
corresponding to quite different intermediate states. The problem of
isolating these contributions can be generically referred to as
``global'' duality.

Both contributions thus represent a contamination for real decays.
Fortunately, this contamination is irrelevant for our analysis.
Indeed, to address the contamination due to the ``lower'' cut,
let us choose the ``reference" point of $q_0$ between
the cuts (see Fig. 5),  close to $M_B-M_D$,
\begin{equation}
q_0 = M_B-M_D -\epsilon
\label{qeps}
\end{equation}
where $\epsilon$ is a negative number,
$$
m_D\;\gg \;-\epsilon \;\gg\; \Lambda_{\rm QCD}
$$
When calculating the functions $h_i$ at the quark level from
the operator product expansion we get a similar dispersion relation
for the OPE coefficients (with the meson masses replaced by the
quark ones).
We then use duality concepts in identifying the physical relevance of
these
cuts. Local duality of QCD means that there is a
one-to-one correspondence between the part of the OPE coefficients
originating from the lower cut and the corresponding hadronic
contribution in the phenomenological (hadronic) representation of
$h_i$. The validity of local duality can be verified by itself
by choosing $q_0$ in the complex plain close to the particular remote
cut, the lower one for the
case at hand. Therefore,
we can systematically discard the contribution of that cut
simultaneously, in the theoretical expression for
$h_i$ and in the ``phenomenological'' saturation. In this
way we arrive at the relations
\begin{equation}
\frac{1}{2\pi}\int_0^{\mu }w_i^{quark}\epsilon^n d\epsilon
=
\frac{1}{2\pi}\int_0^{\mu}w_i\epsilon^n d\epsilon
\label{ww}
\end{equation}
with $\Lambda_{\rm QCD}\ll \mu \ll M_B$,
where $\epsilon$ is the same variable as defined in Eq. (\ref{qeps}),
but on the cut it is positive. It represents the
excitation energy. If, additionally, $\mu < M_B-M_D$ the region of integration
in Eq.~(\ref{ww}) lies completely in the physical domain.
The left hand side of Eq. (\ref{ww}) includes the
perturbative
corrections as well as the power-like nonperturbative terms.
The local duality we have invoked to discard the contribution
from the ``lower'' cut has an accuracy of the type
$\exp \{ -\mbox{ Const}\cdot \mu /\Lambda_{\rm QCD}\}$.

An analogous analysis can be repeated almost verbatim
for the contribution of the
distant cuts to address the question of ``global duality''. This
duality is even more transparent physically and
is explicit in all perturbative calculations and
for calculations in ``soft'' external fields. On the other hand, in
principle, its accuracy is generally determined by the similar factor
depending on $m_c$, namely
$\exp \{ -\mbox{Const}\cdot m_c/\Lambda_{\rm QCD}\}$, because $2m_c$ determines
the distance to the remote cut.
It is important to remember
that it is the ratio $m_c/\Lambda_{\rm QCD}$, not
$m_b/\Lambda_{\rm QCD}$ that enters, at least at zero recoil.
In the real world $m_c/\Lambda_{\rm QCD}$ numerically is
not so large, and since the constant in the exponent is unknown
one may be afraid of an insufficient accuracy of the local
duality for $D$'s. At present theory provides us with no clues
as to the value of the constant in the exponential; the degree of
possible violations of the duality should be established
empirically. With this caveat in mind
we still believe that heavy quark expansion must work well in
$B\rightarrow
X_ce\nu$.

\subsection{Sample calculation}

After these more general remarks we return to concrete calculations
of the
theoretical part of the sum rules.
To find $w_i^{quark}$ we take the discontinuity of the
transition operator (\ref{tree}),
$$
\frac{1}{i}\mbox{disc}\, {\hat T}_{\mu\nu}
= 2\pi \bar b \Gamma_\mu (k_0\gamma_0+m_c+\not\!\pi )
\times
$$
\begin{equation}
\sum_{n=0}^\infty \frac{1}{n!}\delta^{(n)}(k_0-m_c)
\frac{(2k_0\pi_0+\pi^2+(i/2)\sigma
G)^n}{(m_c+k_0)^{n+1}}\Gamma_\nu b\;\;.
\label{discn}
\end{equation}
where $\delta^{(n)}(x)=(d/dx)^n \delta(x)$.
Using this discontinuity in the left hand side of Eq.
(\ref{ww}) it is not difficult to calculate all
moments of the structure functions in the leading approximation.
For definiteness we will first consider $w_1^{AA}$, the first
structure function (see Eq. (\ref{five})) in the transition
induced by the axial current. (The corresponding ``elastic"
contribution is $B\rightarrow D^*$.) At zero recoil
$h_1^{AA}$ is singled out merely by considering the
spatial components of the axial current,
$$ h_1^{AA}=\frac{1}{3} h_{kk}^{AA}\; ; \;\;\; (i=1,2,3).$$
All other
structure functions and other currents can be treated
in a similar manner.

Technically it is convenient to carry out the computation in
two steps: first one obtains auxiliary moments convoluted
with the quark value of $\epsilon$,
\begin{equation}
\tilde\epsilon = m_b -m_c -q_0 = k_0-m_c ,
\label{quarkeps}
\end{equation}
and then, at the second stage, these results are converted
into the true $\epsilon$ moments. Notice that in the
case at hand $\epsilon$ should be defined as
$$
\epsilon = M_B-M_{D^*}-q_0
$$
since the lowest-lying state produced is $D^*$, not $D$. Then, according to
Eqs.~(\ref{13m})-(\ref{24c})
$$
\delta_A \equiv \epsilon -\tilde\epsilon=(M_B-m_b)-(M_{D^*}-m_c)
=-(\mu_\pi^2-\mu_G^2)\left(
\frac{1}{2m_c}-\frac{1}{2m_b}\right)
-\frac{2}{3m_c}\mu_G^2 \;+
$$
\begin{equation}
+\;\left(-\rho_D^3+\rho_{\pi\pi}^3+\rho_{S}^3\right)
\left(\frac{1}{4m_c^2}-\frac{1}{4m_b^2}\right)
\;+\;\left(\rho_{LS}^3-\rho_{\pi G}^3-\rho_{A}^3\right)
\left(\frac{1}{12m_c^2}+\frac{1}{4m_b^2}\right)
\;+\;{\cal O}\left(\frac{\Lambda_{\rm QCD}^4}{m^3}\right)
\label{mshift}
\end{equation}
where $\mu_\pi^2$ and $\mu_G^2$ defined in Eq.~(\ref{defme}) are the asymptotic
expectation values of the kinetic and chromomagnetic operators. In the
estimates of the neglected terms $m$ generically denotes both $m_c$ and $m_b$.
The mass shift $\delta_A$
determines the difference in the threshold energy between the real
hadrons at
zero recoil and the quark mass difference, {\em i.e.} it is
a direct zero
recoil analog of
$\Delta$ in Eq. (\ref{delta}).

Let us define moments of the structure functions $w_i$ as
\begin{equation}
I_n^{(i)}\;=\;\frac{1}{2\pi}\int \; \epsilon^n w_i(\epsilon)\;d\epsilon\;\;.
\label{95a}
\end{equation}
In the leading non-trivial approximation we get for the moments of $w_1$
$$
I_0^{(1)AA} =\langle \bar b\gamma_0
\left\{ 1-\frac{{\vec\pi}^2+\vec\sigma\vec B}{4}
(\frac{1}{m_c^2}+\frac{1}{m_b^2}+\frac{2}{3m_cm_b})
+\frac{1}{3m_c^2}\vec\sigma\vec B\right\}b \rangle +
{\cal O}(\Lambda_{QCD}^3/m^3) \;\;,
$$
$$I_1^{(1)AA}= \langle\bar b \left\{
(\vec\pi^2+\vec\sigma\vec B)
\left(\frac{1}{2m_c}-\frac{1}{2m_b}\right) - \frac{2}{3m_c}\vec\sigma\vec B
\; +\right.
$$
$$
\left.
+
\;
\left(-(\vec D \vec E)+
2 \vec\sigma\cdot \vec E \times \vec \pi \right)
\left(\frac{1}{8m_c^2}-\frac{1}{8m_b^2}\right)
- \frac{1}{3m_c^2} \vec\sigma\cdot \vec E \times \vec \pi
\right\}b \rangle\; +\;
\delta_A\;+\;{\cal O}(\Lambda_{QCD}^4/m^3)\;,
$$
$$
I_2^{(1)AA}=\frac{1}{3}\langle\bar b \left\{\left[
(\vec\pi^2+\vec\sigma\vec B)(\frac{1}{2m_c}-
\frac{1}{2m_b})\sigma_k
+\frac{1}{2m_c}[\sigma_k,\vec\sigma\vec B]\right]\times\right.
$$
$$
\left.
\left[
\sigma_k({\vec\pi}^2+\vec\sigma\vec B)(\frac{1}{2m_c}-
\frac{1}{2m_b})
-\frac{1}{2m_c}[\vec\sigma\vec B, \sigma_k]
\right] \right\}b \rangle\;-\;\delta_A^2\;+\;
{\cal O}(\Lambda_{QCD}^{5}/m^3)\;\;,
$$
$$
I_n^{(1)AA}=\frac{1}{3}\langle\bar b \left\{\left[
({\vec\pi}^2+\vec\sigma\vec B)(\frac{1}{2m_c}-
\frac{1}{2m_b})\sigma_k
+\frac{1}{2m_c}[\sigma_k,\vec\sigma\vec B]\right]\pi_0^{n-
2}\times\right.
$$
\begin{equation}
\left.
\left[
\sigma_k({\vec\pi}^2+\vec\sigma\vec B)(\frac{1}{2m_c}-
\frac{1}{2m_b})
-\frac{1}{2m_c}[\vec\sigma\vec B, \sigma_k]
\right]\right\}b \rangle+
{\cal O}(\Lambda_{QCD}^{n+3}/m^3)
\label{moments}
\end{equation}
where $\vec B$ denotes the chromomagnetic field, and the last
equation refers
to $n > 2$. It is worth noting that to
this order in $1/m_Q$ only chromomagnetic field appears explicitly for $n\ge 2$
although
in the original expansion (\ref{discn}) the field tensor $G$ contains both
chromomagnetic and chromoelectric components; the chromoelectric
field
cancels out. The underlying reason for that will
become apparent
shortly (see Sect. 6).
Averaging $\langle ... \rangle$ over the initial
hadron state is understood as $(2M_B)^{-1}\langle B|...|B \rangle$
in all expressions in the right hand side. In the derivation of
the above predictions  for the moments we used the QCD equations of motion
(\ref{N7}).

The structure of the solution of Eqs. (\ref{moments}) for the
excitation
function
$w_1(\epsilon)$ is quite transparent. The first
equation tells us that the sum of all
probabilities is
equal to unity up to small corrections $\sim {\cal O}
(\Lambda_{\rm QCD}^2/m^2)$. On the
other hand,
all higher moments $I_n$ explicitly start with  terms of
order
$\Lambda_{\rm QCD}^n \cdot (\Lambda_{\rm QCD}/m)^2$.
Since the scale for the excitation
energies
$\epsilon_i$ is given \footnote{Effectively the same
refers even to
the thresholds associated with the $D^{(*)}+\mbox{pion(s)}$ in
the chiral
limit which, strictly speaking, have no  excitation gap.
This is true due to the fact
that the corresponding amplitudes are proportional to
the pion momentum;  they can produce only chiral logarithms
and do not
change powers of mass in the analysis, see Ref. \cite{SUV}.}
 by $\Lambda_{\rm QCD}$ one
immediately
concludes that the
probabilities of transitions to the excited states all
 scale like $\Lambda_{\rm
QCD}^2/m^2$.
To saturate the first sum rule one then
needs state(s)
which do not contribute to the higher moments $I_n$ and are
produced with
practically unit
probability; the only way to satisfy this constraint
 is to saturate by the final
states $D$
and $D^*$ with the masses
$$M_B-(m_b-m_c)\;,$$
up to corrections vanishing in the limit $m_b, m_c \rightarrow
\infty$.
Moreover,
these ``elastic" transition amplitudes must be
equal to unity up to terms inversely
proportional
to
the {\em square} of the heavy quark masses -- a fact observed
originally
in Ref. \cite{14}
and now known as Luke's theorem \cite{Luke}.
 Although the
expressions above are derived for the axial current,
similar results are valid for the vector currents
as well, where the lowest-lying final state is $D$, not $D^*$.
In this way one obtains also the statement of the
heavy flavor symmetry in the spectrum of hadrons.

The second of Eqs. (\ref{moments}) corresponding
to $n=1$ has a special status. When written in terms of the excitation
energies
$\tilde\epsilon$, counted from the quark threshold, the right-hand
side does
contain terms of  order ${\cal O}(\Lambda_{QCD}^2/m)$ which are
expressed in
terms of $\mu_\pi^2$ and $\mu_G^2$; on the other hand the
structure of the
solution described above implies that the contribution of the excited
states in
the phenomenological part is
only of the order of $\Lambda_{\rm QCD}^3/m^2$. Therefore, this
leading term is
to be completely saturated by the elastic peak that resides not
 at
$\tilde\epsilon=0$ but is rather shifted by the amount
$\delta_A=\epsilon
-\tilde\epsilon$.
This condition unambiguously determines the leading, ${\cal
O}(\Lambda_{QCD}^2/m)$,
correction to the masses of the heavy flavor hadrons which has
been anticipated in
Eq.\hspace*{.1em}(\ref{mshift}). Similarly, to
order ${\cal O}(\Lambda_{QCD}^3/m)$ the
sum rule for the second moment yields the ``local'' $1/m^2$ term in the
effective Hamiltonian, Eq.~(\ref{hamil}), and expresses the non-local
correlators $\rho^3$ as the sum over inelastic probabilities.

It is not difficult to obtain a general expression for the function
$w_1^{AA}$
as it emerges  from its moments $I_n^{(1)AA}$. The inelastic part of
$w(\epsilon)$ appears at the
$1/m^2$ level and is given by the Fourier transform of the
time-dependent correlation
functions of the leading operators in the effective
Lagrangian:
$$
\epsilon^2 w_1^{AA}(\epsilon )=
$$
$$
\frac{1}{3}\int \frac{dt}{2\pi} {\rm e}^{-i t \epsilon}\frac{1}{2M_B}
\langle B|\bar b(t=0,\vec{x} =0)
\left[ (\vec{\pi}^2+\vec{\sigma}\vec{B})(\frac{1}{2m_c}-
\frac{1}{2m_b})\sigma_k+\frac{1}{2m_c}[\sigma_k,\vec\sigma\vec
B]\right]\times
$$
\begin{equation}
\mbox{T}\,{\rm e}^{-i\int_0^t A_0(t',0)dt'}\;
\left[ \sigma_k (\vec{\pi}^2+\vec{\sigma}\vec{B})(\frac{1}{2m_c}-
\frac{1}{2m_b})-\frac{1}{2m_c}[\sigma_k, \vec\sigma\vec B ]\right]
b(t,\vec x =0)|B\rangle '\;\; .
\label{correl}
\end{equation}
The prime here means subtraction of the ``elastic"
 contribution (see the expression for $I_2$ and the discussion
below).

Equation (\ref{correl}) has a very
transparent quantum-mechanical meaning. It corresponds
to the following picture. At time $t=0$ the Hamiltonian of
the system under consideration is suddenly changed by adding
a perturbation
\begin{equation}
\left(\frac{1}{2m_c}-\frac{1}{2m_b}\right)\;
(\vec{\pi}^2+ \vec{\sigma}\vec{B})\;\;.
\label{pert}
\end{equation}
Simultaneously the original wave function is changed due
to the spin flip. This effect is represented by the term
$(-1/2m_c)[\sigma_k,\vec\sigma\vec B]$ where $k$ marks the
spatial
component of the axial current. Had we considered
the time component of the vector currents the latter term
would be absent, and the entire perturbation would
reduce to Eq. (\ref{pert}). At time $t$ the perturbation is
switched off. The second order term in this combined perturbation
yields
us the excitation probabilities. This interpretation brings
us very close to Lipkin's quantum-mechanical formalism which will
be discussed in some detail in Sect. 6.

Transitions induced by the time component of the vector current
are given by the following combination of the structure functions:
\begin{equation}
w^{VV}\equiv -w_1^{VV}+w_2^{VV}+q_0^2 w_4^{VV}+q_0 w_5^{VV}\;\;.
\label{102a}
\end{equation}
The moments $I_n^{VV}$ of $w^{VV}$ defined in analogy to Eq.~(\ref{95a})
look even simpler:
$$
I_0^{VV} =\langle \bar b\gamma_0
\left\{ 1-\frac{{\vec\pi}^2+\vec\sigma\vec B}{4}
(\frac{1}{m_c}-\frac{1}{m_b})^2 \right\}b \rangle +
{\cal O}(\Lambda_{QCD}^3/m^3) ,
$$
$$I_1^{VV}=
\langle\bar b \left\{
(\vec\pi^2+\vec\sigma\vec B)
\left(\frac{1}{2m_c}-\frac{1}{2m_b}\right)\;+\;
\left( -(\vec D \vec E) +2\vec \sigma \cdot \vec E \times \vec\pi)
\right)\left(\frac{1}{8m_c^2}-\frac{1}{8m_b^2}\right)
\right\}b \rangle\; +
$$
$$
+\;
\delta_V\;+\;{\cal O}(\Lambda_{QCD}^4/m^3)\;,
$$
$$
I_2^{VV}=\langle\bar b \left\{\left[
({\vec\pi}^2+\vec\sigma\vec B)(\frac{1}{2m_c}-\frac{1}{2m_b})
\right]\times
\left[
({\vec\pi}^2+\vec\sigma\vec B)(\frac{1}{2m_c}-\frac{1}{2m_b})
\right] \right\}b \rangle \;-
$$
$$
-\;
\delta_V^2 \;+\;
{\cal O}(\Lambda_{QCD}^{5}/m^3),
$$
\vspace{0.2cm}
$$
I_n^{VV}=\langle\bar b \left\{\left[
({\vec\pi}^2+\vec\sigma\vec B)(\frac{1}{2m_c}-\frac{1}{2m_b})
\right]\pi_0^{n-2}\times\right.
$$
\begin{equation}
\left.
\left[
({\vec\pi}^2+\vec\sigma\vec B)(\frac{1}{2m_c}-\frac{1}{2m_b})
\right]\right\}b \rangle+
{\cal O}(\Lambda_{QCD}^{n+3}/m^3)\;\;\;.
\label{vmoments}
\end{equation}
Note that for the vector current the lowest charmed state is the $D$
meson and,
therefore, $\epsilon$ is to be defined now as $\epsilon = M_B-M_D-
q_0$;
the value of
$\delta_V$ then differs from $\delta_A$, and the second sum rule
(the expression for the
first moment) yields now
$$
\delta_V \equiv \epsilon -\tilde\epsilon=(M_B-m_b)-(M_{D}-m_c)
=-(\mu_\pi^2-\mu_G^2)\left(
\frac{1}{2m_c}-\frac{1}{2m_b}\right)\;+
$$
\begin{equation}
-\;\left(\rho_D^3 + \rho_{LS}^3-\rho_{\pi\pi}^3 -\rho_{\pi G}^3
-\rho_{S}^3 -\rho_{A}^3\right)
\left(\frac{1}{4m_c^2}-\frac{1}{4m_b^2}\right)\;
+\;{\cal O}(\Lambda_{\rm QCD}^4/m^3)
\label{vmshift}
\end{equation}
instead of Eq.\hspace*{.15em}(\ref{mshift}).

It is not difficult to trace a straightforward parallel
between the derivation given above, Eq.\hspace*{.15em}(\ref{correl}),
and the analysis of the effects of the heavy quark motion in the
SV limit carried out in Ref.~\cite{9} that led us to the temporal
distribution function. In
that
case, however, inelastic excitations appeared proportional to the
velocity squared of
the final quark, and it was the correlator of the spatial
momentum operators
$\vec{\pi}$ that emerged.

A remark is in order here. Obviously, one can, in principle,
calculate higher
order power corrections to the sum rules (\ref{moments}), (\ref{vmoments}).
It is possible
to see that -- in what concerns the inelastic contributions to the
structure
function -- $w(\epsilon)$ will order by order reproduce the
successive
terms of the
ordinary quantum-mechanical perturbation theory corresponding to
both next
order iterations of the leading terms as well as to perturbations
representing
subleading power operators in the effective heavy quark Lagrangian; the local
relativistic corrections in currents appear as well.
Similarly, one can consider the sum rules  in the case of nonzero
recoil. This correspondence to the  perturbation theory in quantum
mechanics will be discussed in Sect. 6.

To conclude this section, let us say a few words elucidating
why this general analysis based
on the sum rules  is
important although
it might seem  to lead  to no new results beyond the
picture of the
standard perturbation theory in quantum
mechanics.

First, we shall see in the subsequent sections that this approach
allows
one to obtain useful bounds on the transition amplitudes
and on the `static' matrix elements governing nonperturbative
corrections.

Second, it demonstrates in a transparent and unambiguous way the
necessity of
introducing the infrared normalization point in addressing
nonperturbative
effects and clarifies both the physical meaning and the qualitative
trend of the $\mu$ dependence. This
is important in view of the existing misinterpretation of this
problem
in applications of HQET.

The last, but not the least aspect:
we clearly see here that such assumptions about
QCD as the validity of global duality discussed
above, that
sometimes are naively believed to be specific only for
the OPE-based approach to
inclusive transitions, are actually the most general requirement
inherent
to any consistent consideration. If for some particular physical
reason
the quark-hadron correspondence used above were modified -- for
example there
were a sizable ``leak" in dispersion integrals from distant cuts, it
would
immediately lead to new contributions in the right-hand side of the
sum rules;
this would
result
in physically observable corrections not contained in the
expansions that can be
obtained in HQET.

\subsection{A bound on the form factor at zero recoil
from the sum rules}

The general theory developed above is applied in this section
to derive a lower bound on the deviation of the
``elastic" form factor at zero recoil from unity. To this end we
analyze the first sum rule.

We  consider, for definiteness,
 transitions generated by the axial current,
$$
A_\mu = \bar c\gamma_\mu\gamma_5 b\; ,
$$
 i.e. the transitions
of the type $B\rightarrow D^*$ and $B\rightarrow $ excitations
of the vector mesons.  Practically they are most important in
the exclusive approach to a determination of $|V_{cb}|$.
These transitions are induced by the axial-vector current $A_\mu$
and as was mentioned above it is most convenient
to focus on the spatial component of this current.
For the spatial components of the current
only $h_1$ survives.

To get the first sum rule at zero recoil
we use the first of Eqs.~(\ref{moments}) which to order $1/m^2$ reads as
\begin{equation}
\frac{1}{2\pi}\int \;d\epsilon\,w_1^{AA}(\epsilon)\;=\;
1 -\frac{1}{3}\frac{\mu_G^2}{m_c^2}
-\frac{\mu_\pi^2-\mu_G^2}{4}\left(
\frac{1}{m_c^2}+\frac{1}{m_b^2}+\frac{2}{3m_cm_b}
\right) \;\;.
\label{1epsilon}
\end{equation}
The chromomagnetic parameter is known experimentally:
$$
\mu_G^2 \simeq \frac{3}{4}(M_{B^*}^2 - M_B^2)= 0.37\,\mbox{GeV}^2
\;\;.
$$
The sum rule stemming from Eq. (\ref{1epsilon}) obviously takes
the form
$$
F_{B\rightarrow D^*}^2 + \sum_{i=1,2,...}F_{B\rightarrow
\rm {excitations}}^2\;\;=
$$
\begin{equation}
=\; 1 -\frac{1}{3}\frac{\mu_G^2}{m_c^2}
-\frac{\mu_\pi^2-\mu_G^2}{4}\left(
\frac{1}{m_c^2}+\frac{1}{m_b^2}+\frac{2}{3m_cm_b}
\right)
\label{SR!}
\end{equation}
where the sum on the left-hand side runs over all excited states
and all form factors are taken at the zero recoil point. This is a
perfect
analog of eq. (\ref{sr1}). The form factor $B\rightarrow D^*$ at zero
recoil is
defined as
$$
\langle B|A_k|D^*\rangle =\sqrt{4M_BM_{D^*}}\;F_{B\rightarrow
D^*}\;e_k^*
$$
where $e_k$ is the polarization of $D^*$ meson.

If all terms ${\cal O}(\Lambda_{\rm QCD}^2)$ are switched off higher
states
cannot be excited at zero recoil -- only
the elastic $B\rightarrow D^*$ transition survives -- and we arrive
at the
well-known
result that
$$
F_{B\rightarrow D^*}= 1 \;\,\,\, \mbox{ (zero recoil) }\;,
$$
the statement of the heavy quark (or Isgur-Wise \cite{Isgur-Wise})
symmetry  first
noted in the SV limit in Ref. \cite{14} (see also \cite{nussinov}).
Including
 ${\cal O}(\Lambda_{\rm QCD}^2)$ terms we start exciting higher
states; all
transition form factors squared are proportional to $\Lambda_{\rm
QCD}^2/m^2$.
Simultaneously
the form factor of the elastic transition shifts from unity.

Both power corrections on the right-hand side are negative. What is
crucial is
the fact that the contribution of the excited states is strictly
positive.
Transferring them to the right-hand side we arrive at the following
lower bound
\begin{equation}
1-F_{B\rightarrow D^*}^2
>\frac{1}{3}\frac{\mu_G^2}{m_c^2}
+ \frac{\mu_\pi^2-\mu_G^2}{4}\left(
\frac{1}{m_c^2}+\frac{1}{m_b^2}+\frac{2}{3m_cm_b}
\right) .
\label{1-F}
\end{equation}

Since $\mu_\pi^2 >\mu_G^2$ (see below)
 we find that the (absolute value of) the
deviation of the elastic form factor
$ F_{B\rightarrow D^*}$
from unity at zero recoil is definitely larger than
$$
\frac{M_{B^*}^2-M_B^2}{8m_c^2}\sim 0.035 \;\;.
$$

As was mentioned, the second term on the right-hand side
of eq. (\ref{1-F}) is also positive, $\mu_\pi^2 >\mu_G^2$.
We will rederive this inequality within the framework of sum rules
themselves in the next
section. Previously it was obtained in this form in
 \cite{newvol} where the quantum-mechanical
argument
of Ref. \cite{9} was extended. The quantum-mechanical line
of reasoning is applicable at a low normalization point.

The inequality $\mu_\pi^2 >\mu_G^2$ is in perfect agreement
with the most refined QCD sum rule calculation of
$\mu_\pi^2$ \cite{Braun2} which lead to $\mu_\pi^2 =0.6\pm
0.1\:\mbox{GeV}^2$; the recent updated analysis along the same line of
calculations yielded (see Ref.~\cite{BBBG})
$$
\mu_\pi^2 =0.5\pm
0.1\:\mbox{GeV}^2\;\;.
$$
If this estimate is accepted then the second term in
eq.(\ref{1-F})
amounts to $\sim 1/2$ of the first one,
and the lower bound for the deviation becomes
$1-F_{B\rightarrow D^*}^2>0.1$. The actual deviation is probably
twice as large.
First,  the sum rule derived above neglects
perturbative $\alpha_s$ corrections. The first order correction
to the elastic form factor was calculated in Ref. \cite{14}.
If in zeroth  order in $\alpha_s$ the
$\bar{b}\gamma_\mu\gamma_5c$
axial-vector vertex at zero recoil is
unity, the first order correction renormalizes it to
$$
\eta_A = 1 +\frac{\alpha_s}{\pi}\left( \frac{m_b+m_c}{m_b-
m_c}\ln\frac{m_b}{m_c}
-\frac{8}{3}\right)\;\; .
$$
Numerically $\eta_A\simeq 0.97$ if one uses $\alpha_s$ normalized
at the point
$\mu=\sqrt{m_cm_b}$ (for the recent discussion see Ref.~\cite{ural}).
For the axial-vector
current the perturbative correction is
negative, so that unity in eq. (\ref{SR!})
is replaced by approximately \footnote{For numerical estimate we
use somewhat larger
number than the literal value of $\eta_A^2$
above, see discussion in Sect.~7 and
Eq.~(\ref{mucorrect}).} 0.95.  Then, the contribution of the excited
states
in eq. (\ref{SR!}) is strictly positive, and this also reduces
$F_{B\rightarrow D^*}^2$. This contribution may be as large as,
roughly, the
power correction on the right-hand side. An estimate of the excited
state contribution supporting this statement is given in Ref.
\cite{SUV}
where
a more detailed numerical discussion of all corrections is given.
Notice that in our approach the excited state contribution
replaces a non-local contribution of Ref. \cite{mannel2}.

We conclude that $1-F_{B\rightarrow D^*}^2$ is definitely larger than
0.1, somewhat
beyond the window obtained in Ref. \cite{Falk-Neubert}.
The phenomenological impact of this observation is discussed
in Ref. \cite{SUV}.

Let us note that in order to use the second sum rule similar to
Eq. (\ref{ineq2}) we would need to know
${\cal O}(\Lambda_{\rm QCD}^3)$ terms both in the transition
operator
and in the relation between $\Delta M$ and $\Delta m$.
The corresponding expressions are provided by Eq.~(\ref{mshift}) and the
second of Eqs.~(\ref{moments}); however the relevant hadronic parameters
$\rho_D^3$, $\rho_{LS}^3$ and non-local correlators
$\rho_{\pi\pi, \pi G, S,A}^3$
are not known yet.

In a very similar way one can obtain the bound and estimate for
the vector
form factor of the $B\rightarrow D$ transition $F_{B\rightarrow D}$
at zero
recoil.
Here only
the timelike component of the current contributes, and for this
reason the
full semileptonic decay amplitude is proportional to the lepton
masses.
Therefore this mode is not advantageous; the corresponding
form factor is
measurable (in principle)  at zero recoil in the $B\rightarrow
D+\tau\nu_\tau$ decays.
Taking $\Gamma_\mu=\Gamma_\nu=\gamma_0$ we obtain for this
case the sum rule
\begin{equation}
F_{B\rightarrow D}^2 + \sum_{i=1,2,...}F_{B\rightarrow excit}^2
= 1
-\frac{\mu_\pi^2-\mu_G^2}{4}\left(
\frac{1}{m_c}-\frac{1}{m_b}
\right)^2\;\;\;.
\label{SRV}
\end{equation}
Perturbative corrections also differ and now look as follows
\cite{14}:
\begin{equation}
1 +\frac{\alpha_s}{\pi}\left( \frac{m_b+m_c}{m_b-
m_c}\ln\frac{m_b}{m_c}
-2\right) \;\;.
\label{pertv}
\end{equation}
The corrections in the case of the vector form factor obviously
vanish
at
$m_b=m_c$, as they have to in
view of the exact conservation of the current in this
limit.
Numerically therefore they are expected to be smaller for vector
transitions
than for axial ones.

It is worth mentioning that the excitation probabilities entering the
sum
rules (\ref{SR!}) and (\ref{SRV}) are generated separately by the
axial or the vector current, respectively,
but not by the $V-A$ current that directly
produces the
experimental widths. Actually at zero velocity transfer the axial
and vector currents cannot interfere. Therefore for the $V-A$
semileptonic
transitions into massless leptons one has just to add to Eq.
(\ref{SR!})
(assuming that no final
state identification is attempted) the contribution of the
$\bar{c}\gamma_i
b$ current. The sum rule for this current is obtained in the next
subsection.
Combining the two sum rules one gets
\begin{equation}
F_{B\rightarrow D^*}^2 = 1 -
\frac{\mu_\pi^2-\mu_G^2}{3m_bm_c} - \int_{\epsilon>M_{D^*}-
M_D}\;d\epsilon\;
\frac{w_1^{V-A}(\epsilon)}{2\pi}
\label{V-A}
\end{equation}
where the last term representing the inelastic contribution is
expressed via
the differential semileptonic width at zero recoil:
\begin{equation}
\frac{w_1^{V-A}(\epsilon)}{2\pi}=
\frac{8\pi^3}{G_F^2|V_{cb}|^2 q_0^2} \cdot \frac{1}{|\vec{q}\,|}
\frac{d^2\Gamma_{SL}}{d\vec{q}\,^2\,dq_0}\vert_{_{\vec{q}=0,\;q_0
=M_B-M_D-\epsilon}}
\label{exper}
\end{equation}
($q$ is the momentum of the lepton pair in the process). Equation
(\ref{V-A}) is
much less useful as an upper bound because the main part of the
correction
to the elastic form factor
must now come from the presently unknown contribution of
excitations.

\subsection{Lower bound on $\mu_\pi^2$}

Reversing the line of reasoning used to derive the upper bound on
the
form factors (the lower bound on $1-F$), we can exploit the very
same
idea
to get a constraint on the matrix elements $\mu_G^2$ and
$\mu_\pi^2$. At zero recoil there are only two independent structure
functions
for the correlator of the $V-A$ currents; similar functions can be
introduced
for other weak vertices as well. Choosing a particular current one
projects
out a certain combinations of the structure functions. It is
important
that
for the Hermitean conjugated currents in the two weak vertices one
gets a
definitely positive structure function expressed as a sum over
certain
transition probabilities.
If
an appropriate channel is found where the elastic peak is
kinematically absent, the theoretical side of the
corresponding sum rule will contain no unity term, and start with
the leading $1/m^2$
corrections given just by a linear
combination
of $\mu_G^2$ and $\mu_\pi^2$. The ``phenomenological'' side, a sum
over
the excited states, is positive-definite. In this way we arrive at
a constraint on the linear combination of $\mu_G^2$ and
$\mu_\pi^2$ at hand.

It is not difficult to find a specific example. Indeed, let us
consider
the vector current, $\bar q \gamma_\mu Q$,
and in particular its {\em space-like} components.
It is obvious that $-(1/3)h^{VV}_{kk} =-
h_1^{VV}$
is populated
only by the excited states -- if the initial state $H_Q$ is the
ground
state pseudoscalar, the final one, $H_q^*$, must be the axial-vector
meson,
non-degenerate with $H_Q$ in the symmetry limit. (In the quark
model language
one would say that $H_q^*$ is a $P$-wave state.) Using the
results of
Ref.\cite{8}  (Eq. (A1) ) we find
\begin{equation}
-h_1^{VV}(q_0) = \frac{1}{\epsilon}
\left[ \frac{\mu_\pi^2 -\mu_G^2}{4}
\left( \frac{1}{m_Q^2} +  \frac{1}{m_q^2} -\frac{2}{3m_Qm_q}\right)
+\frac{\mu_G^2}{3}\frac{1}{m_q^2}\right] \;\;.
\label{hVV}
\end{equation}
The expression in the square brackets is equal to the sum over
excitations:
\begin{equation}
\frac{\mu_\pi^2 -\mu_G^2}{4}
\left( \frac{1}{m_Q^2} +  \frac{1}{m_q^2} -\frac{2}{3m_Qm_q}\right)
+\frac{\mu_G^2}{3}\frac{1}{m_q^2}\; = \;\sum_i
F^2_{H_Q\rightarrow
H_q^*} \;\;,
\label{repres}
\end{equation}
and, hence, is always positive, for any values of $m_Q$ and $m_q$.
Being
interested in the ``static'' properties of the initial state only, it
is
convenient, according to the comment in the end of Sect. 3.5, to
consider the
theoretical limit $m_q\gg m_Q$ where only initial state effects
survive \footnote{The very same bound can be obtained by
considering,
say, the
correlator of two $i\gamma_5$ currents.
In this case one gets directly
the difference $\mu_\pi^2 -\mu_G^2$ with the coefficient
$(1/m_c+1/m_b)^2$.}.
Requiring the positivity of Eq. (\ref{hVV}) at $m_q\gg m_Q$ we
conclude
that
\begin{equation}
\mu_\pi^2 - \mu_G^2 \;=\; 4m_Q^2 \left\{ \sum_i
|F_{H_Q\rightarrow
H_q^*}|^2\right\}_{m_q=\infty}
\label{srule}
\end{equation}
and, therefore,
\begin{equation}
\mu_\pi^2 > \mu_G^2 \;\;.
\label{inequ}
\end{equation}
This is literally the same inequality that has been obtained
previously
\cite{newvol} in the quantum-mechanical language along the lines
suggested in Ref. \cite{9}. The argument presented
above can be viewed as a consistent and transparent field-theoretic
reincarnation. To be fully consistent,
according to the general discussion of Sect. 3.4, one must
introduce
the cutoff in the ``phenomenological'' integral over the decay
probabilities
from
the upper side of excitation energies $\epsilon$. It is most
important
that
the integral in the right-hand side of Eq. (\ref{srule})
$$
\sum_{\epsilon_i\le \mu} |F_{H_Q\rightarrow H_q^*}|^2
$$
is positive for {\em any} normalization point $\mu$, and therefore
ensures
the validity of the inequality (\ref{inequ}) for operators normalized
at
arbitrary values of $\mu$, provided that the normalization point is
consistently
introduced in this particular way. For high enough $\mu$ the
contribution of
the excited states is given by perturbative expressions and the
$\mu$ dependence is explicitly calculable; of course, for very large
$\mu$
the inequality becomes trivial. We will discuss the issue in more
detail in
Sect.~7.

It is instructive to note that the zero recoil sum rule for the
$V_k\times V_k$
transitions (or similar ones where there is no elastic peak) provides
us with
the direct way of determining the evolution of the kinetic
energy
operator.
One proceeds here along the same line of reasoning as has been
outlined in Sect.~3.4 where we considered the small velocity
kinematics with
$\vec{q}\ne 0$.
In this case the elastic peak identically vanishes
for purely kinematical reasons.
Therefore the only possible impact of
introducing the infrared normalization point (say, {\em via} the
gluon
``mass")
can emerge from the gluon emission probability.

Technically, we can consider the relation (\ref{repres}) in the perturbation
theory. Then the hadron state $H_Q$ must be replaced by the heavy quark $Q$,
and the excited states are $q+\mbox{gluon}$ states. For such initial state the
expectation value of operator $\sigma G$ vanishes and the sum rule
(\ref{repres}) converts into the relation for the perturbative part of
$\mu_\pi^2$:
\begin{equation}
\frac{\left(\mu_\pi^2\right)_{\rm pert}}{4}
\left( \frac{1}{m_Q^2} +  \frac{1}{m_q^2} -\frac{2}{3m_qm_Q}\right)
\; = \;\frac{1}{3}\sum_k\;\int_{\omega<\mu} \;\frac{d^3k}{2\omega(2\pi)^3}\;
\frac{1}{4m_qm_Q} |\langle qg | \bar q \gamma_k Q| Q\rangle |^2\;\;;
\label{121a}
\end{equation}
the sum over gluon polarizations is implied. Following the procedure used
throughout this paper, we have introduced the renormalization point $\mu$ as
the upper limit for excitation energy. The calculation of the amplitude in the
right hand side is very simple:
\begin{equation}
\langle qg | \bar q \vec \gamma Q| Q\rangle \;=\; g_s\, \bar q \,t^a\left[
\vec\epsilon\,^a\left(\frac{1}{2m_q}+\frac{1}{2m_Q}\right)\,-\,i\left(\vec
\epsilon\,^a
\times \vec\sigma \right)\,
\left(\frac{1}{2m_q}-\frac{1}{2m_Q}\right)\right]\,Q\;\;.
\label{121b}
\end{equation}
Taking the square of this amplitude and summing over the gluon polarizations
$\vec\epsilon\,^a$ we get the same dependence on the masses $m_q,\;m_Q$ as in
the
left hand side of Eq.~(\ref{repres}), which is expected on general
grounds. In this way we obtain
\begin{equation}
\left(\mu_\pi^2\right)_{\rm pert}\;=\;\frac{4\alpha_s}{3\pi}\,\mu^2\;\;.
\label{121c}
\end{equation}
This result can be rewritten as
\begin{equation}
\frac{d}{d\mu^2} \bar Q\, (i\vec{D})^2\,Q =
\frac{4\alpha_s(\mu)}{3\pi}\;
\bar Q Q
\label{truemixing}
\end{equation}
which coincides with Eq.\hspace*{.15em}(\ref{mixing}). It is easy to
check that the same evolution
law is obtained for any suitable current and even for a case when
heavy quarks were spinless.

Throughout this paper we have phrased our discussion of real QCD in
terms
of transitions where initial states were heavy flavor {\em mesons},
namely $B$.
It is clear that exactly the same reasoning
can be applied for the transitions
of heavy baryons, for example when the initial hadron is
$\Lambda_b$. The
matrix elements are different, of course. In particular, the
expectation
value
of the chromomagnetic operator vanishes for $\Lambda_b$.
Moreover,
contrary to the meson case no non-trivial lower bound on the kinetic
term
emerges. Therefore it is natural to
expect smaller deviations from the symmetry
limit for both vector and axial-vector form factors in baryons than
for mesons.
\vspace*{.2cm}

An analysis of corrections to $F_{B\rightarrow D^*}$ resembling ours
in spirit, but
not technically, has been carried out
recently in Ref. \cite{mannel2}. There $1-F_{B\rightarrow D^*}$ is
expressed
in terms of some local and non-local expectations values; the latter
are unknown. In our analysis the role of the non-local expectation
values is played by the contribution of the excited states.
What is crucial, this contribution is always positive. In
\cite{mannel2}
$1-F_{B\rightarrow D^*}$ is found to be positive (good!), and a
numerical
estimate
is presented relating $1-F_{B\rightarrow D^*}$ to $\mu_\pi^2$,
a parameter that is somewhat more uncertain than $\mu_G^2$.
As a matter of fact, the numerical values of $\mu_\pi^2$
accepted in \cite{mannel2} under the influence of some recent claims
\cite{claims} are in contradiction with the
inequality (\ref{inequ}).

\section{Sum rules at $\vec q\neq 0$}

We leave now the point of zero recoil and discuss
the sum rules in general situation. The first obvious complication is that
there are more independent structure functions that enter separately
the decay rate
for any particular current,
and each depends on two
variables for which we choose $q_0$ and $\vec{q}\,^2$. Inelastic processes
corresponding to the transition to the states other than $D$ and $D^*$ are now
not necessarily suppressed by powers of $\overline\Lambda/m_Q$. The $n$-th
moments of the structure functions are therefore proportional to
$\overline\Lambda^n$. If nonperturbative effects are calculated explicitly
through
${\cal O}(\Lambda_{\rm QCD}^2)$ terms, we obtain non-trivial corrections
for the first three moments with $n=0,\;1$ and $2$. We use below some results
of the forthcoming paper \cite{leva} where more details will be given.

At $\vec q\neq 0$ the variable $\epsilon$ is defined as \footnote{When
vector current is considered, the lowest state appearing is
$D$; we still use the $D^*$ energy as a reference point keeping in mind that
explicit account for the $D$ contribution in the phenomenological part of the
first moments $I_1^{VV}$
is necessary. The fact that the $D$
contribution to semileptonic width vanishes at $\vec{q}=0$ for massless leptons
justifies such a choice.}
\begin{equation}
\epsilon=M_B-\sqrt{M_{D^*}^2+\vec{q}\,^2}-q_0
\label{q1}
\end{equation}
and moments of the structure functions are
\begin{equation}
I_n^{(i)}(\vec{q}\,^2)=  \frac{1}{2\pi}
\int \;d\epsilon \,\epsilon^n\,w_i (\epsilon,\: \vec{q}\,^2)
\label{q2}
\end{equation}
where $i$ labels the structure function; they can be considered separately for
axial current ($AA$), vector current ($VV$) and for the interference of the
two ($AV$).

Using Eqs. (A1-A6) obtained in Ref.~\cite{8} and expanding the corresponding
hadronic invariant functions $h_i$ in powers of $1/\epsilon$ through terms
$1/\epsilon^3$ one arrives at the set of relations which are valid up
to terms ${\cal O}(\Lambda_{\rm QCD}^3)$.
We present here these relations
only for the
structure functions $w_{1,2,3}$ which contribute to semileptonic decays with
massless leptons (the most general case will be considered in
Ref.~\cite{leva}).
We have the following sum rules for the zeroth moments:
\begin{eqnarray}
&&I_0^{(1)AA}  =
 \frac{E_c+m_c}{2E_c}-\frac{\mu_\pi^2-\mu_G^2}{4E_c^2}
\frac{m_c}{E_c}\left[\frac{m_c^2}{E_c^2}+\frac{E_c^2}{m_b^2}+
\frac{2}{3}\frac{m_c}{m_b}\right]-\frac{\mu_G^2}{3E_c^2}
\frac{m_c}{E_c} \frac{E_c^2+3m_c^2}{4E_c^2} \nonumber \\
&& \label{m0a1}\\
&& I_0^{(2)AA} = \frac{m_b}{E_c} \left\{ 1-\frac{\mu_\pi^2-\mu_G^2}{3E_c^2}
\left[2-\frac{5}{2}\frac{E_c^2}{m_b^2}+ \frac{3}{2}\frac{m_c^2}{E_c^2}\right]
- \frac{\mu_G^2}{3E_c^2}
\left[\frac{1}{2}+\frac{m_c}{m_b}+
\frac{3}{2}\frac{m_c^2}{E_c^2}\right]\right\}\nonumber \\
&&\label{m0a2}\\
&& I_0^{(3)AV} = -\frac{1}{2E_c} \left\{ 1-\frac{\mu_\pi^2-\mu_G^2}{3E_c^2}
\left[1+\frac{3}{2}\frac{m_c^2}{E_c^2}\right]
- \frac{\mu_G^2}{2E_c^2} \left[1+ \frac{m_c^2}{E_c^2}\right]\right\}\;\;.
\label{m0av3}
\end{eqnarray}
Expressions for the $VV$ functions are obtained from the axial ones by
replacing $m_c\rightarrow -m_c$; the structure functions $w^{(1,2)AV}$ and
$w^{(3)AA,VV}$ vanish.

The above equations are analogs of the Bjorken sum
rule; however they incorporate nonperturbative effects which appear at
$1/m_Q^2$ level; the corrections
are not universal and differ explicitly for different currents and
structure functions.

The first moments look as follows:
\begin{eqnarray}
&&I_1^{(1)AA} =  \frac{E_c+m_c}{2E_c}
\left\{ \frac{\mu_\pi^2-\mu_G^2}{2E_c}
\left[1-\frac{E_c}{m_b}- \frac{1}{3}\left(1-\frac{m_c}{E_c}\right)
\left(1+3\frac{m_c}{E_c}
+2\frac{E_c}{m_b}\right)\right]+
 \right. \nonumber \\
&&~~~~~~~~~~~~~~~~~~~
  + \; \left. \frac{\mu_G^2}{2E_c}\left[1-
\frac{2}{3}\frac{m_c}{E_c} + \frac{m_c^2}{E_c^2}\right]+
\left[(M_B-m_b)-(E_{D^*}-E_c)\right] \right\}
\label{m1a1} \\
&& I_1^{(2)AA} = \frac{m_b}{E_c} \left\{\frac{\mu_\pi^2-\mu_G^2}{3E_c}
\left[2-\frac{7}{2}\frac{E_c}{m_b} + \frac{3}{2}\frac{m_c^2}{E_c^2}\right]
+
\frac{\mu_G^2}{3E_c}
\left[\frac{1}{2}-\frac{E_c-m_c}{m_b}+ \frac{3}{2}\frac{m_c^2}{E_c^2}\right]+
\right. \nonumber \\
&&~~~~~~~~~~~~~~~~~~~~~~~
  +  \; \left.
\left[(M_B-m_b)-(E_{D^*}-E_c)\right] \right\}
\label{m1a2}  \\
&& I_1^{(3)AV} = -\frac{1}{2E_c} \left\{\frac{\mu_\pi^2-\mu_G^2}{3E_c}
\left[1-\frac{5}{2}\frac{E_c}{m_b} + \frac{3}{2}\frac{m_c^2}{E_c^2}\right]
+
\frac{\mu_G^2}{2E_c}
\left[1+\frac{m_c^2}{E_c^2}\right]+ \right. \nonumber \\
&&~~~~~~~~~~~~~~~~~~~~~~~
\left.
+ \; \left[(M_B-m_b)-(E_{D^*}-E_c)\right] \right\}
\label{m1av3}
\end{eqnarray}

At $\vec q=0$ these relations determine $1/m$ terms in the masses of heavy
mesons. Their derivatives with respect to $\vec q\,^2$ near zero recoil give
the Voloshin's ``optical'' sum rule. Here they are obtained with better
accuracy for arbitrary, not necessary small, velocity and incorporate
$1/m_Q$ relative corrections. The latter appear to be quite
sizable when $\vec q$ is not particularly large.

The third sum rules, which are relations for the second moments of the
structure
functions, are calculated only in the leading non-trivial approximation. They
look rather simple and manifestly satisfy the heavy quark symmetry relation
\cite{grozin2}:
$$
\frac{2E_c}{E_c+m_c}I_2^{(1)AA}= \frac{2E_c}{E_c-m_c}I_2^{(1)VV}=
\frac{E_c}{m_b} I_2^{(2)AA} =
\frac{E_c}{m_b} I_2^{(2)VV}
= -2E_c I_2^{(3)AV} =
$$
\begin{equation}
=\;\frac{\mu_\pi^2}{3}\:\frac{E_c^2-m_c^2}{E_c^2}
+\overline\Lambda^2\left(1-\frac{m_c}{E_c}\right)^2
\;\;\;.
\label{second}
\end{equation}
All higher moments vanish in our approximation.
We used above the notation $E_{D^*}$ for the energy of $D^*$ and
$E_c$ for the energy of the $c$ quark in the free quark decay:
\begin{equation}
E_{D^*}=\sqrt{M_{D^*}^2+\vec{q}\,^2}\;\;\;,
\;\;\;E_c=\sqrt{m_c^2+\vec{q}\,^2}\;\;.
\label{Ec}
\end{equation}
The quantity
$$
(M_B-m_b)-(E_{D^*}-E_c)\simeq
$$
\begin{equation}
\simeq
\overline\Lambda\left(1-\frac{m_c}{E_c}\right)-(\mu_\pi^2-\mu_G^2)
\left(\frac{1}{2E_c}-\frac{1}{2m_b}\right)-\frac{2\mu_G^2}{3E_c}-
\frac{\overline\Lambda^2}{2E_c}\left(1-\frac{m_c^2}{E_c^2}\right)
\;+\;{\cal O}\left(\frac{1}{m^2}\right)
\label{Delta}
\end{equation}
that enters first and second moments, is similar to $\delta_A$ we have
discussed in the case of zero recoil; at $\vec q \ne 0\,$ however,
it is of the order of $\Lambda_{\rm QCD}$.

It is important for the analysis of the inclusive widths
that the nonperturbative expansion of the invariant hadronic functions $h_i$,
and, therefore, of the moments of the structure functions always runs in the
inverse powers of $E_c$ rather than $m_c$; it is correlated with the fact that
non-physical singularities in $q_0$ representing distant cuts are also
separated from the physical cut by $\sim E_c$.
It implies that one can use the same
expansion literally even if $m_c\rightarrow 0$ as long as $|\vec q\,|\gg
\Lambda_{\rm QCD}$. In turn, this means that at $E_c \sim m_b$ dominating the
inclusive width for $m_b\gg m_c\,$, nonperturbative corrections scale
like the inverse square of the mass of the heavier decaying quark \cite{6,6a}.

Using the same fact one may hope to decrease $1/m_c$ corrections to the
determination of $\overline{\Lambda}$
and $\mu_\pi^2$ from the second and third sum rules, which are
rather significant numerically in the SV limit, considering decay processes
with {\em large} recoil.
A more detailed analysis of the sum rules
at non-zero recoil is given in Ref. \cite{leva}.

\subsection{The second sum rule at $\vec q\neq 0$; measuring
$\overline{\Lambda}(\mu)$}

We discuss now
a few useful applications of the sum rules presented above
in the the simpler case of the SV kinematics when we can expand moments in
$\vec q\;^2/m_c^2$.

If at $\vec q =0$ the second sum rule requires the knowledge
of ${\cal O}(\Lambda_{\rm QCD}^3)$ terms, at $\vec q\neq 0$ (i.e.
$\Lambda_{\rm QCD} \ll |\vec q \,|\ll M_D$) a non-trivial prediction,
an analog of Voloshin's sum rule \cite{voloshin}, arises in order
$\Lambda_{\rm QCD}$. Higher order corrections are written in
Eqs.~(\ref{m1a1})-(\ref{m1av3}); they will be discussed elsewhere \cite{leva}.
We need to consider the average value of $q_{0{\rm max}}-q_0$. If the value of
$q^2$, rather than $\vec q\,^2$, is fixed,
this quantity is related to
$M_{X_c}^2$, the average invariant mass squared of the hadronic
system
produced~\footnote{The corresponding analysis of the
average
invariant hadronic mass presented in Ref. \cite{4} was
incorrect, see Ref. \cite{WA}. The average
invariant mass of the final state hadrons is {\em
not} given directly by matrix elements of local heavy quark operators at the
level of nonperturbative
corrections, contrary to claims in Ref. \cite{4}.}. Indeed, then
\begin{equation}
M_{X_c}^2 = M_{D}^2 +2M_B(q_{0{\rm max}}-q_0)\;\;,\;\;\;
q_{0{\rm max}} =
\frac{M_B^2-M_{D}^2+q^2}{2M_B}\;\; .
\label{MXc}
\end{equation}
If $\vec q\,^2$ is fixed then
\begin{equation}
M_{X_c}^2 = M_{D}^2 +2E_c(q_{0{\rm max}}-q_0)+(q_{0{\rm max}}-q_0)^2\;\;,\;\;\;
q_{0{\rm max}} =M_B-\sqrt{M_{D}^2+\vec q\,^2}\;\;;
\label{MXc1}
\end{equation}
in this case the second moment also contributes, but only at the level of
${\cal O}(\Lambda_{\rm QCD}^2)$. We will use the energy of $D^*$ in what
follows, and therefore need to define the corresponding threshold energy
$q^*_{0{\rm max}}$:
\begin{equation}
q^*_{0{\rm max}} =M_B-\sqrt{M_{D^*}^2+\vec q\,^2}\;\;.
\label{q*}
\end{equation}

The basic idea
is the same as that demonstrated in Sect. 3.4 in the toy model:
to order $\Lambda_{\rm QCD}$ in the SV limit
the weighted integral over the excited states
is proportional to $\overline{\Lambda} (\mu )$. In the previous
section
we considered the point of zero recoil; now we have to shift
from this point and consider terms proportional
to the square of the $c$ quark velocity. The
SV limit will be ensured by choosing
$|\vec q\,|\ll M_D$.

Let us assume first that all structure functions in eq. (\ref{five})
are known separately. Then it is most convenient to consider
the function $h_1$ for axial current transitions.
If we are aiming at effects linear in $\Lambda_{\rm QCD}$, all
$\mu_\pi^2$
and $\mu_G^2$ corrections can be neglected; we can include them, however.
Using Eq.~(\ref{m1a1}) and the similar one for the vector current as well as
the expansion (\ref{Delta}) we get
$$\frac{1}{2\pi}
\int_{q_{0{\rm max}}-\mu}
dq_0(q^*_{0{\rm max}}-q_0) w_1^{V-A}
= \frac{\vec q\,^2}{2M_D^2}\left\{\overline{\Lambda}+
\frac{\overline{\Lambda}^2}{m_c}-\frac{\mu_\pi^2}{3m_c}-
2\frac{\mu_\pi^2-\mu_G^2}{3m_b}\right\} +
$$
\begin{equation}
+\;{\cal O}\left(\frac{\vec q \,^4}{M_D^4},\, \Lambda_{\rm QCD}^3\right)
\label{q-qb}
\end{equation}
where the hadronic tensor considered is that induced by the
$V-A$ current. In this equation $\vec q$ is supposed to be fixed  and small as
compared to $m_c^2$.

Let us note the explicit presence of the normalization point $\mu$ in the lower
limit of integration. Contrary to the naive quantum mechanical
description, in real QCD the structure functions do not vanish when excitation
energy becomes larger than a typical hadronic mass scale, but rather contain a
long tail associated with the gluon emission.
To reiterate the conclusion of Sect.\hspace*{.15em}3.5: we introduce a
normalization point
$\mu$ in such a way that all frequencies smaller than
$\mu$ can be considered as ``soft" or
inherent to the bound state wave function; at the same time
$\alpha_s(\mu )$ has to be sufficiently small for
the perturbative
expansion in $\alpha_s(\mu )/\pi$ to make sense. We then draw a
line
at $q_0 =q^*_{0{\rm max}}-\mu$ (the picture
is similar to that of Fig. 4, with $d\Gamma /dE_\phi$
replaced by $w_1^{V-A}$ and $E_\phi$ by $q_0$). The integral (\ref{q-qb}) taken
over the range
$q^*_{0{\rm max}}-\mu$ to $q^*_{0{\rm max}}$ represents
$(\vec v\,^2/2)\overline{\Lambda} (\mu)$
modulo corrections of higher order in $v$ and in $\Lambda_{\rm
QCD}$.
The running mass is then defined as $m_b(\mu )
=M_B - \overline{\Lambda} (\mu )$.

Practically it may be not so easy
to separate different structure functions from each other, which would require
the triple decay distribution over $q_0$, $q^2$ and $E_\ell$. A similar
prediction can be given for double differential distribution in the
semileptonic decay when the integral over the energy of lepton is considered.
The explicit form of the sum rule depends again on whether
one fixes $q^2$ or $\vec q\,^2$ in the process.
Below we assume that $\vec q\,^2$ is kept fixed.

Using the relation
\begin{equation}
\frac{d^2\Gamma}{dq_0 d\vec q\,^2}=|V_{cb}|^2\frac{G_F^2}{16\pi^4}
|\vec q\,| \left[(q_0^2-\vec q\,^2) w_1 -
\frac{\vec q\,^2}{3} w_2\right]
\label{double}
\end{equation}
and Eqs.~(\ref{m1a1}), (\ref{m1a2}), (\ref{second}) and (\ref{Delta})
one arrives at
$$
\int dq_0 (q^*_{0{\rm max}}-q_0) \frac{d^2\Gamma}{dq_0 d\vec q\,^2}=
\frac{G_F^2}{8\pi^3}|V_{cb}|^2(m_b-m_c)^2\frac{|\vec q\,|^3}{2M_D^2}\;\times
$$
$$
\times\;
\left\{\overline{\Lambda} \,+\, \frac{\overline{\Lambda}^2}{m_c}\,
-\,\frac{\mu_\pi^2-\mu_G^2}{3}\left[\frac{1}{m_c}+\frac{20}{3(m_b-m_c)}
+\frac{2}{m_b}\right]\;-\right.
$$
\begin{equation}
\left.
-\;\frac{\mu_G^2}{3}\left[\frac{1}{m_c}+\frac{4}{m_b-m_c}-
\frac{4m_c}{3(m_b-m_c)^2}\right]\right\}\;+
\;{\cal O}\left(|\vec q\,|^5, \Lambda_{\rm QCD}^3\right)\;.
\label{q-q}
\end{equation}

Again, one can consider radiative corrections. For simplicity we will neglect
all terms that are $\sim \Lambda_{\rm QCD}^2$ and higher (then the difference
between $q^*_{0{\rm max}}$ and $q_{0{\rm max}}$ becomes unimportant), and
limit ourselves
only by terms $\sim \vec q\,^2$. Including the
radiative tail we have
$$
\frac{8\pi^3}{G_F^2|V_{cb}|^2}
\int_{|\vec q\,|}^{q_{0{\rm max}}}
dq_0 (q_{0{\rm max}}-q_0) \frac{1}{|\vec q\,|}
\frac{d^2\Gamma}{dq_0 d\vec q\,^2}\;=
$$
\begin{equation}
\frac{8\pi^3}{G_F^2|V_{cb}|^2}\:\left\{
\int_{q_{0{\rm max}}-\mu}^{q_{0{\rm max}}}
dq_0 (q_{0{\rm max}}-q_0) \frac{1}{|\vec q\,|}
\frac{d^2\Gamma}{dq_0 d\vec q\,^2}\;+\;
\int_ {|\vec q\,|}^{q_{0{\rm max}}-\mu} dq_0 (q_{0{\rm max}}-q_0)
\frac{1}{|\vec q\,|}
\frac{d^2\Gamma}{dq_0 d\vec q\,^2} \right\}\;\;.
\label{ddr}
\end{equation}
The first term in the right hand side, up to the known factor,
has the meaning of the running value of
$\overline{\Lambda}$
if $\mu$ is much smaller than $|\vec q\,|$:
\begin{equation}
\frac{8\pi^3}{G_F^2|V_{cb}|^2}\:
\int_{q_{0{\rm max}}-\mu}^{q_{0{\rm max}}}
dq_0 (q_{0{\rm max}}-q_0) \frac{1}{|\vec q\,|}
\frac{d^2\Gamma}{dq_0 d\vec q\,^2}\;=
\;(m_b-m_c)^2 \frac{|\vec q\,|^2}{2m_c^2} \:\overline{\Lambda}(\mu)
\label{ddrlam}
\end{equation}

If $q_{0{\rm max}}-q_0$ is still sufficiently large
the integrand coincides with
the result obtained in the perturbative calculation. To
the first order in
$\alpha_s$ it is given by the probability to emit gluon with energy
$\omega\approx q_{0{\rm max}}-q_0$. It looks particularly simple
when $q_{0{\rm max}}-q_0\ll |\vec q\,|$.
As compared to the case of zero recoil we
have discussed previously, now the dipole
gluon emission
appears which is proportional to $\vec v\,^2$, and for small $\omega$ one can
neglect
$1/m$ suppressed amplitude considered before.
In the SV limit, when $\vec q\,^2 \ll m_c^2\,$, the results for the radiative
corrections obtained in Section 3.4 for the toy model are directly applicable
to the
semileptonic decays.  Indeed, the gluon can be emitted
(absorbed) either by the color charge of the $c$ quark or by its
magnetic
moment. Moreover, there is no interference -- the gluon emitted by
the magnetic moment has to be absorbed by the magnetic moment.
As long as we consider gluons with momenta much less than $\vec q$,
we can disregard the gluon interaction
with the magnetic moment. Then we are left with the charge
interaction only which is the same for spin-0 bosons and spin-1/2
fermions.  As a result, the expression for the radiative correction
obtained previously in the toy model
is modified in a minimal way, only due to a
slightly different kinematics, and for $q_{0{\rm max}}-q_0
\ll |\vec q\,|$ we get~\footnote{Virtual corrections lead to
the term $\sim \delta(q_{0{\rm max}}-q_0)$ which vanishes in the moment we
consider, but for the zeroth moment {in the first sum rule} it cancels the
logarithmic singularity following from expression (\ref{q-qa}).}
\begin{equation}
\frac{8\pi^3}{G_F^2|V_{cb}|^2}\:
\frac{1}{|\vec q\,|} \frac{d^2\Gamma_{\rm pert}}{dq_0 d\vec q\,^2}\;=
\;(m_b-m_c)^2 \,\frac{|\vec q\,|^2}{2m_c^2}\;\frac{16\alpha_s}{9\pi(q_{0{\rm
max}}-q_0)}\;\;.
\label{q-qa}
\end{equation}
The above equation shows how the the value of $\overline{\Lambda}$ obtained
from the sum rules depends on
the parameter $\mu$; it coincides with Eq.~(\ref{della}).
If one does not introduce this explicit cutoff, the
value of $\overline{\Lambda}$ would generally scale like $\alpha_s\cdot m_Q$.

Thus, the  sum rules  (\ref{q-qb}) and (\ref{ddrlam}), (\ref{q-qa}) can be
used to elucidate what is  actually meant by the heavy quark mass.
This question is rather subtle since the heavy quark mass is a purely
theoretical parameter which is not directly measurable. On the other
hand, it is a very important parameter,  crucial in a wide range
of questions.

The sum rules (\ref{q-qb}) or (\ref{ddrlam})  express
$\overline{\Lambda}$ in terms of the integral over
the physically observable quantities. Therefore,
we have a suitable phenomenological definition of $\overline{\Lambda}$
and,
through this quantity, the heavy quark
mass. Of course, both of them depend explicitly on the
normalization point.

Although the above equations yield $\overline{\Lambda}(\mu )$
in terms of the excitation distributions which are, in principle,
experimentally measurable, this does not mean that
it can be easily measured in practice. Needless to say
that it has not been measured so far. Still, the expression (\ref{q-qb}),
combined with the Bjorken sum rule, implies \cite{grozin2} a lower
bound
on $\overline{\Lambda}(\mu )$ (see also \cite{voloshin}) which turns out to be
quite restrictive,
\begin{equation}
\overline{\Lambda}> 2\Delta_1\, (\rho^2-\frac{1}{4}) ,
\label{11v}
\end{equation}
where $\Delta_1$ is the mass difference between the first
excitation of $D$ and the $D$ meson;
$\rho^2$ is the
slope of the Isgur-Wise function \cite{12,13}. Numerically
the right-hand side of Eq. (\ref{11v}) is close to 0.5 GeV. Further
details can be found in Ref. \cite{grozin2}.

\subsection{The third sum rule in the SV kinematics}

In a similar manner one can use the third
sum rule to relate the kinetic
operator to the average value of
$(q_{0max}-q_0)^2$. The value of $\mu_\pi^2$ can be extracted in a
model-independent way
from the sum rule similar to Eq. (\ref{muphys}) if the double
differential
measurements are used; say,  for small velocity events
\begin{equation}
\mu_\pi^2 (\mu ) = 3\tilde \Gamma^{-1}v^{-2}\int_{q_{0_{max}}-
\mu}^{q_{0_{max}}}
\;\;\frac{d^2\Gamma}{dq_0 d\vec q\,^2} \;(q_{0_{max}}-q_0)^2\; dq_0
\label{muphys2}
\end{equation}
where  $v=|\vec q\,|/m_c$ or $|\vec q\,|/M_D$ (which particular mass
is used in the denominator does not matter in the approximation
considered); the
normalization $\tilde\Gamma$ is defined as
\begin{equation}
\tilde\Gamma=\int_{q_{0max}-
\mu}^{q_{0max}}
\frac{d^2\Gamma}{dq_0  d\vec q\,^2} \, dq_0 \;.
\label{norm}
\end{equation}
This determines the value of
$\mu_\pi^2$ normalized at point $\mu$.

Again, similarly to the situation with
$\overline\Lambda$, one can get a lower bound on $\mu_\pi^2$
without waiting till measurements of the double differential
distributions are done. To this end one combines
the third sum rule (\ref{muphys2}) with the Bjorken
sum rule and gets \cite{grozin2}
\begin{equation}
\mu_\pi^2 > 3\Delta_1^2\left(\rho^2 -\frac{1}{4}\right)
\label{mu}
\end{equation}
where the quantities on the right hand side are the same
as in Eq. (\ref{11v}). Numerically the right hand side
is close to 0.5 GeV$^2$, see Ref. \cite{grozin2}.

\section{Quantum-mechanical interpretation}

A quantum-mechanical approach to the derivation of the sum rules
for the heavy quark transitions has been suggested by Lipkin
\cite{lipkin}. The formalism he exploits is similar to that used
in the theory of the M\"{o}ssbauer line shape;  it is discussed in
detail in Lipkin's text-book \cite{lipkin1}.

Some of our results can be readily understood within the framework
of Lipkin's approach, and, therefore, it seems instructive to provide
a dictionary allowing one to translate (where possible) the
field-theoretic consideration in the language of quantum mechanics.

As a matter of fact the expressions for the moments of the
distribution
functions (i.e. averages of powers of the excitation energy) obtained
in Sect. 4.2 have very transparent quantum-mechanical interpretation,
which
was
already mentioned in brief. Namely, the $b\rightarrow c$ transition
in the
semileptonic
decay is an instantaneous replacement of the $b$ quark by $c$:
\begin{equation}
|X_c\rangle\;=\;\int d^3x \,{\rm e}^{i\vec{q}\vec{x}} \;\bar
c(0,\vec{x})\Gamma b(0,\vec{x})
| B \rangle \;= \;j_{\vec{q}}|B\rangle
\label{L1}
\end{equation}
where $\Gamma$ is some Dirac matrix ($\gamma_\mu$ for the
vector current and
$\gamma_\mu\gamma_5$ for the axial one). Using the fact that both
$b$ and $c$
quarks are heavy we can use non-relativistic expansion for the
current
$j_{\vec{q}}$.

For definiteness, we will consider here the zero recoil limit which was
discussed
above in most detail. Generalization to $\vec q \neq 0$ is
straightforward. As an example, for the vector and axial-vector
currents
we get
\begin{equation}
j^V_0(\vec{q}=0)=\int d^3x\;
\varphi^+_c(\vec{x})\left[1-\frac{1}{8}\left(\frac{1}{m_c}-
\frac{1}{m_b}\right)^2(\vec\sigma
\vec\pi)^2+{\cal O}\left(\frac{1}{m^3}\right)\right] \varphi_b(\vec{x})
\label{L2a}
\end{equation}
$$
j^A_k(\vec{q}=0)=\int d^3x\;
\varphi^+_c(\vec{x})\left[\sigma_k-\frac{1}{8m_c^2}(\vec\sigma
\vec\pi)^2\sigma_k-\frac{1}{8m_b^2}\sigma_k(\vec\sigma
\vec\pi)^2\;+ \right.
$$
\begin{equation}
\left.
+\;\frac{1}{4m_cm_b}(\vec\sigma\vec\pi )\sigma_k(\vec\sigma\vec
\pi )+
{\cal O}\left(\frac{1}{m^3}\right)\right]\varphi_b(\vec{x})
\label{L2b}
\end{equation}
where $\varphi_{c,b}$ are two-component nonrelativistic spinor fields
defined in Eq.~(\ref{18a}).
Other components of the currents do not
contain the leading $(1/m)^0$ terms. Eqs.~(\ref{L2a}), (\ref{L2b})
are still written in
the second-quantized form where $\varphi_{c,b}$ are field operators.
However,
since the number of heavy quarks does not change in the process at
hand it is
convenient to proceed to the first-quantized form. It corresponds to
an
ordinary nonrelativistic description  by a two component
wave function
$$
\Psi_\alpha(\vec{x}_Q,\{x_{\rm light}\})\;,\;\;\;\alpha=1,2
$$
where $\vec{x}_Q$ is the heavy quark coordinate, $\alpha$ is the
heavy quark
spinor index and $\{x_{\rm light}\}$ represents an infinite number
of light
degrees of freedom. Note that it does not imply any non-relativistic
approximation for the light cloud; $\{x_{\rm light}\}$ are still
field-theoretical
coordinates of QCD.  From now on we will not write out explicitly
these
coordinates in the argument of the wave functions.

Using these notations Eq.\hspace*{.15em}(\ref{L1}) takes the form (at
$\vec{q}=0$)
\begin{equation}
\Psi^{(X_c)}(\vec{x}_Q,\{...\})=\left[1-\frac{1}{8}\left(\frac{1}{m_c}-
\frac{1}{m_b}\right)^2(\vec\sigma
\vec\pi)^2+{\cal
O}\left(\frac{1}{m^3}\right)\right]\Psi^{(B)}(\vec{x}_Q,\{...\})
\label{L3a}
\end{equation}
for $j=j_0^V$, and
$$
\Psi_k^{(X_c)}(\vec{x}_Q,\{{...}\})=\left[\sigma_k-
\frac{1}{8m_c^2}(\vec\sigma
\vec\pi)^2\sigma_k-
\frac{1}{8m_b^2}\sigma_k(\vec\sigma\vec\pi)^2 \;+ \right.
$$
\begin{equation}
\left.
+\;\frac{1}{4m_cm_b}(\vec\sigma\vec\pi)\sigma_k(\vec\sigma\vec\pi
)+
{\cal O}\left(\frac{1}{m^3}\right)\right]
\Psi^{(B)}(\vec{x}_Q,\{{...}\})\;\;\;\;\mbox{ for
}\, \; \; j=j_k^A\, .
\label{L3b}
\end{equation}
The derivative appearing in the momentum operators
$\pi^i=-i(\partial/\partial x_Q^i)+A^i(x_Q)$
acts on the coordinate $x_Q$ of the wave function $\Psi_\alpha$
while
$\sigma$
matrices act on its spin index.

In relativistic and nonrelativistic theories the expectation values
$\langle B|...|B \rangle$ are normalized differently. In relativistic
theory
$\langle B|B \rangle=2M_BV$ (in the $B$ meson rest frame) where
$V$ is the
volume of the ``large box'', while $\langle \Psi^{(B)}|\Psi^{(B)}
\rangle= 1$;
the degree of freedom associated with motion of a system ($B$
meson) as a
whole is usually not considered being treated separately.
Correspondingly, the
transition rule is as follows:
\begin{equation}
\langle B(\vec{p}\,)|\int\,d^3x\;\bar b O b (x)|B(\vec{p}\,)\rangle_{\rm
second\;
quant.}\;\rightarrow \langle \Psi^{(B)}|O|\Psi^{(B)} \rangle_{\rm QM}
\label{transit}
\end{equation}
where $O$ is some local operator, e.g. $\gamma_0$, $\pi_i$, etc.

The characteristic feature of the currents considered above is that the
states
produced by them from $\Psi^{(B)}$ are close to $\Psi^{(D)}$ and
$\Psi^{(D^*)}\approx \sigma_k \Psi^{(D)}$, respectively. Moreover, at
$\vec{q}=0$ the currents do not contain terms linear in $1/m\,$;
corrections
start with operators that explicitly contain $1/m^2$. Therefore,
$$
\Psi^{(X_c)}(\vec{x}_Q)=\left[1-\frac{1}{8}\left(\frac{1}{m_c}-
\frac{1}{m_b}\right)^2\langle B|(\vec\sigma\vec\pi)^2|B\rangle
\right]
\Psi^{(B)}(\vec{x}_Q)
$$
\begin{equation}
-\frac{1}{8}\left(\frac{1}{m_c}-\frac{1}{m_b}\right)^2\sum_n
\Psi^{(B_n)}(\vec{x}_Q)\,\langle B_n|(\vec\sigma\vec\pi)^2| B\rangle
\;+\;
{\cal O}\left(\frac{1}{m^3}\right)
\label{L4}
\end{equation}
$$
\Psi_k^{(X_c)}(\vec{x}_Q)=
$$
$$
\left[1+\frac{1}{6m_c^2}
\langle B|\vec\sigma\vec B |B\rangle
-\frac{1}{8}\left(\frac{1}{m_c^2}+\frac{1}{m_b^2}+
\frac{2}{3m_cm_b}\right)\langle B|(\vec\sigma\vec\pi)^2|B\rangle
\right]
\sigma_k\Psi^{(B)}(\vec{x}_Q)\;+
$$
\begin{equation}
+\sum_n \;\Psi^{(B_n)}(\vec{x}_Q)\,\langle B_n |\,
\left(-\frac{1}{8m_c^2}(\vec\sigma
\vec\pi)^2\sigma_k-\frac{1}{8m_b^2}\sigma_k(\vec\sigma
\vec\pi)^2
+\frac{1}{4m_cm_b}(\vec\sigma\vec\pi )\sigma_k(\vec\sigma\vec
\pi )\right) | B\rangle +
{\cal O}\left(\frac{1}{m^3}\right)\; .
\label{L5}
\end{equation}
Here $|B_n\rangle$ are excitations of $B$ meson, $\Psi^{(B_n)}$ are
the
corresponding wave functions.
(Here and below in this section all matrix elements are written down
in the
nonrelativistic normalization; the factor $(2M_B)^{-1}$ characteristic
of the
relativistic normalization is absent.)

Let us emphasize that $\Psi^{(B)}$,
$\Psi^{(B_n)}$ are the eigenfunctions of the Hamiltonian ${\cal
H}(m_Q)$
with
$m_Q=m_b$, and {\em not} the one with $m_Q=m_c$. In the final
state the
observable
hadrons are described by the eigenstates of ${\cal H}(m_Q=m_c)$.
Therefore the
amplitude of an individual state $|D_n\rangle$ is obtained by
projecting
$|\Psi^{(X_c)}\rangle\,$, $\,|\Psi_k^{(X_c)}\rangle$ onto
these eigenfunctions. The projection onto the ground state is equal to
unity up
to $1/m^2$ terms, whereas the excited states are produced with the
amplitudes
which scale like $1/m$. This reexpansion is the only source of $1/m$
terms in
the amplitudes.

With this information at hand let us return to the sum rules. We start
from the
first sum rules (see the expressions for $I_0^{VV}$, $I_0^{(1)AA}$ in
Eqs.~(\ref{vmoments}), (\ref{moments})). These sum rules have the
meaning of
the total probability of hadronic production at zero recoil if it is
assumed
that arbitrary energy can be carried away by the lepton pair
produced in the
decay process
(see
Eqs.~(\ref{SRV}), (\ref{SR!})). In the nonrelativistic language these
total
probabilities are nothing but the norms of the states $\Psi^{(X_c)}$
and
$\Psi_k^{(X_c)}$. There is no need in reexpansion in order  to find
these
norms.
As a matter of fact, the result is given by the coefficient in front of
$\Psi^{(B)}$
(or $\sigma_k\Psi^{(B)}$); other terms contribute only at the level of
$1/m^4$.
The concrete expressions, thus, are
\begin{equation}
\langle \Psi^{(X_c)}|\Psi^{(X_c)}\rangle =
\langle B|\,(j_0^V)^+\,j_0^V\,|B \rangle =
1-\frac{1}{4}\left(\frac{1}{m_c}-\frac{1}{m_b}\right)^2
\,\langle B|(\vec\sigma\vec\pi)^2|B\rangle +
{\cal O}\left(\frac{1}{m^3}\right)\, ,
\label{L6}
\end{equation}
$$
\frac{1}{3}\sum_{k=1}^3\langle \Psi_k^{(X_c)}|\Psi_k^{(X_c)}\rangle =
\frac{1}{3}\langle B|\,(j_k^A)^+\,j_k^A\,|B\rangle =
1+\frac{1}{3m_c^2}\langle B|\vec\sigma\vec B|B\rangle -
$$
\begin{equation}
-\frac{1}{4}\left(\frac{1}{m_c^2}+\frac{1}{m_b^2}+
\frac{2}{3m_cm_b}\right)
\,\langle B|(\vec\sigma\vec\pi)^2|B\rangle +
{\cal O}\left(\frac{1}{m^3}\right)\, ,
\label{L7}
\end{equation}
which exactly coincides with the results presented in
Eqs.~(\ref{SRV}),
(\ref{SR!}).

It is natural to ask at this point how could it happen that the
relativistic
current $\bar b \gamma_\mu c$ or $\bar b \gamma_\mu
\gamma_5 c$ that was
seemingly normalized to unity (through equal time commutators)
lead us to
nonrelativistic current with the normalization different from unity.
The
answer is
that in the
nonrelativistic consideration one excludes the states whose excitation
energy $\sim
m_Q$. In fact, adding such states would restore the normalization
back
to unity. This explains, in particular, the negative sign of $1/m^2$
corrections to the nonrelativistic normalization.
(Let us mention in passing that Lipkin in his analysis did not
consider
these $1/m^2$ relativistic corrections altogether.)

The fact that the states
$\Psi^{(B)}(\vec{x}_Q)$ and $\sigma_k\Psi^{(B)}(\vec{x}_Q)$ are not
the
eigenstates
of the Hamiltonian ${\cal H}(m_c)$ becomes important for higher
moments $I_n$,
Eq.\hspace*{.15em}(\ref{95a}), which in the quantum-mechanical
language take the form
\begin{equation}
I_n^{VV}=\langle \Psi^{(B)}(\vec{x}_c)|\left\{ {\cal H}(m_c)-
\langle D|{\cal
H}(m_c)| D \rangle \right\}^n|\Psi^{(B)}(\vec{x}_c)\rangle
\,\left(1+{\cal
O}\left(\frac{1}{m}\right)\right)
\label{L8}
\end{equation}
and
\begin{equation}
I_n^{AA}=\frac{1}{3}\langle \Psi^{(B)}(\vec{x}_c) |\sigma_k\left\{
{\cal
H}(m_c)-
\langle D^*|{\cal H}(m_c)| D^*\rangle \right\}^n \sigma_k|
\Psi^{(B)}(\vec{x}_c)\rangle \,
\left(
1+ {\cal O}\left(\frac{1}{m}\right)\right)\, .
\label{L9}
\end{equation}
The $1/m^2$ corrections to currents do not show up in the leading
term
presented above. From now on we will abbreviate $\langle D|{\cal
H}(m_c)|D\rangle =
\langle{\cal H}(m_c)\rangle_D$, and likewise for $D^*$.

As we have  already seen, all excitation probabilities are proportional
to
$1/m^2$ and, therefore,
$$
I_n \sim \frac{\Lambda_{\rm QCD}^{n+2}}{m_Q^2}\;\;\;\;\mbox{ for
}\;\;n\ge
1\;\;.
$$
The very same $1/m^2$ behavior is also readily seen from
Eqs.~(\ref{L8}), (\ref{L9}). Indeed, ${\cal H}(m_c)-\langle{\cal
H}(m_c)\rangle$
acting on the states $\Psi^{(B)}(\vec{x}_c)\,$, $\;\sigma_k
\Psi^{(B)}(\vec{x}_c)$
produces a nonvanishing result only at the level $1/m$:
$$
\left({\cal H}(m_c)-\langle{\cal H}(m_c)\rangle_D\right)
|\Psi^{(B)}(\vec{x}_c)\rangle\;=
$$
$$
\;=\left({\cal H}(m_b)-\langle{\cal
H}(m_c)\rangle_D+
{\cal H}(m_c)-{\cal H}(m_b)\right) |\Psi^{(B)}(\vec{x}_c)\rangle=
$$
\begin{equation}
\left((M_B-M_D)-(m_b-
m_c)+(\vec\sigma\vec\pi)^2\left(\frac{1}{2m_c}-
\frac{1}{2m_b}\right) \right)|\Psi^{(B)}(\vec{x}_c)\rangle + {\cal
O}\left(\frac{1}{m^2}\right)
\label{L10}
\end{equation}
and, by the same token,
$$
\left({\cal H}(m_c)-\langle{\cal H}(m_c)\rangle_{D^*}\right)
\sigma_k
|\Psi^{(B)}(\vec{x}_c)\rangle=
\left[ \sigma_k
\left( (M_B-M_{D^*})-(m_b-m_c)\,+\right.\right.
$$
\begin{equation}
\left.\left.
+\,(\vec\sigma\vec\pi)^2\left(\frac{1}{2m_c}-
\frac{1}{2m_b}\right) \right) - \frac{1}{2m_c}
[\sigma_k,\vec\sigma\vec B]\,\right]|\Psi^{(B)}(\vec{x}_c)\rangle +
{\cal
O}\left(\frac{1}{m^2}\right)\;\;.
\label{L11}
\end{equation}

Using these equations it is easy to obtain the predictions for the
moments with
$n\ge2\,$. To this end we act by the left-most and the right-most
operators
${\cal H}(m_c)-\langle{\cal H}(m_c)\rangle$ onto $\langle\Psi^{(B)}|$
and
$|\Psi^{(B)}\rangle\,$, respectively. This produces an overall
$1/m^2$
suppression. For vector currents we have
$$
I_n^{VV}=\langle \Psi^{(B)} |\left({\cal H}(m_c)-\langle{\cal
H}(m_c)\rangle_D\right)^n| \Psi^{(B)}\rangle =
$$
$$
\langle\Psi^{(B)}|\,\left((\vec\sigma
\vec\pi)^2\left(\frac{1}{2m_c}-
\frac{1}{2m_b}\right)+\delta_V\right)
\left({\cal H}-\langle{\cal H}\rangle\right)^{n-2}\left((\vec\sigma
\vec\pi)^2\left(\frac{1}{2m_c}-
\frac{1}{2m_b}\right)+\delta_V\right)
|\Psi^{(B)}\rangle
$$
\begin{equation}
+ \;\;{\cal O}\left(\frac{1}{m^2}\right)\;\; ,
\label{L12}
\end{equation}
while for the axial-vector currents
$$
I_n^{AA}=\frac{1}{3}\langle \Psi^{(B)}|\sigma_k\left({\cal H}(m_c)-
\langle{\cal
H}(m_c)\rangle_{D^*}\right)^n\sigma_k|\Psi^{(B)}\rangle =
$$
$$
\frac{1}{3}
\langle \Psi^{(B)}|\,\left(\left((\vec\sigma
\vec\pi)^2\left(\frac{1}{2m_c}-
\frac{1}{2m_b}\right)+\delta_A\right)\sigma_k + \frac{1}{2m_c}
[\sigma_k,\vec\sigma\vec B]\,\right)
\left({\cal
H}-\langle{\cal H}\rangle\right)^{n-2} \times
$$
\begin{equation}
\times
\left(\sigma_k
\left((\vec\sigma
\vec\pi)^2\left(\frac{1}{2m_c}-
\frac{1}{2m_b}\right)+\delta_A\right)- \frac{1}{2m_c}
[\sigma_k,\vec\sigma\vec B]\,\right)\,|\Psi^{(B)}\rangle
+ {\cal O}\left(\frac{1}{m^2}\right)
\label{L13}
\end{equation}
where
$$
\delta_V=(M_B-M_D)-(m_b-m_c)\;\;\;,\;\;\;\delta_A=(M_B-M_{D^*})-
(m_b-m_c)
$$
are given in Eqs.~(\ref{vmshift}), (\ref{mshift}). Note that in the
remaining
operators $({\cal H}-\langle{\cal H}\rangle)^{n-2}$ one can already
neglect the
explicit heavy quark mass dependence.
For $n>2$ the terms with
$\delta$ lead only to corrections higher than $1/m^2$ and can be
omitted.
Indeed, picking up $\delta$ rather than $(\vec\sigma\vec\pi)^2$
allows one to
act by $({\cal H}-\langle{\cal H}\rangle)$ directly on $\Psi$ or
$\sigma_k\Psi$ which results, in turn,  in an
additional $1/m$ suppression. The case $n=1$ is
somewhat special and will be discussed shortly.

One can easily see that Eqs.~(\ref{L12}), (\ref{L13}) coincide with our
previous results for the moments, Eqs.~(\ref{vmoments}),
(\ref{moments}),
provided that one identifies $\pi_0^{(n-2)}$ in the latter with $({\cal
H}-
\langle{\cal H}\rangle)^{(n-2)}$ in the quantum-mechanical relations.
This correspondence
has a transparent physical meaning and can be readily proven.
Indeed, both play the role of the
generator of the
time evolution, $({\cal H}-\langle{\cal H}\rangle)$ in the
first-quantized approach and
$\pi_0=i\partial_0+A_0$ in the second-quantized formalism.

To see that this is indeed the case, let us consider the state
\begin{equation}
|\Psi_O(t)\rangle\;\;=\;\;\bar b(\vec{x}=0,t)\, O \,b(\vec{x}=0,t)\,|B\rangle
\label{L14}
\end{equation}
where $O$ is an arbitrary operator. Then
\begin{equation}
\langle \Psi_O(0)|\Psi_O(t)\rangle=\langle B| \,\bar b(\vec{x}=0,t=0)\,O^+
\:{\rm T}\,{\rm
e}^{-i\int_0^t A_0(\vec{x}=0,\tau)d\tau}\:O\,
b(\vec{x}=0,t)\,|B\rangle
\label{L15}
\end{equation}
in the leading in $1/m$ approximation
when the heavy quark Green function reduces to the $T$-exponent above.
Expanding the right-hand side in $t$ one
gets
\begin{equation}
\langle \Psi_O(0)|\Psi_O(t)\rangle\;=\;\langle B|\bar b(\vec{x}=0,t=0)\,O^+\:
\sum_{n=0}^{\infty} (i\partial_0+A_0)^n \frac{(-it)^n}{n!} \:O \,
b(\vec{x}=0,t)|B\rangle \; ,
\label{L16}
\end{equation}
that proves the assertion above.

Let us return now to the discussion of the special case of the $n=1$
moments (the second sum rule). This
case is singled out because, on one hand,  $I_1$ must be
proportional to $\Lambda_{\rm QCD}^3/m^2$ on general grounds,
and, on the
other hand, this
suppression does
not immediately show up as it happens for $n\ge 2$, since now we
have only one power of ${\cal H}-\langle{\cal H}\rangle$ sandwiched
between the bra and ket states. To
calculate
$I_1^{VV}$ and $I_1^{AA}$ we need to account for all $1/m^2$ terms
that have
been neglected in Eqs.~(\ref{L10}), (\ref{L11}). Including these terms
we get
$$
I_1^{VV}=\left((M_B-M_D)-(m_b-m_c)\right) +
\langle
\Psi^{(B)}(\vec{x}_c)|
\left(\frac{1}{2m_c}-\frac{1}{2m_b}\right) (\vec \sigma \vec \pi)^2
\;+
$$
\begin{equation}
+\frac{1}{8}\left(\frac{1}{m_c^2}-\frac{1}{m_b^2}\right)
\left( -(\vec D \vec E) + 2\vec\sigma \cdot \vec E\times\vec \pi
\right)|\Psi^{(B)}(\vec{x}_c)\rangle +
{\cal O}\left(\frac{\Lambda^4_{\rm QCD}}{m^3}\right)\;\; ,
\label{L17}
\end{equation}
and
$$
I_1^{AA}=\left((M_B-M_{D^*})-(m_b-m_c)\right) +
\langle
\Psi^{(B)}(\vec{x}_c)|
\left(\frac{1}{2m_c}-\frac{1}{2m_b}\right) (\vec \pi\vec
\sigma)^2-\frac{2}{3m_c}\vec\sigma\vec B \;+
$$
\begin{equation}
+\frac{1}{8}\left(\frac{1}{m_c^2}-\frac{1}{m_b^2}\right)
\left(-(\vec D \vec E) + \vec\sigma \cdot \vec E\times\vec \pi
\right)-\frac{1}{3m_c^2}\vec\sigma \cdot \vec E\times\vec \pi
|\Psi^{(B)}(\vec{x}_c)\rangle +
{\cal O}\left(\frac{\Lambda^4_{\rm QCD}}{m^3}\right)
\label{L18}
\end{equation}
where we have used the expression (\ref{hamil}) for the
$1/m^2$ part of the Hamiltonian.

If one neglects the $1/m^2$ terms then $I_1^{VV}\,$, $\,I_1^{AA}$
vanish, and the
known relations Eqs.~(\ref{vmshift}), (\ref{mshift}) for hadron
masses at the
level $1/m$ arise. Eqs.~(\ref{L17}), (\ref{L18}) are extension of these
relations to the order $1/m^2$. There are two sources of $1/m^2$
corrections in
the right-hand side: the first one is explicit $1/m^2$ terms in these
equations; the second source is an implicit $(1/m^1)$ dependence of
the
expectation values of operators $\vec \pi^2$ and $\vec\sigma \vec
B$ over the
state $\Psi^{(B)}(\vec{x}_c)$. This dependence is due to the fact that
$\Psi^{(B)}(\vec{x}_c)$ is the eigenfunction of the Hamiltonian ${\cal
H}(m_b)$
rather than of the asymptotic  one ${\cal H}(m_Q=\infty)$. It is
obvious, by
the way, that these matrix elements do not produce $1/m_c$
dependence.

Let us discuss in brief how this quantum-mechanical formalism
works in the case
when $\vec q\ne0$. The instantaneous $b \rightarrow c$ transition
now not
only replaces
$b$ by $c$ but also boosts the $c$ quark providing it with the spatial
momentum
$-\vec q$. As a result the wave function of the system produced
takes the
following form for the vector and axial-vector currents, respectively:
$$
\Psi^{(X_c)}(\vec{x}_Q,\{...\})={\rm e}^{-i\vec q\vec x_Q}\left(1-
\frac{\vec
v\,^2}{8} \right) \cdot\Psi^{(B)}(\vec{x}_Q,\{...\})\;+
$$
\begin{equation}
+\;\;{\cal O}\left(\vec v\,^4,\; \frac{\Lambda_{\rm QCD}|\vec v\,|}{m},
\;\frac{\Lambda^2_{\rm QCD}}{m^2}\right)\;\;,
\label{L36}
\end{equation}
$$
\Psi_k^{(X_c)}(\vec{x}_Q,\{...\})={\rm e}^{-i\vec q\vec x_Q}\left(1-
\frac{\vec
v\,^2}{8} \right) \cdot\sigma_k \Psi^{(B)}(\vec{x}_Q,\{...\})\;+
$$
\begin{equation}
+\;\;{\cal O}\left(\vec v\,^4,\; \frac{\Lambda_{\rm QCD}|\vec v\,|}{m},
\;\frac{\Lambda^2_{\rm QCD}}{m^2}\right)
\label{L37}
\end{equation}
where $\vec v=-\vec q/m_c$ is the velocity of the final $c$ quark in
the SV
limit. The factor $1-\vec v\,^2/8$ reflects the overall normalization of
the
currents; it results in the factor  $1-\vec v\,^2/4$ in the $n=0$
moments
(the Bjorken sum rule). In
higher
moments it is not essential at the level of accuracy we accept here.

The moments $I_n(\vec q\,)$ are defined in Eq.~(\ref{q2}), but here we define
$\epsilon$ for the vector case as $\epsilon=M_B-q_0-\sqrt{M_D^2+\vec q\,^2}$.
The generic formulae for the moments take the form:
\begin{equation}
I_0^{VV}=I_0^{AA}=1-\frac{\vec v\,^2}{4}\, ,
\label{L40}
\end{equation}
\begin{equation}
I_n^{VV}=\langle \Psi^{(B)}|\left({\rm e}^{i\vec q\vec x_Q}{\cal
H}(m_c){\rm
e}^{-i\vec q\vec x_Q}-\sqrt{M_D^2+\vec
q\,^2}\right)^n|\Psi^{(B)}(\vec{x}_c)\rangle
+{\cal O}\left(\vec v\,^4,
\;\frac{\Lambda^2_{\rm QCD}}{m^2}\right)\;,
\label{L41}
\end{equation}
\begin{equation}
I_n^{AA}=\frac{1}{3}
\langle \Psi^{(B)}|\sigma_k\left({\rm e}^{i\vec q\vec x_Q}{\cal
H}(m_c){\rm
e}^{-i\vec q\vec x_Q}-\sqrt{M_{D^*}^2+
\vec q\,^2}\right)^n \sigma_k|\Psi^{(B)}(\vec{x}_c)\rangle
\;+
\label{L42}
\end{equation}
$$
+\;\;
{\cal O}\left(\vec v\,^4,
\;\frac{\Lambda^2_{\rm QCD}}{m^2}\right)\;\; .
$$
Let us notice that the only effect of the exponent is to replace the
operator $\vec \pi$ in the Hamiltonian by $\vec \pi-\vec q\,$:
$$
{\rm e}^{i\vec q\vec x_Q}{\cal H}(m_c;\,\vec \pi){\rm
e}^{-i\vec q\vec x_Q}\;=\;{\cal H}(m_c;\,\vec \pi-\vec q)\;=
$$
\begin{equation}
=\;{\cal H}(m_c;\,\vec \pi)-\frac{\vec q\vec \pi}{m_c}+\frac{\vec
q\,^2}{2m_c}-
\frac{1}{4m_c^2}\vec q \cdot \vec \sigma\times\vec E + {\cal
O}\left(\frac{1}{m_c^3}\right)\, .
\label{L43}
\end{equation}

As an example, let us consider the $\vec q\,^2$ dependence of the
first few
moments. For $n=0$ the result for $dI_0/d\vec v\,^2$ at $\vec
v\,^2=0$
coincides with the Bjorken sum rules;
it has been derived in Sect.~5 with even better accuracy.

The next moment to consider is $n=1$. Here we get for  $dI_1/d\vec
v\,^2$ at $\vec v\,^2=0$
\begin{equation}
\frac{dI_1^{VV}}{d\vec v\,^2}\mid_{\vec
v=0}\;=\;\frac{dI_1^{AA}}{d\vec
v\,^2}\mid_{\vec v=0}\;=\;\frac{\bar \Lambda}{2}\;\;\;,\;\;\;\bar
\Lambda
\simeq M_D-m_c\simeq M_{D^*}-m_c\;\;.
\label{L25}
\end{equation}
This is Voloshin's ``optical'' sum rule \cite{voloshin}.

Finally, let us mention $n=2$ case, where at
$\vec v\,^2=0$
\begin{equation}
\frac{dI_1^{VV}}{d\vec v\,^2}\mid_{\vec
v=0}\;=\;\frac{dI_1^{AA}}{d\vec
v\,^2}\mid_{\vec v=0}\;=\;\frac{1}{3} \langle \vec\pi\,^2\rangle\;\;.
\label{L26}
\end{equation}

Concluding this section, let us make a few comments on those aspects
of the quantum-mechanical approach which we modified compared
to the
original
Lipkin's
presentation. The central point of the framework suggested in Ref.
\cite{lipkin}  is the fact that the
wave function of the charmed system immediately after the
instantaneous $b
\rightarrow c$ transition is known in terms of the $B$ meson
wave function, see
Eqs.~(\ref{L4}), (\ref{L5}) and (\ref{L36}), (\ref{L37}). These
equations
differ from their counterparts in Ref.~\cite{lipkin} in the
normalization
of the
currents; there it was effectively set equal to unity. In our
expressions the
charm wave function is {\em not} normalized to unity, with the
corrections
appearing
at the level ${\cal O}(\Lambda^2_{\rm QCD}/m^2)$ or ${\cal O}(\vec
v\,^2)$. This correction to normalization affects only the first sum
rules,
viz. $I_0^{VV}\,,\;\;I_0^{AA}$.

Another point of difference is that Lipkin did not account for the
fact that
the spin part of the Hamiltonian depends on $m_Q$ at the same level
as the
kinetic energy. As a result, our sum rules at zero recoil are, strictly
speaking, different from those of Ref. \cite{lipkin}. (Our results
coincide with
those of Lipkin provided the spin
interaction is
switched off.) At the level of accuracy we accept in this section
the spin terms
omitted in Ref. \cite{lipkin} do not affect at all the sum rules at $\vec
q\neq
0$. In
particular, the linear in $\vec q\,^2$ part in the first moments (see
Eq.~(\ref{L25}) for $\,I_1^{VV}$ and $I_1^{AA}$) derived in
Ref.~\cite{lipkin}
identically coincides with Voloshin's sum rule \cite{voloshin},
although surprisingly it was
not recognized in Ref.~\cite{lipkin}.

Our final remark concerns $\bar c b$ currents which vanish in the
nonrelativistic limit. For example,  in Sect.~4.4 we considered  the
spatial
components of the vector current $\bar c \gamma_k b$ at $\vec
q=0$. In the
nonrelativistic limit this current takes the form
\begin{equation}
\bar c \vec\gamma b\mid_{\vec
q=0}=\phi_c^+\left[\left(\frac{1}{2m_c}+\frac{1}{2m_b}\right)\vec\pi
+\left(\frac{1}{2m_c}-
\frac{1}{2m_b}\right)i\vec\sigma\times\vec\pi\right]\phi_b
+{\cal O}\left(\frac{1}{m^2}\right)\;\;.
\label{L30}
\end{equation}
This current produces the charmed state with the following
wave function:
\begin{equation}
\Psi_k(\vec{x}_Q)=\left[\left(\frac{1}{2m_c}+\frac{1}{2m_b}\right)
\pi_k
+\left(\frac{1}{2m_c}-
\frac{1}{2m_b}\right)i[\vec\sigma\times\vec\pi]_k\right]
\Psi^{(B)}(\vec{x}_Q)\;\;.
\label{L29}
\end{equation}
The normalization of this state is
$$
\frac{1}{3}\sum_{k=1}^3\langle\Psi_k(\vec{x}_Q)|\Psi_k(\vec{x}_Q)
\rangle\simeq
$$
\begin{equation}
\frac{1}{4}\left[\left(\frac{1}{m_c^2}+\frac{1}{m_b^2}-
\frac{2}{3m_cm_b}\right)
\langle\vec\pi\,^2\rangle+
\left(\frac{1}{m_b^2}-\frac{1}{3m_c^2}-
\frac{2}{3m_cm_b}\right)\langle
\vec\sigma\vec B\rangle\right]\;\;.
\label{31}
\end{equation}
This expression is identical to the left-hand side of the sum rule
(\ref{repres}) which is the $n=0$ moment for this current. For higher
moments
we get
$$
I_n^{(1)VV}=
\frac{1}{3}\sum_{k=1}^3\langle\Psi_k(\vec{x}_Q)|\left({\cal
H}(m_c)-\langle{\cal H}(m_c)\rangle\right)^n|
\Psi_k(\vec{x}_Q)\rangle\simeq
$$
$$
\frac{1}{3}\langle\Psi^{(B)}|\left[\left(\frac{1}{2m_c}+\frac{1}{2m_b}
\right)
\pi_k
-\left(\frac{1}{2m_c}-
\frac{1}{2m_b}\right)i[\vec\sigma\times\vec\pi]_k\right]
\left({\cal
H}(m_c)-\langle{\cal H}(m_c)\rangle\right)^n\times
$$
\begin{equation}
\times
\left[\left(\frac{1}{2m_c}+\frac{1}{2m_b}\right)\pi_k+
\left(\frac{1}{2m_c}-
\frac{1}{2m_b}\right)i[\vec\sigma\times\vec\pi]_k\right]|
\Psi^{(B)}\rangle\;\;.
\label{L32}
\end{equation}
Now the suppression $1/m^2$ is explicit -- it comes from the current
itself, and all
$1/m$ terms in ${\cal H}(m_c)$ can be omitted.

\section{Impact of the Perturbative Evolution of Operators in OPE}

In this section we shall briefly discuss  practical modifications
that arise in the calculation of nonperturbative corrections in
the heavy
flavor decays if one accounts for the normalization point
dependence of the corresponding operators. Although this
question is
rather standard, we feel the need to dwell on it in view of
apparent
confusion taking place in applications of the heavy quark expansions
existing in the literature.

We have already mentioned above that HQET is nothing else than
a version of the Wilson operator product expansion.
It is well understood (see, for example, a recent discussion in
Ref. \cite{15})
that  consistent incorporation of nonperturbative effects in the
framework of
the Wilson  procedure requires  separating
 momenta below and above a certain normalization
point $\mu$.
 The low-momentum physics is then
attributed to the
matrix elements of the operators whereas the high momentum part
enters the
Wilson expansion coefficients; both, therefore, depend explicitly on
$\mu$ in
such a way
that all observables are $\mu$ independent.

This procedure  is always performed  when the
corresponding
operators undergo logarithmic renormalization, for an obvious
reason:
the Feynman integrals determining the coefficient functions in this
case
logarithmically diverge in the infrared domain,  and one merely
cannot put $\mu$, the
infrared
regularization, to zero. In calculating the power corrections to
the heavy
flavor decays another situation can typically arise --
when the operators under consideration have vanishing anomalous
dimensions and no logarithmic mixing. This is the case, for instance,
with the leading operator $\bar Q Q$ and the kinetic energy operator
$\bar Q{\vec\pi}^2 Q$ (the second one still mixes with the first one
non-logarithmically, through power of $\mu^2$, see below).
 In this case the
coefficient functions possess a safe
infrared
limit, and it is very tempting just to calculate them in
perturbation theory {\em per se}, with no infrared cut off (or, which
is
the same, putting $\mu = 0$. ) This is what is routinely done with
respect to the coefficient of the operator $\bar Q Q$.

 From purely theoretical point of view there is no way one
can justify such a procedure.  It gives rise to questions
which have no consistent answers, e.g. how the sum of all
perturbative terms is to be understood, etc. One of the manifestations
of these inconsistencies is
the  infrared renormalon (see a recent discussion in
Refs. \cite{15,beneke} and references therein).  It should be very
clearly
realized that in the consistent operator product expansion (and,
hence, in HQET) one does not discriminate perurbative contributions
versus nonperturbative, but, rather, large-distance ones versus
short distances (all distances are measured in the scale of $\mu^{-
1}$).

However, in practice no  separation of the infrared part from
the coefficient of $\bar Q Q$ is carried out. The usual routine is as
follows: one  takes the known
expressions for the one-loop (or two-loop) perturbative corrections
for a particular quantity and merely adds to
these perturbative terms
nonperturbative corrections expressed {\em via}  certain matrix
elements. (We also follow this routine for numerical calculations). For
instance, the perturbative
correction
$\eta_A$ to the
the axial-vector  current $\bar c \gamma_\mu \gamma_5 b$
calculated from the standard Feynman graphs with no subtractions
was simply added to the
nonperturbative contribution of Eq. (\ref{1-F}). It is clear however
that the
nonperturbative contribution {\em per se} takes care of  all
relevant
gluon exchanges with momenta below $\mu$; on top of it  the
one-loop
Feynman
integral for the radiative correction $\eta_A$ has some (small) piece
coming
from the same
low-momentum domain.  The question which immediately comes to
one's mind is the menace of double counting.

The answer to this question is that in actual OPE-based calculations,
which are always truncated at some finite order in $\alpha_s$
and keep only a few higher-dimensional operators, one can stick to
what is called the ``practical version'' of OPE in QCD \cite{NSVZ}
and still avoid double counting.

Indeed, it is not possible to define the perturbative part in, say,
$\langle\bar Q{\vec\pi}^2Q\rangle$ to
all orders in $\alpha_s$. On the other hand,
it is quite possible to introduce a ``one-loop perturbative
contribution"
to $\bar Q{\vec\pi}^2Q$ normalized at $\mu$. To this end we,
{\em by definition}, take two gluon lines in $\bar Q{\vec\pi}^2Q$,
contract them to form the gluon loop, use the bare gluon
propagator and calculate the loop cutting the integration off from
above, at $k_g = \mu$.
We then get
$$
\bar Q{\vec\pi}^2Q|_{\rm one-loop}
=\frac{4\alpha_s}{3\pi}\mu^2 \bar Q Q .
$$
Now let us subtract and add this ``one-loop" $\pi^2$ from the
properly and scientifically defined $\bar Q{\vec\pi}^2Q$.
Moreover,
let us combine the added part with $C(\mu )\bar Q Q$;
then we get $\bar Q Q$ times the coefficient coinciding
-- up to corrections of higher order in $\mu^2/m^2$ --
with the coefficient obtained from the
full perturbative one-loop calculation, with no subtractions
whatsoever. Simultaneously, the matrix element of the kinetic
energy operator is replaced by that with the subtracted ``one-loop"
part. Strictly speaking, this quantity does not represent now matrix
element of any operator; hence, factorization inherent to the genuine
OPE is lost. This is unimportant for numerical analysis due to the fact
that the term which we added and subtracted is numerically small,
much smaller than the actual value of $\langle\pi^2\rangle$.
For this reason the replacement of the added term
plus $C(\mu) \bar Q Q$ by merely $C_{\rm one-loop}\bar Q Q$
introduces a very small numerical error, much smaller than
$\pi^2$ correction itself. This fact -- the numerical smallness of the
``one-loop" value of the condensate compared to its actual value --
constitutes the conceptual foundation for the practical version of OPE.
In this version $\mu$ does not appear explicitly in the coefficient
function of the lowest-dimension operator; formally this corresponds
to setting $\mu =0$ there. The fact of the numerical smallness
is not an obvious property of QCD and is to be cross-checked in any
new situation. So far it turns out that it is always valid, for reasons
which are not completely understood, see the reprint volume cited in
\cite{NSVZ}. We repeat, however, that in principle one should
calculate the Wilson
coefficients by evaluating relevant Feynman integrals with an
explicit  infrared cutoff in the propagators.

To further facilitate understanding of this subtle aspect let us discuss
this general strategy, which we usually stick to, in a particular
example, namely the first sum rule, Eq.
(\ref{SR!}). For simplicity we will forget about the operators that
depend on
spin of the heavy
quark since they do not mix with the leading operator and are
irrelevant for the aspect under discussion. The
sum rule (\ref{SR!})
with the perturbative corrections added takes the form
\begin{equation}
F_{B\rightarrow D^*}^2 + \sum_{i=1,2,...}F_{B\rightarrow {\rm excit}}^2
=\xi_A -\frac{1}{3}\frac{\mu_G^2}{m_c^2}
-\frac{\mu_{\pi}^2-\mu_G^2}{4}\left(
\frac{1}{m_c^2}+\frac{1}{m_b^2}+\frac{2}{3m_cm_b}
\right)\;\;;
\label{S2R}
\end{equation}
nonperturbative corrections are then represented only by the
kinetic
operator $\vec{\pi}^2$. For a consistent
calculation of the perturbative correction factor
$\xi_A$ we need to introduce an infrared renormalization point
$\mu\ll
m_c$ and use $\xi_A(\mu)$ in Eq.\hspace*{.1em}(\ref{S2R}); the
operators entering this sum rule are
then normalized at the scale $\mu$. It is worth noting that without
introducing
$\mu$ the sum rule written above, strictly speaking, has little sense
because
the sum over excitation will diverge in the ultraviolet (see below). If
a
renormalization point is set, the sum will run only over states with
excitation energy below $\mu$. Keeping in mind the explicit
$\mu$-dependence of $\xi_A$,
it is clear that it cannot be equal to $\eta_A^2$ which is calculated
without
an infrared cutoff and is thus $\mu$-independent. In reality,
however, the
major part of both $\eta_A^2$ and $\xi_A(\mu)$ comes in the heavy
quark limit
from the momenta $\sim m_Q$ and therefore they must be similar.
Actually,
one even do not need to calculate the perturbative factor $\xi_A$
anew as long
as $\eta_A^2$ is known.

To determine $\xi_A$ one notes that to any particular order in
perturbation
theory the relation holds
\begin{equation}
\eta_A^2=\xi_A(\mu)\mid_{\mu=0}\;\;;
\label{e1}
\end{equation}
in fact it is merely the definition of $\eta_A$. It can be formally
obtained
considering the sum rule (\ref{S2R})
in the perturbation theory in the particular order (we
will assume the first order in $\alpha_s$ in what follows). In this
approximation the spectrum of particles is given by quarks and
gluons, and
matrix elements of operators, in particular $\mu_\pi^2$,
are nothing but ``perturbative one loop" contributions considered
above. The elastic
contribution is given by the exclusive $b\rightarrow c$ probability
whereas the
sum over excitation is represented by the process $b\rightarrow c+g$
whose
probability is proportional to $\alpha_s$. Then, at $\mu=0$ one gets
the
relation (\ref{e1}) because other terms vanish.

To determine $\mu$ dependence of $\xi_A$ we can use the same technique as in
Sect.~4.4, namely, consider the perturbative analog of the sum rule
(\ref{S2R}):
$$
\eta_A^2\,+\,\frac{1}{3}
\sum_k\;\int_{\omega<\mu} \;\frac{d^3k}{2\omega(2\pi)^3}\;
\frac{1}{4m_cm_b} |\langle cg | \bar c \gamma_k\gamma_5 b| b\rangle |^2\;=
$$
\begin{equation}
=\;\xi_A(\mu)-\frac{\left(\mu_\pi^2\right)_{\rm pert}}{4}
\left( \frac{1}{m_c^2} +  \frac{1}{m_b^2} +\frac{2}{3m_cm_b}\right)
\;\;.
\label{183a}
\end{equation}
The gluon emission amplitude is now
\begin{equation}
\langle c g | \bar c \vec \gamma \gamma_5 b| b\rangle \;=\; g_s\,
\bar b \,t^a\left[\left(\vec \epsilon\,^a \times\vec n \right)\times
\vec\sigma\,
\left(\frac{1}{2m_c}+\frac{1}{2m_b}\right)\,+\,i\left(\vec \epsilon\,^a
\times \vec n \right)\,
\left(\frac{1}{2m_c}-\frac{1}{2m_b}\right)\right]\,b\;\;.
\label{184a}
\end{equation}
Here $\vec n=\vec k/|\vec k\,|$. Using the expression (\ref{121c}) for
$(\mu_\pi^2)_{\rm pert}$ we immediately find that
\begin{equation}
\xi_A(\mu)\simeq \eta_A^2 +
\frac{2\alpha_s}{3\pi}\mu^2\left(\frac{1}{m_c^2}+
\frac{1}{m_b^2}+\frac{2}{3m_cm_b}\right)\;\;.
\label{mucorrect}
\end{equation}
Similar situation occurs
in the sum rule for the vector current (the time component)
where both
the coefficient in front of $\mu_\pi^2$ and the gluon emission
probability are
proportional to $(1/2m_b-1/2m_c)^2$.

The scale dependence of the strong coupling $\alpha_s$ in the above
expressions
is not
essential in the one loop calculations; if higher loops are accounted
for,
the sum of all
Feynman integrals with the particular infrared cutoff $\mu$ will
automatically have a suitable form to give $\log$s of the ratio of
$m/\mu$ necessary to convert
$\alpha_s(m)$ appearing in the perturbative calculations of $\xi$,
into
$\alpha_s(\mu)$ in terms responsible for $\mu^2$ corrections.

In principle, one
can now utilize Eq. (\ref{mucorrect}) to use the exact perturbative
value
of $\xi_A(\mu)$ in Eq.~(\ref{S2R}), which differs from $\eta_A^2$ for
$\mu\ne 0$. Obviously, $\xi_A(\mu)-\eta_A^2$ in
Eq.~(\ref{mucorrect})
contains two
pieces, one given by the perturbative one loop sum over excitation
calculated with the upper cutoff $\mu$, and the second representing
the
``perturbative one loop'' contribution to $\mu_\pi^2(\mu)$. It is clear
that
one can formally subtract these two terms from the sum over
excitations in the
left hand side of the sum rule~(\ref{S2R}) and from
$\mu_\pi^2(\mu)$ term in
the right hand side, respectively, and then use as the perturbative
factor
$\xi_A$ its one loop value at $\mu=0$, i.e. $\eta_A^2$. It is just what
one
does following the routine practice of OPE where the perturbative
coefficient is calculated without an explicit infrared cutoff. In such a
case
$\mu_\pi^2$ entering the ``practical'' form of the sum rule can be
merely
understood as
\begin{equation}
\mu_\pi^2 \rightarrow \mu_\pi^2(\mu)-
\mu^2\frac{d\mu_\pi^2}{d\mu^2}
\label{1loop}
\end{equation}
and, simultaneously,
\begin{equation}
\sum_{\epsilon\le \mu} F^2_{B\rightarrow {\rm excit}} \rightarrow
\sum_{\epsilon\le \mu} F^2_{B\rightarrow {\rm excit}}\;-\;
\frac{1}{3}
\sum_k\;\int_{\omega<\mu} \;\frac{d^3k}{2\omega(2\pi)^3}\;
\frac{1}{4m_cm_b} |\langle cg | \bar c \gamma_k\gamma_5 b| b\rangle |^2\;\;.
\label{excit}
\end{equation}
The normalization point $\mu$ is chosen high enough to ensure the
approximate
duality of the parton computed probabilities to real hadronic ones
above
$\mu$. Then one can remove the upper limit in the difference in the
right hand
side of the above equation and assume that the sum runs over all
energies: the
region above $\mu$ cancels out automatically.

The expression in the right hand side of Eq.~(\ref{1loop}) is nothing
but a
linear (in $\mu^2$)
extrapolation
of the matrix element from the point $\mu$ to $\mu=0$. Formally, it
is
independent on $\mu\:$: the dependence appears in terms
proportional to
$\alpha_s^2$. Therefore one can take any value of $\mu$ as long as
$\Lambda_{\rm
QCD}\ll \mu \ll m_c$.
At the same time, the sum over excitations now is to be
understood, strictly speaking,
as the one from which the small one loop perturbative gluon
probability is subtracted.

A similar analysis can be literally extended to the two, three, and in
principle, to
any finite loop calculation of $\xi_A$. We would like to stress once
again, though, that this procedure has no theoretical justification and
does
not carry any physical meaning. We dwell on it only for the reason
that it
literally corresponds to the routine procedure used in numerical
calculations;
using $\mu=0$ not only does not allow to use consistently OPE, but
would lead
also to numerical problems in higher orders.

Now let us turn from this rather general theoretical discussion to
more
practical questions related to heavy flavor decays. Up to now
nonperturbative effects have been discussed in detail through
corrections
of order $1/m_Q^3$. Some effects, like the invariant mass of the final
hadronic state in the decays considered above,
have corrections starting at order $1/m_Q$ and are expressed via the
parameter $\overline{\Lambda}$ (see also Ref.~\cite{WA}). As
pointed out in
Ref. \cite{15}
and
illustrated in the
present paper, their effects are determined by Feynman integrals
which
have a linear behavior in the infrared region; the corresponding IR
effects
are not expressed in terms of matrix elements of any local
operator.

The inclusive widths of heavy flavor particles {\em are} obtained
from an expansion in
local operators, and corrections start with terms scaling like
$1/m_Q^2$.
These are described by two universal operators -- the
chromomagnetic
one, $\bar Q
\,\frac{i}{2}\sigma G \,Q$, and the kinetic operator, $\bar Q
\,(i\vec{D})^2 \,
Q$. The
natural normalization scale for them is given by $\mu\simeq m_Q$,
at least if one neglects the mass of the final
state quark(s) \footnote{It
is important that for the hyperfine splitting in heavy mesons which
allows
one to extract experimentally the chromomagnetic matrix elements,
the same
normalization point emerges.}. One cannot however use directly this
high
normalization point because then the matrix elements of high
dimension
operators will scale like $m_Q$ to the corresponding power due to
purely
perturbative contributions,
and instead of an expansion in $1/m_Q$ one would obtain
the suppression of the higher order terms only as some powers of
$\alpha_s(m_Q)$. To obtain
the real power expansion one must evolve these operators down to
a scale
$\mu$ which is to be much smaller than $m_Q$ but still much larger
than
$\Lambda_{\rm QCD}$. The expansion one arrives at in this way
runs,
strictly
speaking, in
powers of $\mu/m_Q$.

The present state of the art in this kind of calculations is limited only
by one loop corrections
to Wilson coefficients which -- apart from the
chromomagnetic operator that has a logarithmic renormalization --
are
calculated (or even typically borrowed from old QED calculations)
without
an infrared cutoff. The same refers to the corrections to weak
currents used
in the present paper.
The analysis above suggests therefore
that the value of the kinetic term
$\mu_\pi^2$ can be understood in the corresponding expressions as
a linear
extrapolation to $\mu=0$. This problem does not arise at all
for the chromomagnetic
operator $O_G$ whose mixing with the leading one, $\bar Q Q$,
appears only in
the next
order in $1/m_Q$ due to the fact that it is not a spin singlet; no
double
counting
occurs for this reason. From a practical viewpoint, because the
value of
$\mu_\pi^2$ is basically unknown yet, it does not make a big
difference
at present to prefer this or an alternative definition. Let us note
in
passing
that it is quite probable that the QCD sum rule estimates determine a
similar
quantity extrapolated to $\mu=0$ because no
explicit infrared
cutoff in the integrals is introduced; though this question
definitely
deserves a more careful analysis if one wants to go beyond the
accuracy of
``practical'' OPE.
\vspace*{.2cm}

The issue of an accurate understanding of the definition of matrix
elements of
operators becomes important when one turns to the real
practical
bounds on physical observables of the type discussed in this paper.
For example,
extrapolating the operator
$\bar Q \,\vec{\pi}^2\, Q$ to the zero renormalization point
implies a subtraction of a positive
quantity which, in
principle, might have even changed
the sign of the matrix element.
To state it differently: one may be
concerned
whether the inequality (\ref{inequ}) survives the extrapolation of
$\mu_\pi^2$
to a low point,  which is often assumed implicitly. We shall
argue now that
this effect is too small numerically and cannot upset the bound
we used.

To see it, let us consider the reasonably high normalization point
$\mu\simeq
1\,\mbox{GeV}$. Using the explicit estimate of the renormalization
point
dependence
of operator $\bar Q \,\vec{\pi}^2\, Q$ in Eq. (\ref{mixing}) and
assuming $\alpha_s(\mu) \simeq 0.36$ one readily
obtains
that the amount one may need to subtract from $\mu_\pi^2(\mu)$
constitutes at
most $0.15\,\mbox{GeV}^2$, a value that does not exceed the
theoretical
uncertainties in the existing estimates of $\mu_\pi^2$. Most probably
this
number overestimates the real contribution to be subtracted,
because approximate duality of the
perturbative corrections is expected to start earlier; moreover,
the perturbative coefficient functions ($\eta_A$, $\eta_V$,
corrections to
inclusive widths etc.) are evaluated in one loop
using a smaller
value of
$\alpha_s\simeq \alpha_s(m_Q)$.
The second effect, though formally of higher order in
$\alpha_s$, is
too transparent physically to raise doubts that a more realistic
estimate of effects of potential double counting corresponds to
using $\alpha_s(m_c,m_b)$ rather than
$\alpha_s(1\,\mbox{GeV})$ above.

At the same time, as emphasized in Sect.~4.4, if the
normalization
point is introduced {\em via} the upper bound in the integral over
the energy of
the excited states, the inequality $\mu_\pi^2 > \mu_G^2$ holds for
any
normalization point (in other words, such regularization does not
violate the
positivity of the Pauli operator for the spinor quark). Therefore, it
is
legitimate to take the normalization point as low as
$1\,\mbox{GeV}$.
In this
case,
obviously, one deals with $\mu_G^2$ normalized at this low point as
well, and
it is known \cite{fal}
that the perturbative evolution {\em increases} its value
toward
lower $\mu\,$! In our estimates we took  $\mu_G^2$ directly from
the
hyperfine splitting of $B$ and $B^*$, therefore that value
corresponded to
$\mu\simeq 4.5\,\mbox{GeV}$. Apparently its hybrid $\log$
enhancement
would safely
make up for the relatively insignificant subtraction of the
``perturbative''
contribution to $\mu_\pi^2$. Based on these arguments we have
stated in the
previous paper \cite{SUV} that the inequality
$$
\mu_\pi^2 > \mu_G^2
$$
must survive in QCD, in
spite of
recent claims \cite{claims} that it cannot hold true.

It is instructive to trace how this inequality works at different
scales
$\mu$. Most trivially it is fulfilled when $\mu$ is taken
parametrically
large. Then $\mu_\pi^2$ contains large {\em positive} perturbative
piece of
the order of $\frac{\alpha_s}{\pi}\mu^2$ that grows faster than
any possible change in $\mu_G^2$ having no additive
renormalization, even
if the hybrid anomalous dimension of the latter were negative.

A more interesting consideration emerges when one wants to push
$\mu$ toward
lower values. Using the naive one loop expression for the evolution
of
$\mu_G^2$ corresponding to the hybrid anomalous dimension
$\gamma_G=3$
one would obtain an arbitrarily large value for
$\mu_\pi^2$ which, of
course, makes little sense. The answer to this apparent paradox is
rather
obvious, especially if one looks at a hypothetical zero recoil
excitation
curve (for the external current $\bar{c} \gamma_i b\:$)
similar to the one depicted in Fig.~4. The real
evolution of the difference
$\mu_\pi^2 - \mu_G^2$ at $m_q\gg m_Q$,
according to the sum rule (\ref{repres}), is
given by the
decay probability occurring at energy $\epsilon=\mu$; obviously the
latter in
no way is given at low $\epsilon$
by the simple perturbative formulae based on the strong coupling
$\alpha_s$
with the Landau pole, and rather stays finite at any $\mu$. In other
words, the evolution of $\mu_G^2$ is to be smooth even
when one approaches the strong
coupling regime, and no formal contradiction emerges.

\section{Conclusions}

In the present paper we have addressed weak transitions between
heavy
quarks from
the ``inclusive'' side most suitable theoretically for applying the
technique
of the Wilson OPE. This analysis naturally extend the
consideration outlined in Ref. \cite{9}
which concerned the heavy quark
distribution
function relevant for the decays in the limit of small velocity for
the
final
state hadron system. It has been demonstrated that a few sum rules
discussed so far
in the literature are in fact dispersion relations for the
moments of
a single distribution function, considered in different orders in
$1/m_Q$
and in
different kinematical regimes.
We have shown how the expansions of HQET can consistently be
obtained from
QCD using this strategy, and in this way illustrated that such
natural
assumptions as ``global duality'', that usually are attributed only
to
the
inclusive width calculations, are in fact necessary ingredients in
{\em
any}
consistent model-independent treatment and, in particular, are
implicitly
used in HQET as well.

The analysis of the sum rules proved to be very instructive in
elucidating
the important fact that has usually been neglected in HQET -- the
necessity
of introducing an explicit infrared normalization point $\mu$
ensuring true
separation of low and high momentum physics, which {\em cannot}
be set to
zero. The consistent application of this approach leads to the fact
that
all
nonperturbative parameters, including $\overline{\Lambda}$, cannot
be
sensibly
defined as
universal constants, but rather depend explicitly on the
normalization
point.
This has been previously mentioned in our paper \cite{9} and
discussed in
detail in Refs.~\cite{15,beneke}. Here we gave a physical
illustration
of how
it works analyzing possible constructive phenomenological
definitions of
corresponding quantities in the presence of radiative corrections. In
this way we have
supplemented the previous calculations by estimates of the
dependence of the kinetic energy operator
$\bar Q \,\vec{\pi}^2\, Q$ on the renormalization point.

The fact that such ``purely nonperturbative'' objects like the pole
mass of
the heavy quark, $\overline{\Lambda}$, ``purely nonperturbative''
distribution
function of
heavy quarks routinely used in HQET
are incompatible with a consistent OPE-based approach and are
ill-defined theoretically calls for the clarification of how the
known
results on nonperturbative corrections in HQET must be interpreted.
This
does not mean of course that the concrete calculations that have been
done so
far are irrelevant, and we formulated the way in which they are to
be
understood for a few typical examples.

As a practical application of our sum rules we have derived a model
independent
lower bound on the deviation of the exclusive axial form factor
$F_{B\rightarrow
D^*}$ of the $B\rightarrow D^* + l\,\nu$ decay at zero recoil --  a
process that
for a
long time has been believed
to give the best theoretical accuracy to determine $|V_{cb}|$, and
estimated
a reasonable `central' value, $F_{B\rightarrow D^*}(\vec{q}=0)\simeq
0.9$ .
The deviation appears to be essentially
larger than the estimates that had been obtained before from
model
calculations based on standard HQET, and apparently better agree
with quite
general expectations about the size of corrections to the Heavy Quark
symmetry for charmed particles. On the other hand, the theoretical
clarification of the notion of the heavy quark mass made in recent
papers
\cite{15,beneke} suggests that the most accurate theoretically way to
determine the CKM matrix elements for heavy quark decays is using
the
{\em inclusive} semileptonic widths. These results have been
reported in
Ref. \cite{SUV}.

It is worth clarifying in this respect
that in our estimates of the exclusive form factors of the
$b\rightarrow c$
transitions we
consistently took into account terms through order $1/m_Q^2$ and
discarded
effects that scale like $1/m_Q^3$. The parameter $1/m_c$
is actually not
very small and even the second order corrections are as large as
$10\%$
here; therefore one can expect sizable relative corrections for real
form factors due to higher order terms. In particular, this applies
to the
model-independent upper bound for $F_{B\rightarrow D^*}$. Our
result
is
strict in the
sense that it holds for corrections through terms of order $1/m_Q^2$
that
have been
addressed in the literature so far.

One of the sum rules at zero recoil enabled us to rederive a model
independent
lower bound on the value of the kinetic energy operator in $B$
mesons,
$$
\mu_\pi^2 =\frac{1}{2M_B}\langle B|{\bar b \,\vec{\pi}^2\,
b}|B\rangle
{\mathrel{\rlap{\lower4pt\hbox{\hskip1pt$\sim$}}\raise1pt\hbox{$
>$}}}
\frac{3}{4} \left(M_{B^*}^2-M_B^2\right)
$$
in a way that clearly showed its physical relevance even in real QCD,
and not
only in the approximate framework of quantum mechanical
consideration, as it is sometimes stated.

\vspace*{0.5cm}

{\bf ACKNOWLEDGMENTS:} \hspace{.4em} N.U. gratefully
acknowledges
the creative
atmosphere and hospitality at Theoretical Physics Institute of
University of
Minnesota, and the exchange of ideas related to the subject
of this paper with V. Braun. We would like to thank  B. Blok,  A.
Falk, T. Mannel,
M. Savage and especially L.~Koyrakh and M.~Voloshin for useful  discussions.
This work was supported in part by the National Science Foundation
under the grant number
PHY 92-13313 and by DOE under the grant
number DE-FG02-94ER40823.

\section{Appendix}
\renewcommand{\theequation}{A.\arabic{equation}}
\setcounter{equation}{0}

Let us derive the expression (\ref{QML})
$$
\frac{\overline{\Lambda}}{2}=
 \sum_n\;\frac{1}{3}\frac{|\langle B| \pi_i|n\rangle |^2}
{E_n-M_B}
$$
in ordinary quantum mechanics. We will
use the nonrelativistic normalization of states below.

It is easy to see that
in simple potential description of the heavy hadron as a two body
system with
the reduced mass $m_r\simeq m_{\rm sp}$ the
sum on the right hand side is given just by the half of this mass
(which in this
case is
identified with $\bar\Lambda$).
Indeed,
$$
-\;\sum_n\;\frac{|\langle B|\vec\pi\vec v |n\rangle |^2}
{E_n-M_B}
$$
represents the order $\vec v\,^2$ correction to the ground state
energy,
produced by a perturbation
$$
\delta H = \vec{\pi} \vec{v} \;\;.
$$
On the other hand
$$
H + \delta H = H(\vec{\pi}+ \vec{v})- \frac{m_r\vec v\,^2}{2}\;\;.
$$
The eigenvalues of the Hamiltonian given by the first term on the
right hand
side are the same as for the unperturbed one, which leads to the
relation
\begin{equation}
\sum_n\;\frac{|\langle B|\vec{\pi}\vec{v}|n \rangle |^2}
{E_n-M_B}=\frac{m_r \vec v\,^2}{2}
\label{53c1}
\end{equation}
which coincides with Eq.~(\ref{QML}) if $\bar\Lambda=m_{\rm sp}$.
In reality, of course, Eq.\hspace*{.1em}(\ref{QML}) is more general
and accounts
also for the binding energy. To show it we can apply the following
general
consideration.

First, let us note that the relation similar to
Eq.~(\ref{53c1}) can be obtained in a more general
way. Namely, for any system whose Hamiltonian depends on the
heavy quark
momentum $\pi$ in the non-relativistic quadratic way,
\begin{equation}
{\cal H}= \frac{\vec\pi^2}{2m_Q}+ {\cal H}_{\rm light}(\vec x_Q,
\{x_{\rm light}\})
\label{qm4}
\end{equation}
one has the exact commutation relation
\begin{equation}
\vec\pi=im_Q[{\cal H},\vec x_Q]
\label{qm5}
\end{equation}
where $\vec\pi$ and $\vec x_Q$ are the operators of the heavy quark
momentum and coordinate. Then one writes
$$
\sum_n\;\frac{|\langle 0|\pi_k|n\rangle |^2}{E_n-E_0}=
i\frac{m_Q}{2}\sum_n\;\left(\frac{\langle 0|\pi_k |n\rangle
\langle n|[{\cal H}, x^Q_k]|0\rangle}{E_n-E_0}+
\frac{\langle 0|[{\cal H},x^Q_k] |n\rangle \langle n|\pi_k |0\rangle}
{E_n-E_0}
\right)=
$$
$$
=i\frac{m_Q}{2}\sum_n\;\left(\langle 0|\pi_k |n\rangle
\langle n|x^Q_k|0\rangle-\langle 0|x^Q_k |n\rangle \langle n|\pi_k
|0\rangle
\right)=
$$
\begin{equation}
=\frac{m_Q}{2}\sum_n\;\langle 0|i[\pi_k,x^Q_k] |0\rangle =
\frac{m_Q}{2}\;\;.
\label{qm6}
\end{equation}
Note that we do not sum over spatial index $k$ in this equation; it is
valid
for arbitrary $k$. Below in this section we will assume that the
summation over
$k$ is {\em not} performed.
We emphasize that this relation is rigorous for any
system as long as the heavy quark momentum $\vec \pi$ enters the
Hamiltonian quadratically.

Now we apply Eq.~(\ref{qm6}) to $B$ meson. We get
\begin{equation}
\frac{m_Q}{2}=\sum_n\;\frac{|\langle B|\pi_k |n\rangle |^2}
{E_n-M_B}\;\;.
\label{qm7}
\end{equation}
The sum over intermediate states runs over all possible hadronic
states which
are marked, in particular, by their total momentum $\vec\pi$. The
matrix
elements, however, do not vanish only for zero momentum transfer:
\begin{equation}
\langle B(\vec{p}=0)|\pi_k |n(\vec{p})\rangle= \langle
B|\pi_k |n\rangle_{QM} \frac{(2\pi)^3\delta^3(\vec{p})}{\sqrt{V}}
\label{qm8}
\end{equation}
where $V$ is volume.
When squared only the equal momentum matrix elements are
present, and the
factor
$(2\pi)^3\delta^3(0)=V$ cancels against $1/\sqrt{V}$ normalization
of states
associated with the continuous spectrum of total momentum. The
situation
requires more care when the state $n$ is $B$ meson; the energy
denominator
$E_n-M_B=\vec{p}\,^2/2M_B$ has a pole in this case
when integrated over $d^3\vec{p}\,$, and
this singularity must be treated properly. We write therefore
\begin{equation}
\frac{m_Q}{2}=\sum_{n\ne B}\;\frac{|\langle B|\pi_k |n\rangle_{QM}
|^2}
{E_n-M_B} + \int \frac{d^3\vec{p}}{(2\pi)^3}\, \frac{|\langle
B(\vec{p}=0)|\pi_k |B(\vec{p})\rangle|^2}{\vec{p}\,^2/2M_B}\;\;.
\label{qm9}
\end{equation}
The matrix elements in the last term can be represented in the form
\begin{equation}
\langle B(\vec{p}=0)|\pi_k |B(\vec{p})\rangle = \frac{m_Q}{M_Q}
\langle B(\vec{p}=0)|P_k |B(\vec{p})\rangle
\label{qm10}
\end{equation}
where $P$ is the total momentum operator. To calculate the integral
over
the momentum of $B$ meson we can again use Eq.~(\ref{qm6}):
\begin{equation}
\int \frac{d^3\vec{p}}{(2\pi)^3}\, \frac{|\langle
B(\vec{p}=0)|P_k
|B(\vec{p})\rangle|^2}{\vec{p}\,^2/2M_B}=\frac{M_B}{2}\;\;;
\label{qm11}
\end{equation}
this is nothing but the relation (\ref{qm6}) for ``quantum mechanics''
of free
$B$ mesons considered as elementary particles. Combining
Eqs.~(\ref{qm9})-(\ref{qm11}) we finally get
$$
\frac{m_Q}{2}= \sum^{}{'}\;\frac{|\langle B|\pi_k |n\rangle_{QM} |^2}
{M_n-M_B}\;+\;\frac{m_Q^2}{2M_B}\;\;,
$$
or
\begin{equation}
\sum^{}{'}\;\frac{|\langle B|\pi_k |n\rangle_{QM} |^2}
{M_n-M_B}= \frac{m_Q}{M_B}\frac{M_B-m_Q}{2} \simeq
\frac{\bar\Lambda}{2}\;\;.
\label{qm12}
\end{equation}
This equation is clearly the relation (\ref{QML}) we are looking for; if
one
sums over $k$ additional factor $3$ emerges to match the exact
coefficient in
the latter.

\newpage

{\bf Figure Captions}

\vspace{0.5cm}

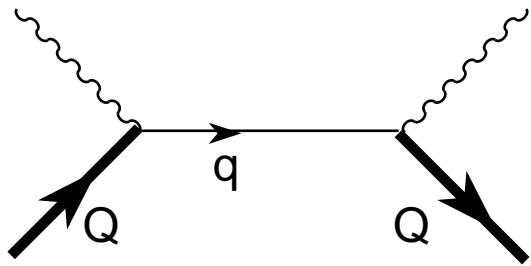
Fig. 1. The tree graph for the transition operator.

\vspace{0.3cm}

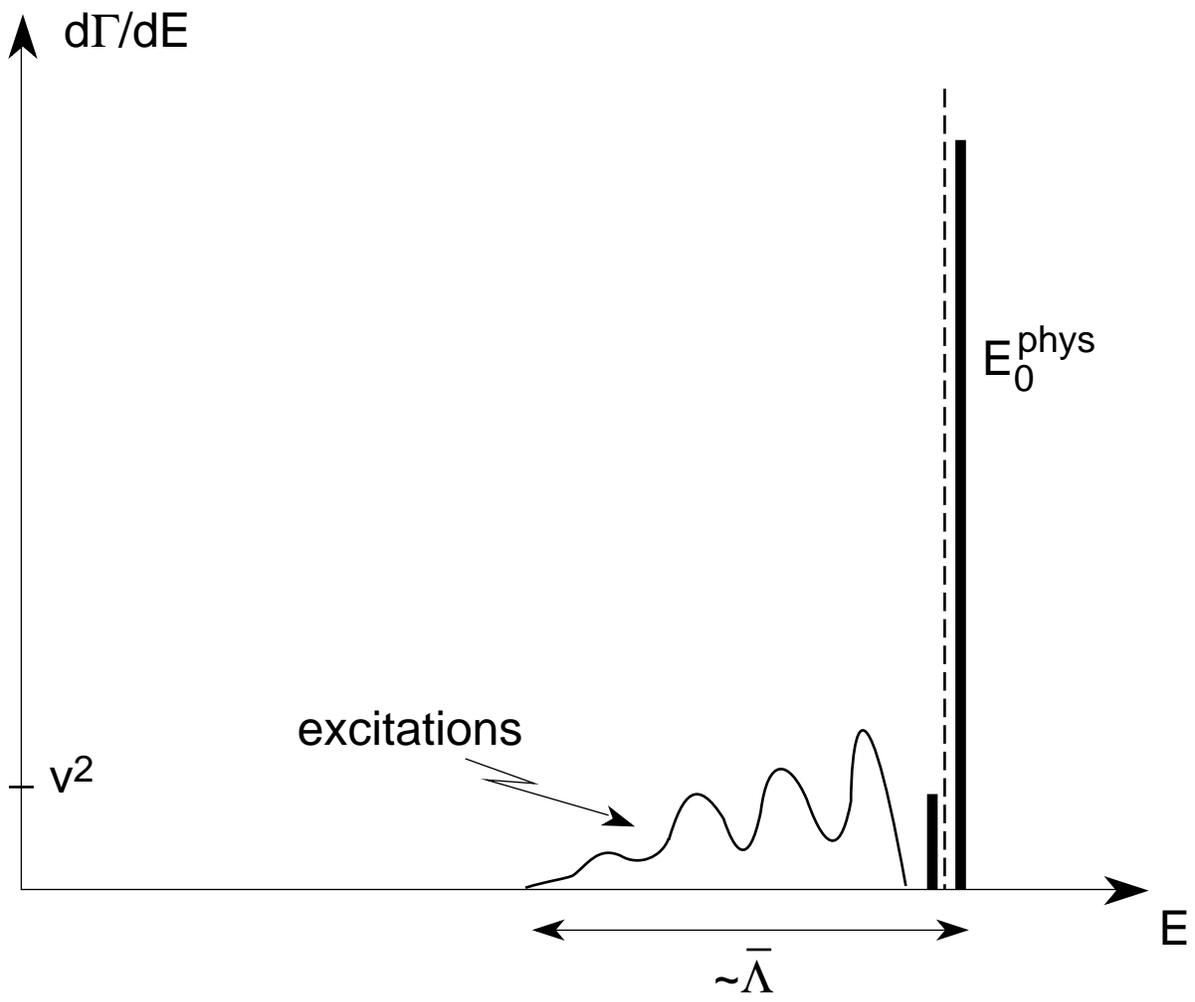
Fig. 2. A qualitative picture of the spectrum $d\Gamma /dE_\phi$ in
the $H_Q\rightarrow\phi X_q$ with ${\cal O}(\vec v\,^2)$ terms included.
The monochromatic line of the quark transition $Q\rightarrow\phi
q$
(the dashed line
at $E=E_0$)
is dual to the physical line corresponding
to the elastic decay $H_Q\rightarrow H_q\phi$ at $E=E_0^{phys}$
plus
a shoulder
due to the transitions to the excited states $H_Q\rightarrow\phi
H_q^*$. The
height of the shoulder is $\sim \vec v\,^2$. Hard gluons are neglected.

\vspace{0.3cm}

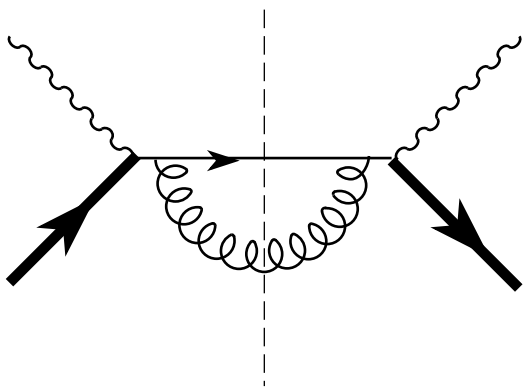
Fig. 3. The diagram responsible for the
one-gluon correction in the energy distribution
$$
\frac{d\Gamma (Q\rightarrow \phi q + \,\, {\rm gluon})}{dE_\phi} .
$$

\vspace{0.3cm}

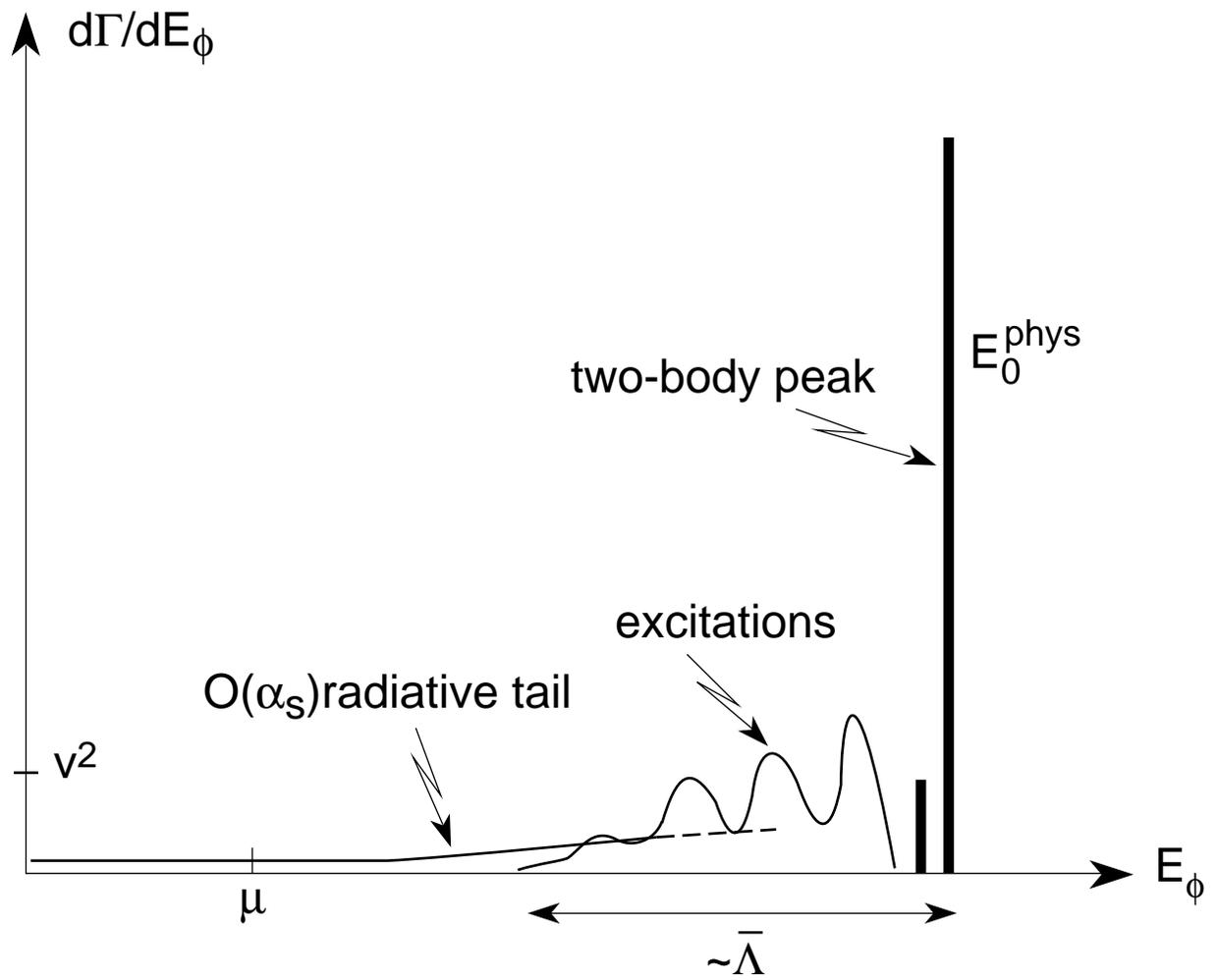
Fig. 4. A sketch of the energy spectrum $d\Gamma /dE_\phi$
with ${\cal O}(\alpha_s)$ radiative tail included.

\vspace{0.3cm}

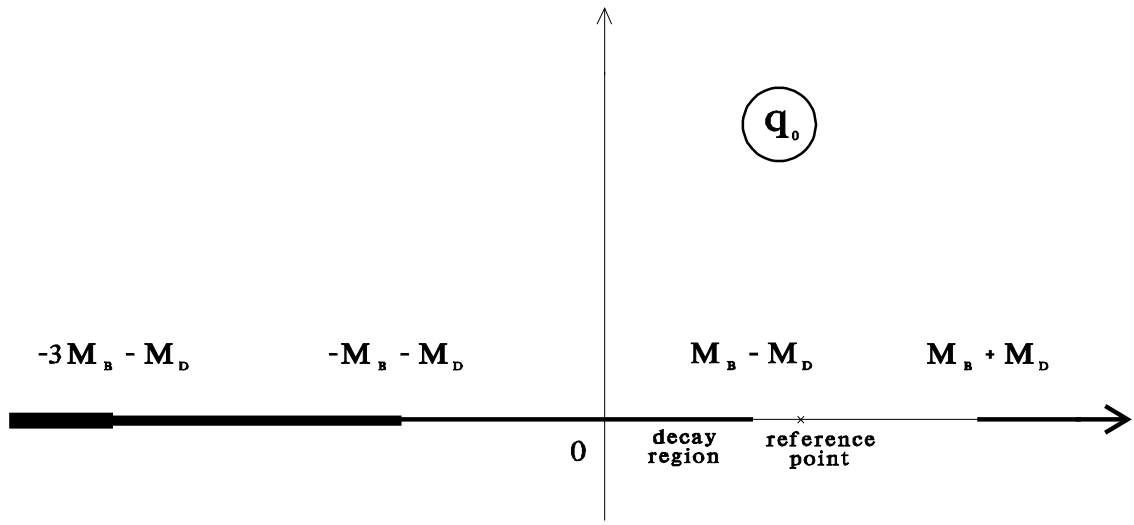
Fig. 5. The cuts of $h_i(q_0)$ in the complex $q_0$ plane at $\vec q=0$.
The point
$q_0 \simeq M_B+M_D$
marks the beginning of three cuts: two of them originate at the mass of $B_c$
meson and the third starts at $M_B+M_D$. The three cuts can be differentiated
considering $\vec q\ne 0$.
Only a part of one cut at
$0<q_0<M_B-M_D$ is relevant to the decay process
under consideration. The masses are not shown in a real scale.

\end{document}